\documentclass[fleqn,usenatbib]{mnras}

\usepackage{newtxtext,newtxmath}

\usepackage[T1]{fontenc}

\DeclareRobustCommand{\VAN}[3]{#2}
\let\VANthebibliography\thebibliography
\def\thebibliography{\DeclareRobustCommand{\VAN}[3]{##3}\VANthebibliography}

\usepackage{graphicx}	
\usepackage{amsmath}	
\usepackage{anyfontsize}

\newcommand{\mpc}{\,h^{-1}{\rm{Mpc}}}
\newcommand{\gpc}{\,h^{-1}{\rm{Gpc}}}
\usepackage{dutchcal}
\usepackage{comment}
\usepackage{hyperref}
\DeclareMathAlphabet{\mathcal}{OMS}{cmsy}{m}{n}

\usepackage{glossaries}
\glsdisablehyper
\newacronym{paus}{PAUS}{Physics of the Accelerating Universe Survey}
\newacronym{sdss}{SDSS}{Sloan Digital Sky Survey}
\newacronym{cfhtls}{CFHTLS}{Canada-France-Hawaii Telescope Legacy Survey}
\newacronym{cfhtlens}{CFHTLenS}{Canada-France-Hawaii Telescope Lensing Survey}
\newacronym{kids}{KiDS}{Kilo-Degree Survey}
\newacronym{des}{DES}{Dark Energy Survey}
\newacronym{hsc_ssp}{HSC SSP}{Hyper Suprime-Cam Subaru Strategic Program}
\newacronym{hsc}{HSC}{Hyper Suprime-Cam}
\newacronym{lsst}{LSST}{Legacy Survey of Space and Time}
\newacronym{desi}{DESI}{Dark Energy Spectroscopic Instrument}
\newacronym{wht}{WHT}{William Herschel Telescope}
\newacronym{fwhm}{FWHM}{Full Width at Half Maximum}
\newacronym{nb}{NB}{narrow-band}
\newacronym{bb}{BB}{broad-band}
\newacronym{ra}{R.A}{Right ascension}
\newacronym{gama}{GAMA}{Galaxy And Mass Assembly}
\newacronym{vipers}{VIPERS}{VIMOS Public Extragalactic Redshift Survey}
\newacronym{vvds}{VVDS}{VIMOS VLT Deep Survey}
\newacronym{paudm}{PAUdm}{PAU data management}
\newacronym{pic}{PIC}{Port d’Informació Científica}
\newacronym{esa}{ESA}{European Space Agency}
\newacronym{dune}{DUNE}{Dark Universe Explorer}
\newacronym{space}{SPACE}{Spectroscopic All-Sky Cosmic Explorer}
\newacronym{jwst}{JWST}{James Webb Space Telescope}
\newacronym{lss}{LSS}{large-scale structure}
\newacronym{fom}{FoM}{Figure of Merit}
\newacronym{de}{DE}{Dark Energy}
\newacronym{dm}{DM}{dark matter}
\newacronym{vis}{VIS}{Visible Imaging Instrument}
\newacronym{nisp}{NISP}{Near Infrared Spectrometer and Photometer}
\newacronym{ccd}{CCD}{Charge-Coupled Device}
\newacronym{cdpu}{CDPU}{Control and Data Processing Unit}
\newacronym{fov}{FoV}{field of view}
\newacronym{ews}{EWS}{Euclid Wide Survey}
\newacronym{eds}{EDS}{Euclid Deep Survey}
\newacronym{edf}{EDF}{Euclid Deep Fields}
\newacronym{eaf}{EAF}{Euclid Auxiliary Fields}
\newacronym{psf}{PSF}{point spread function}
\newacronym{aegis}{AEGIS}{All-Wavelength Extended Groth Strip International Survey}
\newacronym{cdfs}{CDFS}{Chandra Deep Field South}
\newacronym{goods}{GOODS}{Great Observatories Origins Deep Survey}
\newacronym{sxds}{SXDS}{Subaru/XMM-Newton Deep Survey}
\newacronym{ia}{IA}{intrinsic alignments}
\newacronym{gc}{GC}{galaxy clustering}
\newacronym{sed}{SED}{spectral energy distribution}
\newacronym{nircam}{NIRCam}{Near Infrared Camera}
\newacronym{snr}{SNR}{signal-to-noise ratio}
\newacronym{cp}{CP}{Cosmological Principle}
\newacronym{flrw}{FLRW}{Friedmann-Lemaître-Robertson-Walker}
\newacronym{cmb}{CMB}{Cosmic Microwave Background}
\newacronym{cobe}{COBE}{Cosmic Background Explorer}
\newacronym{wmap}{WMAP}{Wilkinson Microwave Anisotropy Probe}
\newacronym{pm}{PM}{particle-mesh}
\newacronym{hod}{HOD}{halo occupation distribution}
\newacronym{sham}{SHAM}{subhalo abundance matching}
\newacronym{2pcf}{2PCF}{two-point correlation function}
\newacronym{rsd}{RSD}{redshift-space distortions}
\newacronym{ap}{AP}{Alcock-Paczyński}
\newacronym{som}{SOM}{Self Organising Map}
\newacronym{uv}{UV}{ultraviolet}
\newacronym{la}{LA}{Linear Alignment}
\newacronym{nla}{NLA}{Non-Linear Alignment}
\newacronym{tatt}{TATT}{Tidal Alignment Tidal Torquing}
\newacronym{mcmc}{MCMC}{Monte-Carlo Markov Chain}
\newacronym{apo}{APO}{Apache Point Observatory}
\newacronym{lco}{LCO}{Las Campanas Observatory}
\newacronym{vlt}{VLT}{Very Large Telescope}
\newacronym{eso}{ESO}{European Southern Observatory}
\newacronym{2dfgrs}{2dFGRS}{Two-degree Field Galaxy Redshift Survey}
\newacronym{hst}{HST}{Hubble Space Telescope}
\newacronym{cigale}{\texttt{CIGALE}}{Code Investigating GALaxy Emission}
\newacronym{imf}{IMF}{initial mass function}
\newacronym{ir}{IR}{infrared}
\newacronym{viking}{VIKING}{VISTA Kilo-degree INfrared Galaxy}
\newacronym{ssfr}{sSFR}{specific star formation rate}
\newacronym{jk}{JK}{jackknife}
\newacronym{svd}{SVD}{singular value decomposition}
\newacronym{cloe}{CLOE}{Cosmology Likelihood for Observables in Euclid}
\newacronym{dv}{DV}{data vector}
\newacronym{eft}{EFT}{effective field theory}
\newacronym{cosmosis}{CosmoSIS}{COSMOlogical Survey Inference System}
\newacronym{photo-z}{photo-$z$}{photometric redshift}
\newacronym{spec-z}{spec-$z$}{spectroscopic redshift}
\newacronym{los}{LOS}{line-of-sight}
\newacronym{ksb}{KSB}{Kaiser-Squires-Broadhurst}

\title[Measuring IA in deep wide fields]{The PAU Survey: Measuring intrinsic galaxy alignments in deep wide fields as a function of colour, luminosity, stellar mass and redshift}

\author[D.~Navarro-Gironés]{D.~Navarro-Giron\'{e}s,$^{1, 2, 3}$\thanks{E-mail: david.navarro.girones@gmail.com}
M.~Crocce,$^{1, 2}$
E.~Gaztañaga,$^{4,1,2}$
A.~Wittje,$^{5}$
M.~Siudek,$^{6, 1}$
H.~Hoekstra,$^{3}$
\newauthor
H.~Hildebrandt,$^{5}$
B.~Joachimi,$^{7}$
R.~Paviot,$^{8}$
C.M.~Baugh,$^{9, 10}$
J.~Carretero,$^{16, 11}$
R.~Casas,$^{1, 2}$
F.~J.~Castander,$^{1, 2}$
\newauthor
M.~Eriksen,$^{12}$
E.~Fernandez,$^{12}$
P.~Fosalba,$^{1, 2}$
J.~Garc\'ia-Bellido,$^{13}$
R.~Miquel,$^{12, 14}$
C.~Padilla,$^{12}$
P.~Renard,$^{15}$
\newauthor
E.~Sánchez,$^{16}$
S.~Serrano,$^{17, 1, 2}$
I.~Sevilla-Noarbe,$^{16}$
P.~Tallada-Cresp\'{i}$^{16, 11}$
\\
$^{1}$Institute of Space Sciences (ICE, CSIC), Campus UAB, Carrer de Can Magrans, s/n, 08193 Barcelona, Spain\\
$^{2}$Institut d'Estudis Espacials de Catalunya (IEEC), E-08860 Castelldefels (Barcelona), Spain\\
$^{3}$Leiden Observatory, Leiden University, Einsteinweg 55, 2333 CC Leiden, the Netherlands\\
$^{4}$Institute of Cosmology \& Gravitation, University of Portsmouth, Dennis Sciama Building, Burnaby Road, Portsmouth PO1 3FX, UK\\
$^{5}$Ruhr University Bochum, Faculty of Physics and Astronomy, Astronomical Institute (AIRUB), German Centre for Cosmological Lensing, 44780 Bochum, \\
Germany \\
$^{6}$Instituto de Astrofísica de Canarias (IAC), Departamento de Astrofísica, Universidad de La Laguna (ULL), 38200, La Laguna, Tenerife, Spain \\
$^{7}$Department of Physics and Astronomy, University College London, Gower Street, London WC1E 6BT, UK \\
$^{8}$Université Paris-Saclay, Université Paris Cité, CEA, CNRS, AIM, 91191, Gif-sur-Yvette, France \\
$^{9}$Institute for Computational Cosmology, Department of Physics, Durham University, South Road, Durham DH1 3LE, UK.\\
$^{10}$Institute for Data Science, Durham University, South Road, Durham DH1 3LE, UK \\
$^{11}$Port d’Informaci\'o Cient\'ifica (PIC), Campus UAB, C. Albareda s/n, 08193 Bellaterra (Barcelona), Spain.\\
$^{12}$Institut de F\'isica d’Altes Energies (IFAE), The Barcelona Institute of Science and Technology, Campus UAB, 08193 Bellaterra (Barcelona), Spain.\\
$^{13}$Instituto de F\'isica Te\'orica CSIC/UAM, Universidad Aut\'onoma de Madrid, 28049 Madrid, Spain \\
$^{14}$Instituci\'o Catalana de Recerca i Estudis Avan\c{c}ats (ICREA), 08010 Barcelona, Spain.\\
$^{15}$Department of Astronomy, Tsinghua University, Beijing 100084, China\\
$^{16}$Centro de Investigaciones Energéticas, Medioambientales y Tecnológicas (CIEMAT), Avenida Complutense 40, 28040 Madrid, Spain.\\
$^{17}$Satlantis, University Science Park, Sede Bld 48940, Leioa-Bilbao, Spain
}

\date{Accepted XXX. Received YYY; in original form ZZZ}

\pubyear{2015}

\begin{document}
\label{firstpage}
\pagerange{\pageref{firstpage}--\pageref{lastpage}}
\maketitle

\begin{abstract}

We present the measurements and constraints of \gls{ia} in the \gls{paus} deep wide fields, which include the W1 and W3 fields from the \gls{cfhtls} and the G09 field from the \gls{kids}. Our analyses cover 51deg$^{2}$, in the \gls{photo-z} range $0.1 < z_{\mathrm{b}} < 1$ and a magnitude limit $i_{\mathrm{AB}}<22$. The precise \glspl{photo-z} and the luminosity coverage of \gls{paus} enable robust \gls{ia} measurements, which are key for setting informative priors for upcoming stage-IV surveys. For red galaxies, we detect an increase in \gls{ia} amplitude with both luminosity and stellar mass, extending previous results towards fainter and less massive regimes. As a function of redshift, we observe strong \gls{ia} signals at intermediate ($z_{\mathrm{b}}\sim0.55$) and high ($z_{\mathrm{b}}\sim0.75$) redshift bins. However, we find no significant trend of \gls{ia} evolution with redshift after accounting for the varying luminosities across redshift bins, consistent with the literature. For blue galaxies, no significant \gls{ia} signal is detected, with $A_{1}=0.68_{-0.51}^{+0.53}$ when splitting only by galaxy colour, yielding some of the tightest constraints to date for the blue population and constraining a regime of very faint and low-mass galaxies.

\end{abstract}



\glsresetall

\section{Introduction}

Weak gravitational lensing is a key cosmological probe that describes the small distortions that light experiences as it travels through the Universe, due to its interaction with matter inhomogeneities in its path \citep{cosmic_shear_review_1, cosmic_shear_review_2}. This phenomenon enables mapping the total mass distribution in the Universe, comprising both \gls{dm} and luminous matter. However, a significant astrophysical systematic that affects the interpretation of weak gravitational lensing measurements is the \gls{ia} of galaxies.

\gls{ia} arises from the preferred orientation of galaxies due to local gravitational interactions with the surrounding \gls{lss}. This phenomenon has gained relevance in the past decades (see \citealt{IA_overview, IA_guide} for an overview on this topic), as it not only provides valuable insights into the formation and evolution of galaxies \citep{galaxy_formation_review}, but also acts as a contaminant of weak gravitational lensing studies, by mimicking the lensing signal, usually with an opposite sign, thus reducing the overall amplitude of the observed gravitational lensing signal. In the era of precision cosmology and, especially, for stage-IV surveys, such as \textit{Euclid} \citep{Euclid_overview}, the Vera C. Rubin Observatory LSST \citep{LSST} and the Nancy Grace Roman Space Telescope \citep{Roman}, it is crucial to accurately model and measure \gls{ia}, in order not to bias cosmological analyses.

The physical mechanisms that drive \gls{ia} are thought to differ between galaxy types, motivating the commonly observed distinction between red and blue galaxies in \gls{ia} studies. Red galaxies, which mainly correspond to elliptical, pressure-supported galaxies, are thought to be governed by the tidal alignment of their major axis with their host \gls{dm} halo. In contrast, blue spiral galaxies, which correspond to rotationally-supported objects, are driven by their angular momentum, with the momentum axis aligned with that of the \gls{dm} halo, generating quadratic alignments \citep{Hirata_2004}. Related to this scheme, the \gls{nla} model \citep{Bridle_2007}, an extension of the \gls{la} model \citep{Catelan_2001, Hirata_2004} in which the linear matter spectrum is substituted by the non-linear one, enables the \gls{ia} to be explained as a function of the tidal field, while the \gls{tatt} model \citep{Blazek_TATT} allows us to incorporate the tidal torquing effect into the equation. As a consequence, it is expected that the \gls{nla} model is enough to explain the \gls{ia} observed in elliptical galaxies, while the \gls{ia} seen in spiral galaxies requires the higher-order terms from the \gls{tatt} model. On smaller scales, a halo model \citep{Halo_model_IA_Schneider, halo_model_IA} has been proposed, which is able to describe \gls{ia} at scales comparable to those of \gls{dm} halos. Finally, more complex models have been proposed, which include higher-order expansions of the tracers of \gls{ia} \citep{Vlah_2020, Vlah_2021, Bakx_model, Maion_2024}. These models aim to describe \gls{ia} more accurately in the weakly non-linear regime, at the expense of introducing additional parameters.

Observational studies of \gls{ia} are also of extreme importance, since they allow us to quantify the signal for diverse galaxy samples. In particular, many studies have focused on the study of red and bright galaxy samples \citep{LandySzalay_wgp, Observation_red_galaxies_Hirata, Observation_red_galaxies_Okumura, Observation_red_galaxies_Joachimi, Observation_red_galaxies_Johnston, IA_harry, Observation_red_galaxies_Zhou, Unions_IA, Georgiou_2025}, finding strong evidence of positive alignments. However, an open question is the \gls{ia} of blue galaxies, where observations \citep{Observation_red_galaxies_Hirata, Mandelbaum_weight_function, Observation_red_galaxies_Johnston, IA_harry, Georgiou_2025} find \gls{ia} amplitudes consistent with zero, while some simulations (e.g. \citealt{Codis_hydro, Chisari_hydro, Samuroff_hydro}) show some degree of alignment. For red galaxies, there is a consensus in the literature \citep{BOSS_LOWZ_2, Observation_red_galaxies_Joachimi, Observation_red_galaxies_Johnston, KiDS_IA,Samuroff_photometric_correlations, Unions_IA} for a dependence of \gls{ia} on luminosity, which is usually described as a double power law, with more luminous galaxies exhibiting a stronger dependence than fainter ones. Recently, \citet{maria_cristina_fortuna_halo_mass} proposed that this double power law is the result of the double power law in the luminosity-to-halo mass relation. Finding that, in the range they explored, they can describe the dependence of \gls{ia} on halo mass using a single power law, which would imply that the halo mass is the driving force of the \gls{ia} amplitude, as already suggested in \citet{mass_dependence_IA_3}. In terms of redshift evolution, no clear evolution has been found in the literature \citep{Observation_red_galaxies_Joachimi, BOSS_LOWZ_2, Observation_red_galaxies_Johnston, KiDS_IA}. However, in weak lensing studies, the \gls{ia} contribution appears stronger at low redshifts relative to the weak gravitational lensing signal, because the latter decreases at low redshifts. While the majority of observations have focused on relatively bright samples, the \gls{ia} impact for low-mass and low-luminosity objects, which constitute an important fraction of the cosmic galaxy population, remains underrepresented in \gls{ia} studies, limiting our understanding of this effect.

Measurements of \gls{gc} and \gls{ia} require precise knowledge of the distance to the objects being correlated and allow us to constrain the \gls{ia} amplitude with the models described above. Even though \glspl{spec-z} allow us to obtain precise distances, obtaining them is expensive, since only a limited number of objects can be observed at once and the completeness of faint galaxy samples is lower than for photometric galaxy surveys. As an alternative to \glspl{spec-z}, \glspl{photo-z} allow us to measure distances of objects with higher completeness, and of all objects with measured fluxes, at the expense of reducing the precision of the distance estimates. In that sense, the precision of the recovered distances strongly depends on the filters used, with most surveys relying on \gls{bb} filters that reproduce the \gls{sed} of galaxies at limited resolution and recover distance estimates with a $\sim5\%$ uncertainty at the mean redshift analysed in this work (e.g., \citealt{Hildebrandt_2012, Hildebrandt_2021}). The precision of \glspl{photo-z} can be further improved by increasing the number of bands and reducing their width, with so-called \gls{nb} surveys, such as the Physics of the Accelerating Universe Survey (\glsentryshort{paus}\glsunset{paus}; \citealt{PAUS_general}), ALHAMBRA \citep{Alhambra}, LAGER \citep{LAGER} or mini-JPAS \citep{miniJPAS}.

\gls{paus} addresses the complexity of \gls{photo-z} estimation using 40 \glspl{nb}, in combination with other \glspl{bb}, which allow us to better reconstruct the \gls{sed} of galaxies by improving the wavelength accuracy with which features, such as the $4000$ \AA\, break, can be recovered and to detect emission lines. This allows us to obtain an order of magnitude better \glspl{photo-z} than typical \gls{bb} surveys for bright and low-redshift objects, where the fluxes of galaxies present high \glspl{snr} in the \glspl{nb} \citep{photo_z_wide_fields}. Additionally, the large number densities measured by \gls{paus} enable the study of the weakly non-linear regime, covering the gap between limited areas and volumes provided by spectroscopic surveys and larger areas with lower redshift precision provided by photometric surveys. Hence, \gls{paus} enables the study of highly dense intermediate areas, with state-of-the-art \glspl{photo-z}. Moreover, due to its precise redshift estimates, \gls{paus} can focus on fainter objects than other stage-III surveys, such as the Dark Energy Survey (\glsentryshort{des}\glsunset{des}; \citealt{DES_overview}), the Kilo-Degree Survey (\glsentryshort{kids}\glsunset{kids}; \citealt{KiDS}) and the Hyper Suprime-Cam (\glsentryshort{hsc}\glsunset{hsc}; \citealt{HSC_overview}), which usually focus on the objects with better \gls{photo-z} estimates. This is of utmost importance for the upcoming stage-IV surveys, which will cover fainter magnitudes and will need a precise description of \glspl{ia} in order not to bias their cosmological analyses.

Here, we focus on measuring and modelling the \gls{gc} and the \gls{ia} as a function of colour, luminosity, stellar mass and redshift using PAUS. For this purpose, we measure the 3-dimensional galaxy-galaxy and galaxy-shape correlation functions of close galaxies and project them along the \gls{los}. We model these measurements assuming a non-linear galaxy bias \citep{non_linear_galaxy_bias} and the \gls{nla} models, for the \gls{gc} and the \gls{ia}, respectively. We provide \gls{ia} amplitude fits for different galaxy samples, allowing us to extend our knowledge towards lower luminosities and masses.

This paper is structured as follows. In Section~\ref{sec:data}, we introduce the data used in this analysis. Next, in Section~\ref{sec:estimators} and Section~\ref{sec:modelling}, we introduce the estimators we use to measure and model, respectively, the \gls{gc} and \gls{ia} correlation functions. Section~\ref{sec:results} gives our results and discussion. We end with our conclusions in Section~\ref{sec:conclusions}. Throughout this paper, for consistency with previous work \citep{IA_harry}, we assume a flat $\Lambda$CDM cosmology, with $\Omega_{\rm m}$ = 0.25, $\Omega_{\rm b}$ = 0.044, $h=0.69$, $n_{\rm s} = 0.95$ and $\sigma_8 = 0.8$.

\section{Data}\label{sec:data}

This section introduces \gls{paus} and describes the \glspl{photo-z}, the galaxy shapes and the colour separation employed for the measurement of galaxy \gls{ia} in \gls{paus}. It also includes a brief description of the MICE simulation, used in this work to validate some of our methods.

\subsection{PAUS}\label{sec:PAUS}

\gls{paus} is a photometric survey that was carried out at El Roque de Los Muchachos, in the Canary Islands. It used PAUCam (\citealt{PAU_cam}), a unique instrument that covers a $\sim \! 1^{\circ}$ diameter \gls{fov} and is equipped with a set of 40 \glspl{nb}, ranging from 4500$\,$\r{A} to 8500$\,$\r{A}, in steps of 100$\,$\r{A}. \gls{paus} complements its \gls{nb} observations using \gls{bb} data from the Canada-France-Hawaii Telescope Lensing Survey (\glsentryshort{cfhtlens}\glsunset{cfhtlens}; \citealt{CFHTLenS_Erben, CHFTLenS_Heymans, Hildebrandt_2012}) and \gls{kids} \citep{Kuijken_2019}. The principal targets \gls{paus} has observed are the W1 and the W3 fields from the Canada-France-Hawaii Telescope Legacy Survey (\glsentryshort{cfhtls}\glsunset{cfhtls}; \citealt{CFHTLS}), the G09 field from \gls{kids} and the COSMOS field (\citealt{COSMOS}), with the latter mainly used for calibration and validation purposes. The total area covered by \gls{paus} in these fields is $\sim \! 51$ deg$^{2}$ (for objects with a minimum coverage of 30\gls{nb}), with a number density of $\sim \! 1.3 \times 10^4$ objects deg$^{-2}$ down to $i_{\mathrm{AB}}<22$.

The \gls{paudm} team is responsible for the treatment of the data \citep{Tonello}. The data reduction involves \gls{nb} image processing, where the photometric calibration and scattered-light correction are accounted for, and forced photometry is performed over a given reference catalogue to optimise the flux measurement. For a detailed description of both \gls{paus} \gls{nb} image photometry and photometric calibration, see \cite{narrow_band_image_photometry} and \cite{Francisco_PAUS}, respectively.

The large number of bands available in \gls{paus} allows for the extraction of excellent \glspl{photo-z}, with a precision that approaches that of \glspl{spec-z}, as indicated in many \gls{paus} works \citep[][]{BCNz_eriksen, Deepz_eriksen, PAUS_alarcon, Soo_photoz_PAUS, Cabayol_photoz_PAUS, photo_z_wide_fields, Deep_z_Daza}{}{}. This, combined with its high number density of objects, makes \gls{paus} a unique survey for many scientific studies, such as the analysis of galaxy properties using \glspl{nb} (\citealt{NB_galaxy_properties,Csizi2024}), the determination of the mean mass of close galaxy pairs (\citealt{Galaxy_pairs}), the study of the D4000 spectral break (\citealt{D4000}), the Ly-$\alpha$ intensity mapping (\citealt{Lyman_alpha}) and a first study of the \gls{ia} of galaxies (\citealt{IA_harry}), amongst others. In particular, \gls{paus} covers a considerable fraction of the redshift and luminosity ranges of stage-IV surveys, allowing it to constrain \gls{ia} for similar galaxy samples as those that will be used in their cosmological analyses.

\subsection{\texorpdfstring{Photo-$z$}{photo-z}}\label{sec:photo-z}

Due to the large number of \glspl{nb} used by \gls{paus}, the resolution of the recovered \gls{sed} is remarkable, with an average spectral resolution of $R\sim65$. This allows us to estimate \glspl{photo-z} with high precision compared to other \gls{bb} photometric surveys, such as \gls{cfhtlens} or \gls{kids}, with 5 and 9 \glspl{bb}, respectively. 

Here, we use the \gls{photo-z} estimates presented in \cite{photo_z_wide_fields}, who measured \glspl{photo-z} for $\sim$ 1.8 million objects down to $i_{\mathrm{AB}}<23$ in the W1, W3 and G09 fields. For a detailed description of the redshift estimation procedure, we refer the reader to that paper, but in the following lines we briefly summarise the important points for our \gls{ia} analysis. The \glspl{photo-z} are computed using a \gls{sed} template-fitting algorithm named \texttt{BCNz} (\citealt{BCNz_eriksen}), which compares the observed fluxes in the \gls{paus} \glspl{nb} and \glspl{bb} against \gls{sed} templates. Then, it estimates a \gls{photo-z} probability distribution for each object, where the mode of the distribution is taken as the point-like redshift estimate. Two different \gls{photo-z} estimates are derived, $z_{\mathrm{b, \texttt{BCNz}}}$ and $z_{\mathrm{b, BCNzw}}$, where the former is the direct result of \texttt{BCNz} and the latter is a weighted estimate between $z_{\mathrm{b, \texttt{BCNz}}}$ and $z_{\mathrm{b, \texttt{BPZ}}}$, with $z_{\mathrm{b, \texttt{BPZ}}}$ the \glspl{photo-z} derived by \gls{cfhtlens} \citep{Hildebrandt_2012} and \gls{kids} \citep{Hildebrandt_2021} using \texttt{BPZ} (\citealt{BPZ}), where only \gls{bb} information is used. 

The performance of $z_{\mathrm{b, \texttt{BCNz}}}$ decreases as the objects become fainter, due to the loss of \gls{snr} in the \glspl{nb}. At $i_{\mathrm{AB}}>22.5$, the accuracies of $z_{\mathrm{b, \texttt{BCNz}}}$ and $z_{\mathrm{b, \texttt{BPZ}}}$ are similar, so the weighted estimate, $z_{\mathrm{b, BCNzw}}$, benefits from the information gained by the different templates in both \gls{photo-z} codes. Besides, $z_{\mathrm{b, BCNzw}}$ helps to break some degeneracies in the $z_{\mathrm{b}}$ vs. $z_{\mathrm{s}}$ scatter plot, as can be seen in Fig. 9 of \cite{photo_z_wide_fields}. As a consequence, in this work we use the weighted \gls{photo-z} estimate, $z_{\mathrm{b, BCNzw}}$. Additionally, in order to remove objects with low \gls{photo-z} quality and potentially catastrophic outliers, we employ the $Q_{z}$ parameter:

\begin{equation} \label{eq:Qz}
    Q_{z} \equiv \frac{\chi_{\mathrm{\texttt{BCNz}}}^{2}}{N_{\rm f}-3}\left (\frac{z_{\textrm{quant}}^{99} - z_{\textrm{quant}}^{1}}{\textrm{ODDS}} \right ),
\end{equation}
where $\chi_{\mathrm{\texttt{BCNz}}}^{2}$ is an estimate of the fit between the \texttt{BCNz} \gls{sed} templates and the observed fluxes, $N_{\rm f}$ is the number of filters, $z_{\textrm{quant}}^{n}$ is the nth percentile of the $p(z)$ posterior distribution and ODDS is a parameter that measures the probability located around the peak of $p(z)$:

\begin{equation}\label{eq:ODDS}
    \textrm{ODDS} = \int_{z_{\textrm{b}}-\Delta z}^{z_{\textrm{b}}+\Delta z} \textrm{d}z \; p(z),
\end{equation}
with $\Delta z = 0.035$. After analysing the $Q_{z}$ distribution, which decreases exponentially with $Q_{z}$, we impose a cut at $Q_{z}<25$ to remove the 10\% of objects with the highest values (and hence which are expected to have the poorest photometric redshifts).

\begin{figure*}
    \centering
    \includegraphics[width=0.98\textwidth]{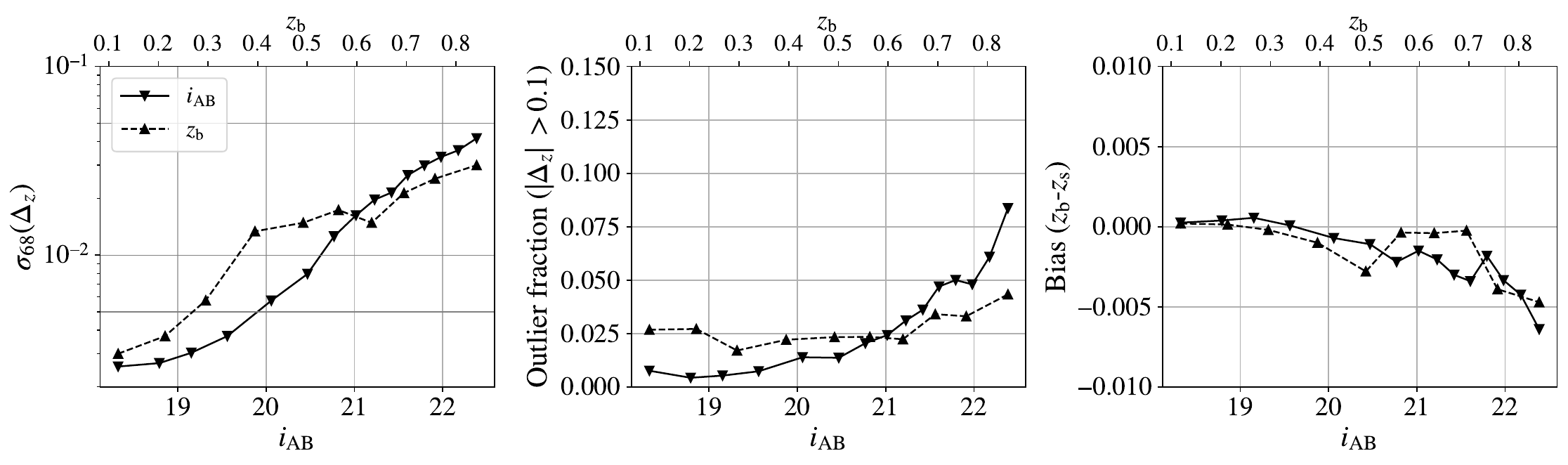}
    \caption{$\sigma_{68}(\Delta_{z})$(left), outlier fraction (centre) and bias (right) as a function of $i_{\mathrm{AB}}$ (triangles down, bottom axis) and $z_{\mathrm{b}}$ (triangles up, top axis) for the \glspl{photo-z} used in this analysis.}
    \label{fig:sigma_68_all_WFs}
\end{figure*}

Fig.~\ref{fig:sigma_68_all_WFs} shows the \gls{photo-z} accuracy in terms of $\sigma_{68}(\Delta_z)$, outlier fraction and bias as a function of $i_{\mathrm{AB}}$ (triangles up) and $z_{\mathrm{b}}$ (triangles down), where $\Delta_z$ is defined as:

\begin{equation}\label{eq:delta_z}
    \Delta_z = \frac{z_\mathrm{b}-z_\mathrm{s}}{1+z_\mathrm{s}},
\end{equation}

$\sigma_{68}(\Delta_z) = (z_{\textrm{quant}}^{84.1}-z_{\textrm{quant}}^{15.9})/2$, the outlier fraction is computed as the fraction of objects with $|\Delta_z|>0.1$\footnote{We use this value following \citealt{photo_z_wide_fields}.} and the bias is determined by the median of $(z_{\mathrm{b}}-z_{\mathrm{s}})$.

The $\sigma_{68}(\Delta_z)$ ranges from $\sim$0.003 to $\sim$ 0.04 (0.03) as a function of $i_{\mathrm{AB}}$ ($z_{\mathrm{b}}$), with an increase towards faint and high-redshift objects. Similar behaviour is seen for the outlier fraction, although the faintest $i_{\mathrm{AB}}$ bins perform worse in this metric than their counterpart $z_{\mathrm{b}}$ bins. Finally, the bias is centred around 0 until $i_{\mathrm{AB}}\sim21$ and $z_{\mathrm{b}}\sim0.7$, after which it begins to gradually deteriorate. The \glspl{spec-z} that have been used for the validation of the \glspl{photo-z} are the ones presented in Table 2 of \cite{photo_z_wide_fields}, which mainly correspond to the Sloan Digital Sky Survey (\glsentryshort{sdss}\glsunset{sdss}; \citealt{SDSS_spectroscopic}), the \gls{gama} survey (\citealt{GAMA_general}), the VIMOS Public Extragalactic Redshift Survey (\glsentryshort{vipers}\glsunset{vipers}; \citealt{VIPERS}), DEEP2 (\citealt{DEEP2_1}) and KiDZ-COSMOS (\citealt{kidz_cosmos}).

\subsection{Galaxy shapes}

Accurate galaxy shapes are crucial for any \gls{ia} analysis. Here, we use the widely used \gls{ksb} algorithm, described in \citet{KSB_1995} and \citet{Luppino_1997}, with corrections presented in \citet{Henk_1998}, which returns an estimate of the shear. Noise in the data, as well as limitations of the approach, result in a multiplicative bias that can be quantified, and corrected for, using realistic image simulations, as discussed in \citet{Henk_2015}. We refer the reader to those papers for a detailed description of this process. As a summary, below we describe the main aspects necessary to understand the shear calibration in our analysis.

The first step is to measure the central second moments of the galaxy images, $I_{ij}$, defined as:
\begin{equation}
    I_{ij} = \frac{1}{I_{0}}\int \mathrm{d}^2\boldsymbol{x} \, x_{i} x_{j} W(\boldsymbol{x})f(\boldsymbol{x}),
\end{equation}
with $I_{0}$ the weighted monopole moment, $x_{i}$ the $i$ coordinate of the galaxy image, $f(\boldsymbol{x})$ the observed galaxy image and $W(\boldsymbol{x})$ a weight function to reduce the sky noise. To choose the width of the weight function, we follow the same configuration used in \citet{IA_harry}, that is, 1.75 times the observed half-light radius of the galaxy.

The shape can then be quantified by combining the weighted quadrupoles into the polarisation:
\begin{equation}\label{eq:polarisation}
    e_{1} = \frac{I_{11}-I_{22}}{I_{11}+I_{22}} \quad \mathrm{and} \quad e_{2} = \frac{2I_{12}}{I_{11}+I_{22}}.
\end{equation}

This is a biased estimator of the shear, because of the weight function, with an uncertainty $\sigma_e$ that can be computed following \citet{Henk_2000}. The shift in the polarisation, $\delta e_\alpha$, due to the shear, $\gamma_\beta$, is quantified by the shear polarisability, $P_{\alpha \beta}^{\mathrm{sh}}$, defined so that:
\begin{equation}\label{eq:shift_shear}
    \delta e_\alpha = P_{\alpha \beta}^{\mathrm{sh}} \gamma_{\beta},
\end{equation}
where we have used the sum convention. Formally, 
$P_{\alpha \beta}^{\mathrm{sh}}$ is a tensor,
but in practice the ensemble average is diagonal, with both elements having the same value, because of symmetry.
\citet{KSB_1995} showed how the shear polarisability can be determined from higher-order moments. 

Accounting for the shear polarisability is, however, not sufficient, because the observed shapes are biased by noise in the data, and the blurring by the \gls{psf}.
If the \gls{psf} is anisotropic, it introduces correlations between the observed polarisations, potentially mimicking a lensing or IA signal. 
This needs to be corrected before applying the shear polarisability in eq.~\ref{eq:shift_shear}. As shown in \citet{KSB_1995}, the change in polarisation due to an anisotropic \gls{psf} can be expressed as:
\begin{equation}\label{eq:shift_PSF}
    \delta e_\alpha = P_{\alpha \beta}^{\mathrm{sm}} p_{\beta},
\end{equation}
where $P_{\alpha \beta}^{\mathrm{sm}}$ is the smear polarisability, which captures the response of an object due to the convolution with an anisotropic \gls{psf}, and $p_{\beta}$ is a measure of the \gls{psf} anisotropy. The latter can be measured by using the observed polarisations and smear polarisabilities of stars:
\begin{equation}
    p_\alpha = \frac{e_{\alpha}^{*}}{P_{\alpha \alpha}^{\mathrm{sm} \, *}},
\end{equation}
where the measurements of the stars use the same weight function as was used for a particular galaxy \citep{Henk_1998}.

Finally, to correct for the circularisation of the galaxy images due to their convolution with the \gls{psf}, the pre-seeing shear polarisability is computed following \citet{Luppino_1997}:
\begin{equation}\label{eq:preseeing_polarisability}
    P^{\gamma} = P^{\mathrm{sh}}-\frac{P_{*}^{\mathrm{sh}}}{P_{*}^{\mathrm{sm}}}P^{\mathrm{sm}}.
\end{equation}

Even though $P^{\gamma}$ is a 2x2 tensor, we assume it is diagonal, with both elements having the same value,
because of symmetry. Moreover, individual estimates for $P^\gamma$ are noisy, and small values increase the variance of the shear measurements. Therefore, we 
only select objects with $P^{\gamma}>0.1$, and account for these selections in the image simulations.

Combining eq.~\ref{eq:shift_shear}, \ref{eq:shift_PSF} and \ref{eq:preseeing_polarisability}, we obtain the shear estimate:
\begin{equation}\label{eq:polarisation_corr}
    \hat\varepsilon_{i} = \frac{e_{i}-P_{ii}^{\mathrm{sm}}p_{i}}{P^{\gamma}}.
\end{equation}
We use the symbol $\hat\varepsilon$ here because we correlate ellipticities for \gls{ia} studies. This is because the ellipticity, defined as $(1-q)/(1+q)$, with $q$ the axis ratio, is an unbiased estimate of the shear, and thus can be compared directly to lensing signals.

Although eq.~\ref{eq:polarisation_corr} provides good estimates, especially for large, bright galaxies, 
an incomplete correction for the \gls{psf} contamination might lead to residual multiplicative ($\mu_{i}$) and additive biases ($c_{i}$). Moreover, shape estimates involve ratios of noisy quantities, resulting in bias. These biases can be captured as \citep{Heymans_multiplicative_bias, Henk_2015}:
\begin{equation}\label{eq:polarisation_multiplicative}
    \varepsilon_{i} = (1+\mu_{i}) \hat\varepsilon_{i} + c_{i},
\end{equation}
where we explicitly indicate that we aim to obtain unbiased ellipticity estimates.

In this paper, to study the IA signal, we correlate positions and shapes. As a consequence, residual additive bias has a minimal impact, because it tends to vanish in the ensemble average over many pairs of galaxies by symmetry. Therefore, following \citet{IA_harry}, we assume that the
correction from \gls{ksb} is sufficient. As for the multiplicative bias, we follow the simulation setup
described  in \citet{Henk_2015} and used in \citet{IA_harry}, and determine $\mu=(\mu_1+\mu_2)/2$ from simulated images. In particular, we capture how the bias depends on the size of the galaxy and its \gls{snr}. We refer interested readers to these papers for further details. We compute the mean multiplicative bias for each field and sample selection we study in Section~\ref{sec:results}, which include splits in red and blue galaxies, in luminosity, stellar mass and redshift.

It is important to note that we aim to perform \gls{ia} measurements with a cut in $i_{\mathrm{AB}}$ magnitude. This poses the challenge that the \gls{cfhtlens} and the \gls{kids} surveys use different bands to measure galaxy shapes, the $i_{\mathrm{AB}}$ band being the one employed by \gls{cfhtlens} and the $r_{\mathrm{AB}}$ that employed by \gls{kids}. This leads to different completenesses in the galaxy shape estimation as a function of the limiting magnitude $i_{\mathrm{AB}}$ for both surveys. In the case of \gls{cfhtlens}, galaxy shape measurements down to $i_{\mathrm{AB}}=22.5$ are available and complete. Nevertheless, for the \gls{kids} sample, galaxy shape measurements with good quality are only complete down to $i_{\mathrm{AB}}=22$. For this reason, we perform measurements of \gls{ia} for two scenarios. The baseline scenario is the measurement of the combined \gls{cfhtlens} (W1 and W3) and \gls{kids} (G09) fields down to $i_{\mathrm{AB}} = 22$. The second scenario extends the measurements to deeper magnitudes ($i_{\mathrm{AB}}=22.5$) but only using the \gls{cfhtlens} fields, and is presented in Appendix~\ref{sec:brighter_vs_fainter}.

\subsection{Restframe magnitudes and colours}\label{sec:colour_split_PAUS}

A common approach taken in the measurements of \gls{ia} is to separate samples into red and blue galaxies. The reason for this is that galaxy colour and morphology tend to be closely related, with red galaxies mainly corresponding to ellipticals and passive types and blue galaxies to spirals and star-forming (active) galaxies \citep{Strateva_colour_separation, Park_colour_separation, Siudek_colour_separation}.

This separation is performed using the \gls{paus} physical properties derived by Siudek et al. (in prep.) and used in \cite{photo_z_wide_fields}, which were estimated by performing \gls{sed} fitting with the Code Investigating GALaxy Emission (\glsentryshort{cigale}\glsunset{cigale}; \citealt{CIGALE}). This code takes into account the dust absorption affecting stellar emission in the UV and optical bands and its re-emission in the IR. As stated in \cite{photo_z_wide_fields}, \gls{cigale} does not perform well for \gls{paus} objects with low \gls{snr} in the \glspl{nb}. This particularly impacts the estimation of the physical properties we use to separate between active and passive galaxies in this work. As a consequence, we use the physical properties estimated with \gls{bb}-only fits, as they perform better in the separation. We also remove objects with a reduced $\chi^{2}_{\nu, \mathrm{\gls{cigale}}}>5$ \citep{Masoura_2018, Buat_2021}, which affects $\sim 0.5\%$ of the objects, where $\chi_{\nu, \mathrm{\gls{cigale}}}^{2}$ is related to the quality of the \gls{cigale} fit.

We split our samples between active and passive populations using a NUV-$r$ vs. $r$-$K$ (NUV$rK$ from now on) diagram, following \citet{Color_cut_Arnouts}. There, the redshift range tested was $0.2 \leq  z \leq  1.3$, which closely matches the redshift range we consider in this analysis, which is $0.1 \leq  z \leq  1$. On the one hand, the NUV-$r$ colour traces the \gls{ssfr}, given the capability of the NUV and the $r$ bands to track young and old populations, respectively \citep{Salim_2005}. On the other hand, the $r-K$ colour accounts for dust attenuation in active galaxies and helps to break degeneracies related to this effect in the NUV-$r$ cut \citep{Color_cut_Arnouts}. The separation of active and passive galaxies following a NUV$rK$ diagram is similar to the one performed in \citet{Williams_2009} following a ($U-V$) vs. ($V-J$) diagram, which was also employed in the separation between active and passive galaxies in \cite{photo_z_wide_fields} in order to study the performance of \glspl{photo-z} as a function of colour. In this case, we opt for the NUV$rK$ diagram, since it allows to expand the covered wavelength range analysed in the separation. We note that the terms ``active/blue'' and ``passive/red'' are used interchangeably throughout the rest of this paper.

Following \citet{Color_cut_Davidzon}, we use 2 cuts in the colour-colour space to separate between active and passive galaxies. In their case, they divide the NUV$rK$ space in 3 regions: active (galaxies fulfilling eq.~\ref{eq:active_cut}), green-valley (galaxies fulfilling eq.~\ref{eq:green_valley_cut} but not eq.~\ref{eq:active_cut}) and passive (the rest of the diagram):

\begin{equation}\label{eq:active_cut}
    \mathrm{NUV}-r > 3.75 \, \, \, \, \& \, \, \, \, \mathrm{NUV}-r > 1.37 (r-K) + 3.2 \, \, \, \, \& \, \, \, \, r-K < 1.3,
\end{equation}

\begin{equation}\label{eq:green_valley_cut}
    \mathrm{NUV}-r > 3.15 \, \, \, \, \& \, \, \, \, \mathrm{NUV}-r > 1.37 (r-K) + 2.6 \, \, \, \, \& \, \, \, \, r-K < 1.3.
\end{equation}

However, in our case, we cannot aim to reach this level of precision when separating active and passive galaxies, since the absolute magnitudes computed by \gls{cigale} are affected by the redshift uncertainty from the \glspl{photo-z}, while \citet{Color_cut_Davidzon} use \glspl{spec-z}. Nevertheless, given that the region located in the space defined by eq.~\ref{eq:active_cut} and \ref{eq:green_valley_cut} corresponds to galaxies which are reducing their \gls{ssfr}, and may be in the transition between active and passive galaxies, we assign objects as active or passive so as to obtain a similar percentage of red and blue galaxies in all \gls{paus} fields. In particular, we assign the W1 and W3 objects that lie in this intermediate colour-colour space as passive and the ones from G09 as active. This way, we find that the percentage of passive galaxies is 18.1\% (16.0\%), 23.8\% (20.5\%) and 19.5\% (17.2\%) for the W1, G09 and W3 fields, respectively, at $i_{\mathrm{AB}}=22.0$ ($i_{\mathrm{AB}}=22.5$). We note that the percentages of red/blue galaxies are still different between fields, specially for the \gls{cfhtlens} and \gls{kids} samples, likely due to the different photometric systems. 

To secure the separation between active and passive galaxies, we also use the $T_{\mathrm{\texttt{BPZ}}}$ parameter employed in \gls{cfhtlens} and \gls{kids}. This parameter allows us to separate objects by their spectral type, and is obtained by performing \gls{sed} fitting using the \texttt{BPZ} \gls{photo-z} algorithm. With this new parameter, we define red objects as those with $T_{\mathrm{\texttt{BPZ}}}\leq 1.5$ \citep{T_bpz} and blue for the rest. For red galaxies, the difference in the number of objects after applying $T_{\mathrm{\texttt{BPZ}}}$ does not change much. The agreement in the objects classified as red by both the NUV$rK$ diagram and the $T_{\mathrm{\texttt{BPZ}}}$ parameter is 99.98\%, 93.57\% (94.63\% at $i_{\mathrm{AB}}=22.5$) and 99.98\% of galaxies, for the W1, G09 and W3 fields, respectively. However, the case for blue galaxies is different, with $\sim87\%$ ($\sim$89\% at $i_{\mathrm{AB}}=22.5$) of objects classified as blue by the NUV$rK$ cut and the $T_{\mathrm{\texttt{BPZ}}}$ criteria for all the fields.

\begin{figure*}
    \centering
    \includegraphics[width=0.98\textwidth]{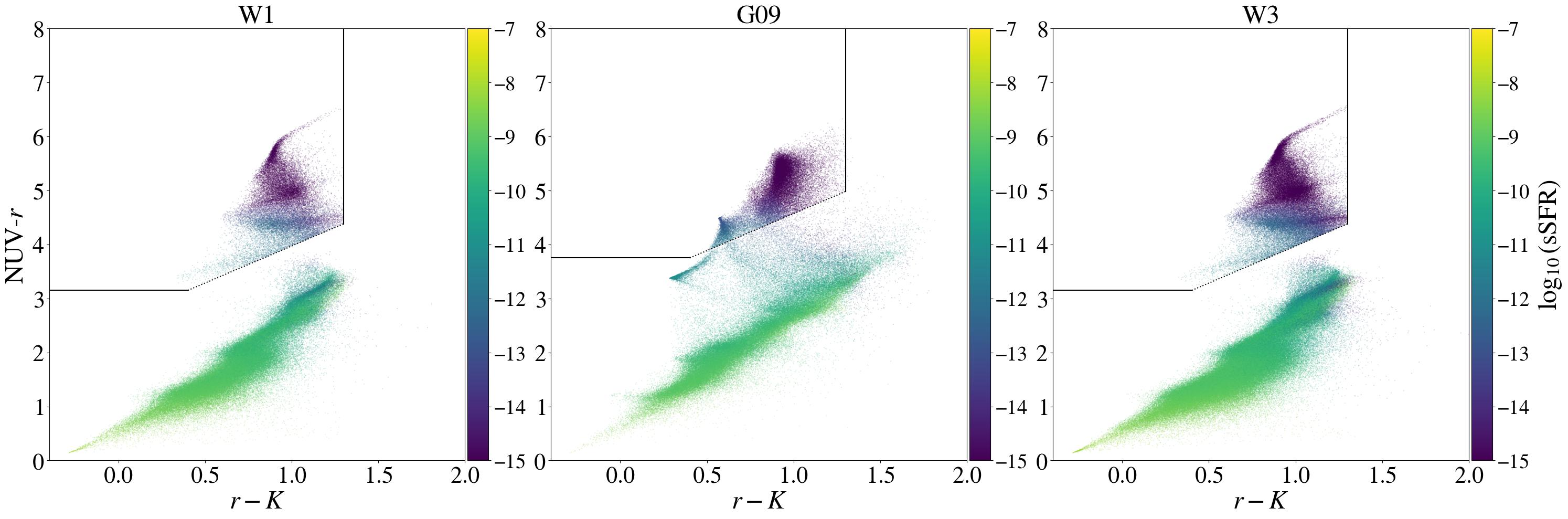}
    \caption{Division in active and passive galaxies following a NUV$rK$ diagram cut and a $T_{\mathrm{\texttt{BPZ}}}$ selection, coloured by the \gls{ssfr} obtained with \gls{cigale}, as indicated by the colour bar to the right of each panel. The top regions delimited by the black lines correspond to passive galaxies, while the complementary regions correspond to active galaxies.}
    \label{fig:PAUS_color_selection}
\end{figure*}

Fig.~\ref{fig:PAUS_color_selection} shows the NUV$rK$ diagram used to separate between red and blue galaxies for the W1, G09 and W3 fields (from left to right). As stated before, the position of the diagonal lines differ between fields, with eq.~\ref{eq:active_cut} delimiting the region for active galaxies in the W1 and W3 fields and eq.~\ref{eq:green_valley_cut} delimiting it for the G09 field. It can be noted that there is a lack of objects just below the diagonal lines, which is produced by the additional cut performed in $T_{\mathrm{\texttt{BPZ}}}$. Confirming our sample selection, we find that active galaxies have a larger \gls{ssfr} (as computed by \gls{cigale}) than passive ones, as indicated by the colour bar on these diagrams. The percentage of red objects after applying both the NUV$rK$ and the $T_{\mathrm{\texttt{BPZ}}}$ cuts is 20.6\% (18.3\%), 26.2\% (25.0\%) and 22.2\% (19.7\%) for the W1, G09 and W3 fields, respectively, at $i_{\mathrm{AB}}=22.0$ ($i_{\mathrm{AB}}=22.5$). In Appendix~\ref{sec:colour_splits_literature}, we show a comparison of the red and blue classification when employing alternative colour-magnitude cuts from the literature.

\begin{figure*}
    \centering
    \includegraphics[width=0.98\textwidth]{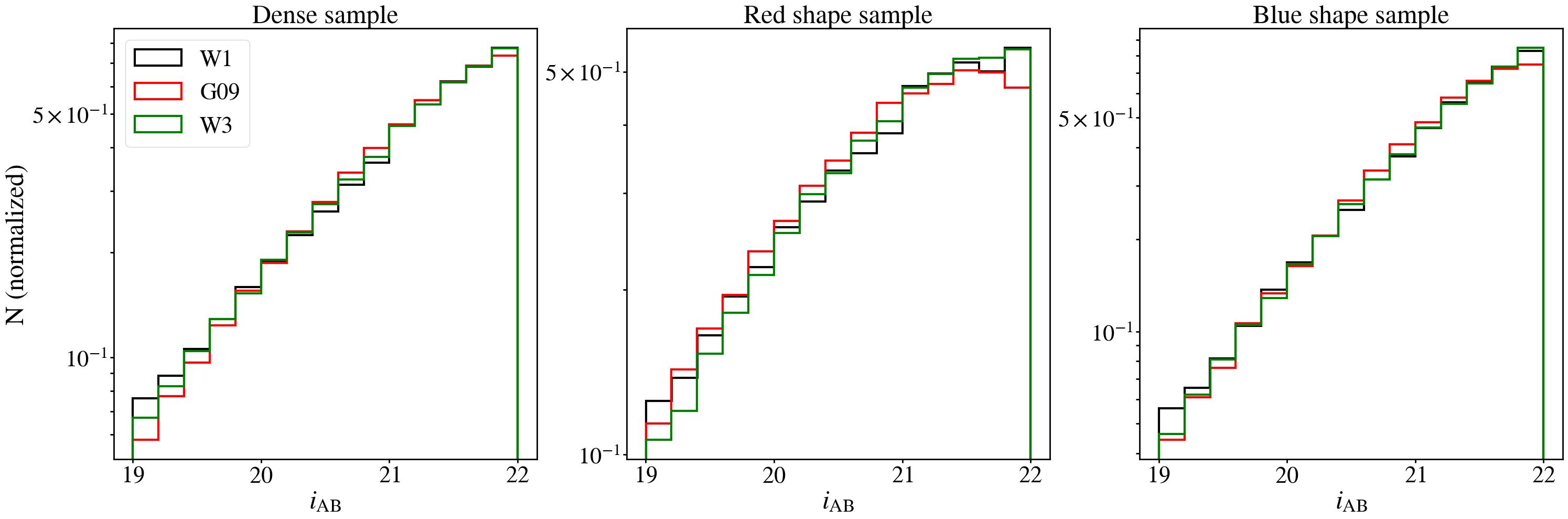}
    \caption{$i_{\mathrm{AB}}$ distribution of the dense sample (left), the red galaxies with shapes (middle) and the blue galaxies with shapes (right) for each of the \gls{paus} wide fields.}
    \label{fig:PAUS_color_distribution}
\end{figure*}

Fig.~\ref{fig:PAUS_color_distribution} shows the $i_{\mathrm{AB}}$ distribution for the dense sample (left), which refers to the full galaxy sample used to define the galaxy positions in the estimators of Section~\ref{sec:estimators}, the red galaxies with shapes (middle) and the blue galaxies with shapes (right) for the three wide fields under study. The observed magnitude distribution is very similar for each of the fields, for both the dense and the shape samples separated by colour. This is an indication that we are selecting similar galaxy populations throughout, which is an important aspect to take into account when combining the measurements of \gls{gc} and \gls{ia} from different fields, as we do in our results in Section~\ref{sec:results}. In this case, we show the shape samples down to $i_{\mathrm{AB}}=22$ for conciseness. Nevertheless, the fainter shape samples of the \gls{cfhtlens} fields also have very similar $i_{\mathrm{AB}}$ distributions.

\subsection{MICE}\label{subsec:MICE}

The Marenostrum Institut de Ciències de l’Espai Grand-Challenge (MICE-GC; \citealt{MICE_1, MICE_2, MICE_3, MICE_4, MICE_5}) is an N-body simulation run using the public code GADGET-2 \citep{Gadget2}. It contains 4096$^3$ \gls{dm} particles in a comoving volume of $(3\gpc)^{3}$ and assumes a flat $\Lambda$CDM cosmology with $\Omega_{\rm m}=0.25$, $\Omega_\Lambda=0.75$, $\Omega_{\rm b}=0.044$, $n_{\rm s}=0.95$, $\sigma_8=0.8$ and $h=0.7$. 

Galaxies are introduced by combining the halo occupation distribution (\glsentryshort{hod}\glsunset{hod}; \citealt{HOD_1, HOD_2, HOD_3}) and \gls{sham} techniques \citep{SHAM_1, SHAM_2, SHAM_3}. The galaxy mock is calibrated to reproduce \gls{sdss} \citep{SDSS} colour distributions, luminosity function and \gls{gc}.

\gls{ia} are introduced in MICE-GC by assigning intrinsic shapes and orientations to the MICE simulation up to redshift $z=1.4$ \citep{MICE_IA}. This is done using a semi-analytic \gls{ia} model, where the intrinsic shapes and orientations are assigned based on the galaxy colour and the galaxy type (central or satellite). In this model, red central galaxies have their 3-dimensional principal axes aligned with their host halo, while blue central galaxies have their minor axis aligned with the angular momentum of the host halo and their major axis randomly oriented in the perpendicular plane of the minor axis. In the case of satellite galaxies, their major axes are oriented towards the host halo and their minor axes are randomly oriented in the perpendicular plane of the major axis. These alignment configurations assume that red central galaxies are pressure-supported objects affected by the same tidal field as their host halo, while blue central galaxies are rotationally supported and do not present \gls{ia}. Finally, both red and blue satellite galaxies are preferentially oriented towards the centre of their host halo. The colour separation between red and blue galaxies in MICE is defined with a $u-r=M_u-M_r>0.94$ cut, where $M_u$ and $M_r$ correspond to the absolute rest-frame magnitudes in the CFHT-$u$ and Subaru-$r$ bands, respectively. The parameters of the semi-analytic \gls{ia} model are calibrated against the COSMOS \citep{COSMOS_MICE_IA_1, COSMOS_MICE_IA_2} and BOSS LOWZ \citep{BOSS_LOWZ} surveys.

This galaxy mock is used to validate and perform some consistency checks. In particular, in Appendix~\ref{sec:randoms}, we describe the generation of the random catalogues (see Section~\ref{sec:estimators}) necessary to perform the measurements and accurately obtain galaxy biases. Next, given that we analyse the data with \glspl{photo-z}, we show in Appendix~\ref{sec:Comparison_zb_zs} that we are able to obtain consistent galaxy biases and \gls{ia} parameters when using \glspl{photo-z} or \glspl{spec-z}. Finally, in Appendix~\ref{sec:error_estimation}, we study the errors (see Section ~\ref{sec:estimators}) associated with the measurements and compare them with the ensemble covariance of MICE, defined from a collection of realisations from the simulation.

Our first goal is to construct a subsample of the MICE-GC, so that it resembles the \gls{paus} wide fields. First, we cut the galaxy mock such that $19<i_{\mathrm{AB}}<22.5$ using the MICE synthetic observed magnitude in the COSMOS CFHT-$i$ band, which is the one that most closely resembles the \gls{cfhtlens} and \gls{kids} $i_{\mathrm{AB}}$ magnitudes. We adopt the cut of the fainter samples ($i_{\mathrm{AB}}=22.5$), instead of the brighter samples cut at $i_{\mathrm{AB}}=22$, since the validations we want to perform are dependent on the \gls{photo-z} quality, which worsens with fainter magnitudes. Thus, if we are able to validate our method for fainter magnitudes, the brighter case is also expected to work. Later, we introduce noise in the \glspl{spec-z} available in the MICE catalogue to generate \gls{paus}-like \glspl{photo-z} in the simulation. For that, we use the galaxy mock designed by Wittje et al. (in prep.),  which is created using the Flagship simulation \citep{Flagship}. This galaxy mock generates objects with \gls{paus}-like fluxes and computes the \gls{photo-z} using \texttt{BCNz} and \texttt{BPZ}. We combine both of these redshift estimates to obtain the weighted \gls{photo-z} presented in \cite{photo_z_wide_fields} and inject them into the MICE simulation. To do so, we bin the galaxy mock presented in Wittje et al. (in prep.) into \gls{spec-z} bins of width $\Delta z_{\mathrm{s}} = 0.01$. Then, for each bin of \glspl{spec-z}, we have a distribution of \glspl{photo-z}, which we use to sample and assign \glspl{photo-z} to the corresponding \gls{spec-z} bins in the MICE simulation.

\begin{figure*}
    \centering
    \includegraphics[width=0.98\textwidth]{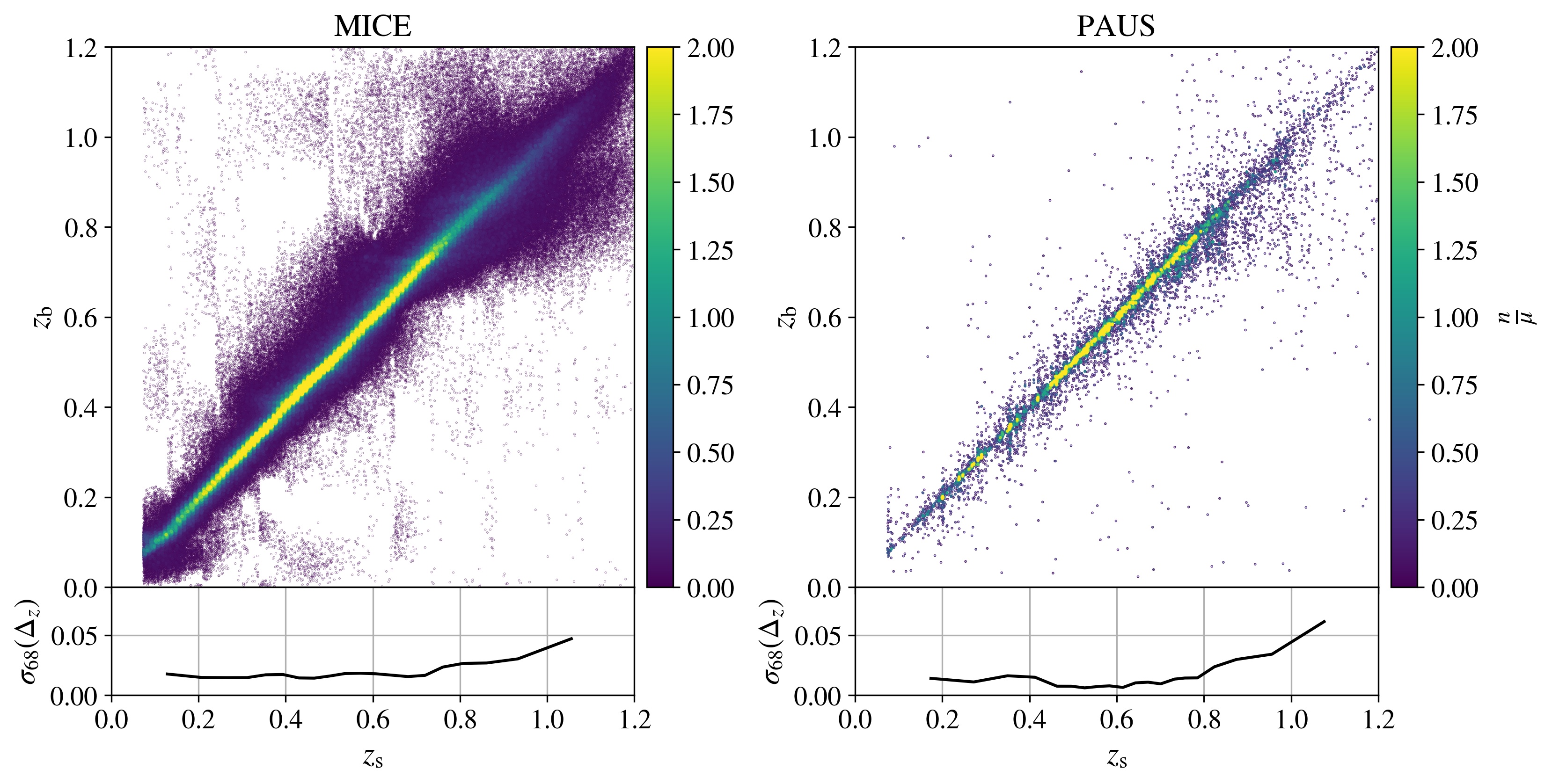}
    \caption{Top: \Gls{photo-z} vs. \gls{spec-z} of the galaxy mock catalogue from the MICE simulation (left) and of the objects from the \gls{paus} W3 field that have \glspl{spec-z} (right), coloured by the number of objects in each pixel normalised by the median value. Bottom: $\sigma_{68}(\Delta_{z})$ vs. \gls{spec-z} for the MICE simulation (left) and \gls{paus} (right).}
    \label{fig:zb_vs_zs_MICE_full_octant}
\end{figure*}

The upper panels of Fig.~\ref{fig:zb_vs_zs_MICE_full_octant} show the distribution of \gls{photo-z} versus \gls{spec-z} for the MICE simulation (left) and for the \gls{paus} objects in the W3 field that have \glspl{spec-z} (right). The resemblance of both distributions and, in particular, the spread along the diagonal line at large redshifts indicate the similarity between both \gls{photo-z} cases. This is further seen in the lower panels of Fig.~\ref{fig:zb_vs_zs_MICE_full_octant}, where we find comparable values between MICE and \gls{paus} in terms of $\sigma_{68}$ versus \gls{spec-z}.

\begin{figure*}
    \centering
    \includegraphics[width=0.85\textwidth]{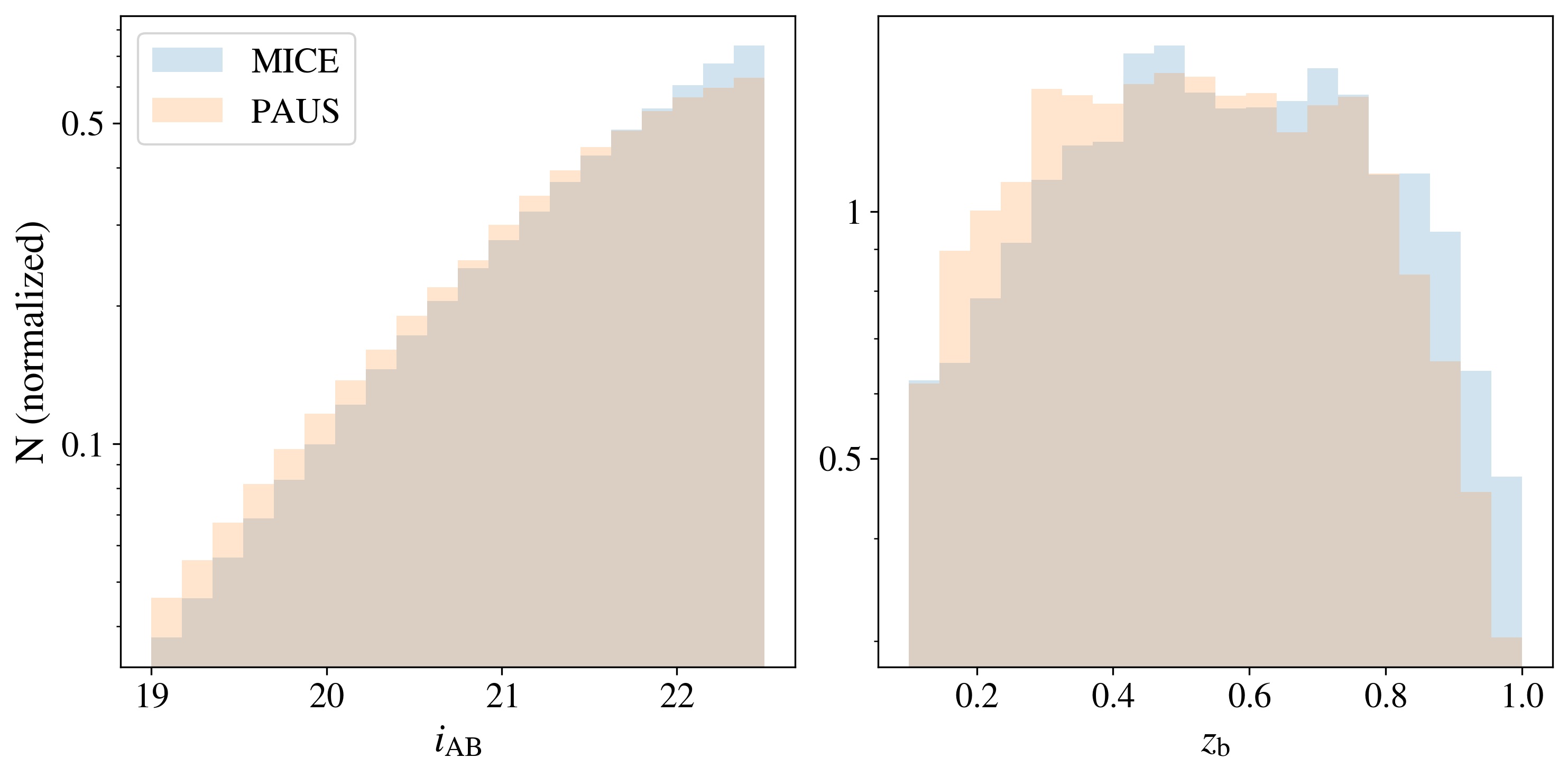}
    \caption{Comparison of the distribution of $i_{\mathrm{AB}}$ (left) and $z_{\mathrm{b}}$ (right) for the \gls{paus} wide fields and the MICE galaxy mock, indicating the similarity of the galaxy populations for both cases. This enables the use of MICE in order to perform some consistency tests, detailed in Appendix~\ref{sec:randoms}, \ref{sec:Comparison_zb_zs} and \ref{sec:error_estimation}.}
    \label{fig:PAUS_MICE_distributions}
\end{figure*}

Fig.~\ref{fig:PAUS_MICE_distributions} shows the comparison of the $i_{\mathrm{AB}}$ and $z_{\mathrm{b}}$ distributions between \gls{paus} and the MICE mock, which shows that the redshift and magnitude distributions are very similar, indicating that the populations selected for \gls{paus} and MICE agree and that the mock is suitable for performing validation and consistency checks.

In particular, when analysing the data, we perform measurements by combining the 3 \gls{paus} wide fields. Thus, to have statistically significant results and ensure a realistic representation, we divide the full octant of MICE into 180 patches, such that we create 60 combinations of 3 patches with the same area as W1, G09 and W3. In this way, we can obtain 60 realisations in MICE of the \gls{gc} and \gls{ia} measurements, which we compare to the \gls{paus} measurements.

\section{Estimators}\label{sec:estimators}

We compute the 3-dimensional position-position correlation function using the Landy-Szalay estimator \citep{LandySzalay_wgg}, binning in transverse and \gls{los} separations, $r_{\rm {p}}$ and $\Pi$, respectively:

\begin{equation}\label{eq:galaxy_clustering_measurement}
    \xi_{\rm{gg}}(r_{\rm {p}}, \Pi) = \frac{{DD - 2DR + RR}}{{RR}},
\end{equation}
where $D$ traces the galaxy positions of the dense samples and $R$ corresponds to random samples that follow the angular and radial distributions of the dense samples. The random catalogues are set to be 50 times larger than the dense sample, in order to reduce shot noise. Even though we perform our \gls{ia} analysis by separating the shape sample in red and blue galaxies, we measure the position-position correlation function without a division by colour, since we have checked that this reduces the errors in the \gls{ia} measurements, given that the number of objects being correlated is larger.

The 3-dimensional galaxy position-intrinsic shear correlation function is a generalisation of the Landy-Szalay estimator for \gls{gc} \citep{LandySzalay_wgp} and is defined as:
\begin{equation}\label{eq:intrinsic_alignment_measurement}
    \xi_{\rm{gp}}(r, \Pi) = \frac{{S_{+}D-S_{+}R_{D}}}{{R_{S}R_{D}}},
\end{equation}
with 
\begin{equation}\label{eq:weight_sum_ellipticities}
    S_{+}D = \sum_{i\neq j\mid r_{\rm {p}}, \Pi} \varepsilon_{+}(j|i).
\end{equation}
$S_{+}$ corresponds to the shape sample, $D$ is the dense sample and $R_{S}$ and $R_{D}$ are the random catalogues that trace the shape and dense sample distributions, respectively. Here, $\varepsilon_{+}=\mathrm{Re}(\varepsilon e^{2i\varphi})$, where $\varepsilon=\varepsilon_{1}+i\varepsilon_{2}$ and $\varphi$ is the polar angle connecting the pair of galaxies being correlated. As in the case of the position-position correlation function (eq. \ref{eq:galaxy_clustering_measurement}), only the shape sample is separated by galaxy colour. The expression in eq.~\ref{eq:weight_sum_ellipticities} is the sum of the ellipticity components of galaxies $i$ with respect to the galaxies in the dense sample $j$. Note that we do not normalise eq.~\ref{eq:weight_sum_ellipticities} by the shear responsivity, $\mathcal{R} \sim 1-\sigma_{\varepsilon}^{2}$, with $\sigma_{\varepsilon}$ the shapes' sample dispersion. The reason for this is that the estimate we obtain in eq.~\ref{eq:polarisation_multiplicative} is an ellipticity, rather than a polarisation, and so it already accounts for the effect of the shear.

As indicated in \citet{Henk_2015}, for each galaxy we weight the ellipticities in eq.~\ref{eq:weight_sum_ellipticities} with $w_{i}$, defined as:

\begin{equation}
    w_{i} = \frac{1}{\left \langle 0.25 \right \rangle^{2}+\left ( \frac{\sigma_{e, i}}{P^{\gamma}_{i}} \right )^{2}},
\end{equation}
where 0.25 is the value we adopt for the intrinsic variance of the galaxy ellipticity \citep{Henk_2000}, $\sigma_{e}$ is the uncertainty in the measurement of the polarisation (eq.~\ref{eq:polarisation}), $P^{\gamma}$ is the preseeing shear polarisability (eq.~\ref{eq:preseeing_polarisability}) and the subscript $i$ refers to each galaxy in the shape sample.

An analogous estimator of eq.~\ref{eq:intrinsic_alignment_measurement}, $\xi_{gx}$, is defined by rotating the polarisations by 45$^{\circ}$, with $\varepsilon_{x}=\mathrm{Im}(\varepsilon e^{2i\varphi})$. A value different from 0 would imply a preferred direction of curl in the shape sample distribution. Thus, this quantity is usually used to check for systematics.

The 3-dimensional correlation functions defined in eq.~\ref{eq:galaxy_clustering_measurement} and eq.~\ref{eq:intrinsic_alignment_measurement} are projected along the \gls{los}, following:

\begin{equation}\label{eq:wab_measurement}
    w_{\rm{ab}}(r_{\rm {p}}) = \int_{-\Pi_{\mathrm{max}}}^{\Pi_{\mathrm{max}}} \xi_{\rm{ab}}(r_{\rm {p}}, \Pi) {\rm d}\Pi,
\end{equation}
where $\Pi_{\mathrm{max}}$ is the maximum \gls{los} separation and ``ab" corresponds to either $\rm{gg}$, $\rm{gp}$ or $\rm{gx}$.

The correlation functions are measured using \texttt{TreeCorr} \citep{TreeCorr}. We define 12 logarithmically-spaced bins in projected separation ranging from 0.1-18 $\mpc$, considering the size of the fields under study, and employ the radial binning devised by \citet{IA_harry}, who defined eq.~\ref{eq:pi_binning} to have positive and negative values:

\begin{equation}\label{eq:pi_binning}
    |\Pi| = 0,1,2,3,5,8,13,21,34,55,89,144,233 \mpc,
\end{equation}
such that $\Pi_{\mathrm{max}}=233 \mpc$. This expression accounts for the spread of galaxies in the radial direction caused by \gls{photo-z}, by enlarging the radial binning as the distance from the correlated object increases. This choice is justified further in Appendix~\ref{sec:Comparison_zb_zs}. The bin ``slop'' parameter from \texttt{TreeCorr} is set to 0 in our analysis.

The errors associated with our measurements are computed using the \gls{jk} method \citep{Norberg2009}:
\begin{equation}\label{eq:JK_error}
    \mathrm{Cov_{JK}} = \frac{N_{\mathrm{JK}}-1}{N_{\mathrm{JK}}}\sum_{i=1}^{N_{\mathrm{JK}}}(w_{\mathrm{ab}, i}- {\overline{w}_{\rm{ab}}})(w_{\mathrm{ab}, i}- {\overline{w}_{\rm{ab}}})^{T},
\end{equation}
where $N_{\mathrm{JK}}$ corresponds to the number of \gls{jk} regions, $w_{\mathrm{ab}, i}$ is the ``ab'' correlation function after removing the signal of the i-th \gls{jk} region and ${\overline{w}_{\rm{ab}}}$ is the mean of all the \gls{jk} regions.

We define $N_{\mathrm{JK}}$ so that each of our fields is divided into  roughly equal area patches, corresponding to 4 angular regions in W1, 6 in G09 and 8 in W3. This number of angular \gls{jk} regions was defined following \cite{IA_harry} for the W3 field, and adjusted to the W1 and G09 fields based on their area. Additionally, we also define regions by dividing the redshift range. This division in redshift depends on the case under study, since these have different redshift ranges. Table~\ref{tab:IA_cases} includes the $N_{\mathrm{JK}}$ defined in each of our studies.

\section{Modelling}\label{sec:modelling}

Here, we describe the modelling of the joint data vector $w_{\rm{gg}} \cup w_{\rm{gp}}$ and the methodology we follow to constrain the galaxy bias and the \gls{ia} parameters for the \gls{paus} data. Sections~\ref{sec:galaxy_power_spectra} and \ref{sec:IA_power_spectra} describe the galaxy power spectrum and the \gls{ia} power spectrum that enter in our models, respectively. Next, we describe how we compute the projected spectroscopic and photometric correlation functions in Sections~\ref{sec:Correlation_spectroscopic} and \ref{sec:Correlation_photometric}, respectively. Section~\ref{sec:Contaminants_correlation_photometric} describes the contaminants introduced due to magnification and galaxy-galaxy lensing. Lastly, in Section~\ref{sec:likelihood} we discuss the likelihood analysis performed to constrain the galaxy bias and the \gls{ia} parameters, including a description of the scale cuts and the priors.

To compute the various power spectra, we use PyCCL \citep{PyCCL} and FAST-PT \citep{Fast_pt_1, Fast_pt_2}. The linear matter power spectrum is computed using CAMB \citep{CAMB} and the non-linear matter power spectrum is computed with Halofit \citep{halofit}, both implemented in PyCCL.

\subsection{Galaxy power spectrum}\label{sec:galaxy_power_spectra}

The relation between the matter and the galaxy distributions is encoded in the galaxy bias term(s). The simplest model one can use is given by a constant term \citep{Kaiser1984}:

\begin{equation}\label{eq:relation_delta_g_delta_m_linear}
    \delta_{\rm{g}} = b_{1}\delta_{\rm {m}},
\end{equation}
where $b_{1}$ is usually referred to as the linear galaxy bias. Although this model is very simplistic, it is known to work well on large scales (e.g. \citealt{Flagship}). Nevertheless, at small scales, the non-linearities of the \gls{lss} need more sophisticated models.

Here, we express the galaxy overdensity field by expanding the density ($\delta_{\rm {m}}$) and the tidal ($s_{ij}$) fields as \citep{galaxy_power_spectra_1, galaxy_power_spectra_2, galaxy_power_spectra_3}:
\begin{equation}\label{galaxy_bias_expansion}
    \delta_{\rm{g}} = b_{1}\delta_{\rm {m}} + \frac{1}{2}b_{2}(\delta_{\rm {m}}^{2}-\left \langle \delta_{\rm {m}}^{2} \right \rangle)+\frac{1}{2}b_{{s}^{2}}(s^{2}-\left \langle s^{2} \right \rangle)+b_{3\mathrm{nl}}\psi,
\end{equation}
where $s^{2}=s_{ij}s^{ij}$ (assuming Einstein summation convention), $\psi$ corresponds to the sum of third-order non-local terms with the same scaling \citep{galaxy_power_spectra_3}, $b_{1}$ is the linear galaxy bias (eq.~\ref{eq:relation_delta_g_delta_m_linear}), $b_{2}$ the local quadratic bias, $b_{s^{2}}$ the tidal quadratic bias and $b_{3\mathrm{nl}}$ the third-order non-local bias. One can express the galaxy-galaxy power spectrum from eq.~\ref{galaxy_bias_expansion} as \citep{galaxy_power_spectra_4}:
\begin{multline}\label{eq:galaxy_power_spectra}
    P_{\rm{gg}}(k)=b_{1}^{2}P_{\delta \delta}(k)+b_{1}b_{2}P_{b_{1}b_{2}}(k)\\+b_{1}b_{s^{2}}P_{b_{1}s^{2}}(k)+b_{1}b_{3\mathrm{nl}}P_{b_{1}b_{3\mathrm{nl}}}(k)\\+\frac{1}{4}b_{2}^{2}P_{b_{2}b_{2}}(k)+\frac{1}{2}b_{2}b_{s^{2}}P_{b_{2}s^{2}}(k)+\frac{1}{4}b_{s}^{2}P_{s^{2}s^{2}}(k),
\end{multline}
where $P_{\delta \delta}$ is the non-linear matter power spectrum and the power spectrum kernels ($P_{b_{1}b_{2}}$, $P_{b_{1}s^{2}}$, etc.), are defined in \citet{galaxy_power_spectra_3}. We also use the following co-evolution relations $b_{s^{2}}=-4/7(b_{1}-1)$ and $b_{3\mathrm{nl}}=b_{1}-1$ \citep{galaxy_power_spectra_3, galaxy_power_spectra_5} to reduce the parameter space and alleviate model complexity.

The galaxy power spectrum that enters the modelling of $w_{\rm{gg}}$ is the one expressed in eq.~\ref{eq:galaxy_power_spectra}. However, for the case of $w_{\rm{gp}}$, the \gls{ia} model we use assumes linear galaxy bias. Thus, the galaxy-intrinsic power spectrum that enters in $w_{\rm{gp}}$ is given by:
\begin{equation}\label{eq:relation_pgI_PGI}
    P_{\rm{gI}}(k,z) = b_{1}P_{\delta \rm{I}}(k,z),
\end{equation}
where $P_{\delta \rm{I}}$ is defined in Section~\ref{sec:IA_power_spectra}. Nevertheless, we do not expect this to have much effect on the constraints of the galaxy bias parameters, since they are mainly constrained by $w_{\rm{gg}}$, due to its higher \gls{snr}. 

\subsection{IA power spectrum}\label{sec:IA_power_spectra}

We model the \gls{ia} following the \gls{nla} model, which is an extension of the \gls{la} model, with the linear matter power spectrum replaced by the non-linear matter power spectrum. The \gls{la} model works under the assumption that the host \gls{dm} halo is tidally distorted by the gravitational field exerted by the surrounding \gls{lss}. The stellar content of the galaxy follows this distortion at the time of its formation and/or during its evolution, being tidally aligned with its host \gls{dm} halo.

Following \citet{Hirata_2004}, the intrinsic shear of an object can be described as:
\begin{equation}\label{eq:intrinsic_shear_NLA}
    \gamma^{\rm I} = -\frac{\bar{C_{1}}}{4\pi G}\left ( \Delta_{x}^{2}-\Delta_{y}^{2}, 2\Delta_{x}\Delta_{y} \right )S[\psi _{P}],
\end{equation}
where $\bar{C_{1}}=5\times10^{-14}M_{\odot}^{-1}h^{-2}\mathrm{Mpc}^{3}$ is a normalisation constant, whose value was set by \citet{C1_value} for low-redshift \gls{ia} measurements in SuperCOSMOS \citep{SuperCOSMOS}, $G$ is the Newtonian gravitational constant, $x$ and $y$ are the Cartesian coordinates in the plane of the sky, $S$ acts as a smoothing filter for $\psi _{P}$, which is the Newtonian potential at the time of galaxy formation, and $\Delta$ is the comoving derivative.

The matter-intrinsic power spectrum for the \gls{nla} model is defined as:
\begin{equation}\label{eq:NLA_power_spectra_delta_i}
    P_{\delta \rm{I}}(k, z) = C_{1}(z) P_{\delta\delta}(k, z),
\end{equation}

where $P_{\delta\delta}$ is the non-linear matter power spectrum and we define $C_{1}$ following the implementation of PyCCL, such that:
\begin{equation}\label{eq:A_ia_amplitude_early}
    C_{1}(z) = -\frac{A_{1}\bar{C}_{1}\rho_{\mathrm{crit}}\Omega_{\rm m}}{D(z)},
\end{equation}
where $A_{1}$ is the \gls{ia} amplitude, $\rho_{\mathrm{crit}}$ is the critical density, $\Omega_{\rm m}$ is the fractional matter density, and $D(z)$ is the growth factor.

The \gls{tatt} model, which is an extension of \gls{nla} that introduces a tidal torquing term, was also used to constrain the \gls{ia} parameters, but there were strong degeneracies present in our \gls{tatt} constraints, probably due to the low \gls{snr} in most of the cases under study, that made us decide to focus on the \gls{nla} model.

\subsection{Correlation functions with \texorpdfstring{spec-$z$}{spec-z}}\label{sec:Correlation_spectroscopic}

From the power spectra described in the previous sections, one can compute projected correlation functions along the \gls{los}, like the ones presented in Section~\ref{sec:estimators}. If we assume Limber's approximation \citep{Limber_approximation}, we can define these projected correlation functions using Hankel transforms, such that:
\begin{equation}\label{eq:wgg_spectroscopic}
    w_{\rm{gg}}(r_{\rm {p}}) = \int {\rm d}z \mathcal{W}(z) \int \frac{ {\rm d} k_{\perp} k_{\perp}}{2\pi}J_{0}(k_{\perp} r_{\rm {p}})P_{\rm{gg}}(k_{\perp}, z),
\end{equation}
\begin{equation}\label{eq:wgp_spectroscopic}
    w_{\rm{gp}}(r_{\rm {p}}) = -\int {\rm d}z \mathcal{W}(z) \int \frac{ {\rm d} k_{\perp} k_{\perp}}{2\pi}J_{2}(k_{\perp} r_{\rm {p}})P_{\rm{gI}}(k_{\perp}, z),
\end{equation}
where $J_{0}$ and $J_{2}$ are the 0th and 2nd-order Bessel functions of the first kind, $k_{\perp}$ is the perpendicular wavelength and $\mathcal{W}$ is the projection kernel defined in \citet{Mandelbaum_weight_function}:
\begin{equation}\label{eq:weight_function}
    \mathcal{W}(z) = \frac{n^{i}(z)n^{j}(z)}{\chi^{2}(z)d\chi/dz} \left [ \int {\rm d} z \frac{n^{i}(z)n^{j}(z)}{\chi^{2}(z)d\chi/dz} \right ]^{-1},
\end{equation}
with $n^{i}$ the redshift distribution of the $i$ sample and $\chi(z)$ the comoving distance along the \gls{los} at redshift $z$. In the case of eq.~\ref{eq:wgg_spectroscopic}, the redshift distributions are those of the dense sample, so that $n^{i}$=$n^{j}$. Conversely, for eq.~\ref{eq:wgp_spectroscopic}, two samples are used, the dense and shape samples. Note that the expressions in eq.~\ref{eq:wgg_spectroscopic} and eq.~\ref{eq:wgp_spectroscopic} are valid when we know the exact positions of the objects, that is, for the case of \glspl{spec-z}.

\subsection{Correlation functions with \texorpdfstring{photo-$z$}{photo-z}}\label{sec:Correlation_photometric}

The modelling of projected correlation functions with \gls{photo-z} is more complex than when using \gls{spec-z}. The effect \glspl{photo-z} have on the correlation functions is to smear out the signal due to the scattering of galaxies along the \gls{los}. The amplitude of this scattering depends on the precision of the \glspl{photo-z}. Even though \gls{paus} \glspl{photo-z} have better precision than typical \glspl{photo-z} computed from \gls{bb} surveys, the scatter is still not negligible. One approach to account for this scattering is to increase the range of the \gls{los} integration, as we did in eq.~\ref{eq:pi_binning}, so that the scattered objects are brought back into the correlation. However, this is a subtle exercise, since one cannot arbitrarily increase this range given that uncorrelated objects might enter the correlation function and reduce the \gls{snr}. Here, we follow the procedure in \citet{Observation_red_galaxies_Joachimi} and \citet{Samuroff_photometric_correlations} to implement the effect of \gls{photo-z} on the correlation functions.

For the case of the projected photometric \gls{gc} correlation function, $w_{\rm{gg}}$, the Limber integral of the galaxy power spectrum is defined as:

\begin{equation}\label{eq:cl_wgg_photometric}
\begin{split}
        & C_{\rm{gg}}^{ij}(l\mid z_{1}, z_{2}) = \int_{0}^{\chi _{\rm hor}} {\rm d}\chi'\\ & \frac{p_{n}^{i}(\chi'\mid \chi (z_{1})) p_{n}^{j}(\chi'\mid \chi (z_{2}))}{\chi'^{2}}  P_{\rm{gg}}\left (k = \frac{l+0.5}{\chi'}, z(\chi')\right ),
\end{split}
\end{equation}
where $\chi _{\rm hor}$ is the comoving horizon distance and $p_{n}(\chi'\mid \chi (z_{1}))$ quantifies the error distribution of the dense sample, which corresponds to the \gls{spec-z} distribution at the comoving distance $\chi$ determined by the \gls{photo-z} value $z_{\rm b}=z_{1}$. Here, the \glspl{spec-z} are the ones used in \cite{photo_z_wide_fields} to quantify the accuracy of the \gls{photo-z}.

The expression in eq.~\ref{eq:cl_wgg_photometric} is transformed to angular space via:

\begin{equation}\label{eq:xi_gg_photometric}
    \xi_{\rm{gg}}^{ij}(\theta \mid z_{1}, z_{2}) = \frac{1}{2\pi} \int_{0}^{\infty} {\rm d}l l J_{0}(l\theta)C_{\rm{gg}}^{ij}(l\mid z_{1}, z_{2}).
\end{equation}

Finally, as indicated in \citet{Observation_red_galaxies_Joachimi}, a change of coordinates is performed, such that $z_{\mathrm{m}} = (z_{1}+z_{2})/2$, $r_{\rm {p}}= \theta \chi(z_{\mathrm{m}})$ and $\Pi = c(z_{2}-z_{1})/H(z_{\mathrm{m}})$. With this, we can project the expression in eq.~\ref{eq:xi_gg_photometric} along the \gls{los} as:

\begin{equation}\label{eq:wgg_photometric}
    w_{\rm{gg}}^{ij} (r_{\rm {p}})=\int_{-\Pi_{{\rm max}}}^{\Pi_{{\rm max}}} {\rm d}\Pi \int {\rm d}z_{\mathrm{m}} \mathcal{W}^{ij}(z_{\mathrm{m}})\xi_{\rm{gg}}^{ij}(r_{\rm {p}}, \Pi, z_{\mathrm{m}}),
\end{equation}
where now the redshift distributions entering $\mathcal{W}(z)$ correspond to \gls{photo-z} distributions.

To cover the range of $z_{\mathrm{m}}$ and $\Pi$ in eq.~\ref{eq:wgg_photometric}, we generate a 3-dimensional grid of $\xi_{\rm{gg}}^{ij}(r_{\rm {p}}, \Pi, z_{\mathrm{m}})$. This is done by selecting different $\Pi$ and $z_{\mathrm{m}}$ values along our range of study, transforming them into $z_{1}$ and $z_{2}$ and performing the computations from eq.~\ref{eq:cl_wgg_photometric} and eq.~\ref{eq:xi_gg_photometric}. The $\Pi$-binning scheme is the same as the one employed in eq.~\ref{eq:pi_binning}.

The procedure for the computation of $w_{\rm{gp}}$ is very similar and is expressed in eq.~\ref{eq:cl_wgp_photometric}-\ref{eq:wgp_photometric}. In this case, the error distributions in eq.~\ref{eq:cl_wgp_photometric} correspond to the dense ($p_{n}^{i}$) and shape samples ($p_{e}^{j}$):

\begin{equation}\label{eq:cl_wgp_photometric}
\begin{split}
    & C_{\rm{gI}}^{ij}(l\mid z_{1}, z_{2}) = \int_{0}^{\chi _{\rm hor}} {\rm d}\chi'\\ & \frac{p_{n}^{i}(\chi'\mid \chi (z_{1})) p_{e}^{j}(\chi'\mid \chi (z_{2}))}{\chi'^{2}}  P_{\rm{gI}}\left (k = \frac{l+0.5}{\chi'}, z(\chi')\right ),
\end{split}
\end{equation}

\begin{equation}\label{eq:xi_gp_photometric}
    \xi_{\rm{gp}}^{ij}(\theta \mid z_{1}, z_{2}) = \frac{1}{2\pi} \int_{0}^{\infty} {\rm d}l l J_{2}(l\theta)C_{\rm{gI}}^{ij}(l\mid z_{1}, z_{2}),
\end{equation}

\begin{equation}\label{eq:wgp_photometric}
    w_{\rm{gp}}^{ij} (r_{\rm {p}})=\int_{-\Pi_{{\rm max}}}^{\Pi_{{\rm max}}}  {\rm d}\Pi \int {\rm d}z_{\mathrm{m}} \mathcal{W}^{ij}(z_{\mathrm{m}})\xi_{\rm{gp}}^{ij}(r_{\rm {p}}, \Pi, z_{\mathrm{m}}).
\end{equation}

\subsection{Contaminants to the correlation functions via magnification and galaxy-galaxy lensing}\label{sec:Contaminants_correlation_photometric}

The quantities defined in eq.~\ref{eq:wgg_photometric} and eq.~\ref{eq:wgp_photometric} do not take into account other possible two-point correlation functions that act as contaminants to our quantities of interest. In particular, in the case of position-position correlation functions, magnification acts as a contaminant by modifying the galaxy number density. As a consequence, besides the term described in eq.~\ref{eq:cl_wgg_photometric}, the magnification-magnification term ($\rm{mm}$) and magnification-galaxy position terms ($\rm{gm}$ and $\rm{mg}$) need to be taken into account. Thus, the terms contributing to the so-called source-source correlation function are:
\begin{equation}\label{eq:all_terms_Cnn}
    C_{nn}^{ij}(l) =  C_{\rm{gg}}^{ij}(l) + C_{\rm{gm}}^{ij}(l) + C_{\rm{mg}}^{ij}(l) + C_{\rm{mm}}^{ij}(l),
\end{equation}
where $C_{\rm{gm}}^{ij}=C_{\rm{mg}}^{ji}$.

On the one hand, the galaxy-magnification contribution is defined as:
\begin{equation}
\begin{split}
        & C_{\rm{gm}}^{ij}(l\mid z_{1}, z_{2}) = C_{\rm{mg}}^{ji}(l\mid z_{1}, z_{2}) = 2(\alpha^{j} - 1)\int_{0}^{\chi _{\rm hor}} {\rm d}\chi'\\ & \frac{p_{n}^{i}(\chi'\mid \chi (z_{1})) q_{n}^{j}(\chi'\mid \chi (z_{2}))}{\chi'^{2}}  P_{\mathrm{g}\delta}\left (k = \frac{l+0.5}{\chi'}, z(\chi')\right ),
\end{split}
\end{equation}
where $q_{x}$ is the lensing weight function:

\begin{equation}
        q_{x}(\chi) = \frac{3H_{0}^{2}\Omega_{m}}{2c^{2}}\frac{\chi}{a(\chi)}\int_{0}^{\chi _{\rm hor}} {\rm d}\chi' p_{x}(\chi')\frac{\chi'-\chi}{\chi'},
\end{equation}
and we define the magnification bias $\alpha$, following the theory of magnification in magnitude-limited samples \citep{weak_lensing_review, self_calibration_1, KiDS_IA}, to be proportional to the slope of the faint-end of the logarithmic galaxy count ($\log n$) over a given magnitude range, in our case: 
\begin{equation}\label{eq:alpha_magnification}
    \alpha(i_{\mathrm{AB}}) = 2.5 \frac{{\rm d}\log[n(i_{\mathrm{AB}})]}{ {\rm d}i_{\mathrm{AB}}}.
\end{equation}
Table \ref{tab:IA_cases} shows the $\alpha(i_{\mathrm{AB}})$ values for the different configurations studied in this analysis.

On the other hand, the magnification-magnification contribution is defined as:

\begin{equation}
\begin{split}
        & C_{\rm{mm}}^{ij}(l\mid z_{1}, z_{2}) = 4(\alpha^{i} - 1)(\alpha^{j} - 1)\int_{0}^{\chi _{\rm hor}} {\rm d}\chi'\\ & \frac{q_{n}^{i}(\chi'\mid \chi (z_{1})) q_{n}^{j}(\chi'\mid \chi (z_{2}))}{\chi'^{2}}  P_{\delta\delta}\left (k = \frac{l+0.5}{\chi'}, z(\chi')\right ).
\end{split}
\end{equation}

In the case of position-shape correlation functions, the sources of contamination are the magnification and the galaxy-galaxy lensing, which introduces a signal produced by the lensing of a background galaxy by a foreground galaxy. We can decompose all the terms contributing to the source-shape correlation function as:

\begin{equation}\label{eq:contaminants_wgp}
    C_{ne}^{ij}(l) =  C_{\mathrm{gI}}^{ij}(l) + C_{\mathrm{gG}}^{ij}(l) + C_{\mathrm{mI}}^{ij}(l) + C_{\mathrm{mG}}^{ij}(l).
\end{equation}

The contribution from the galaxy-shear is defined via:

\begin{equation}
\begin{split}
        & C_{\mathrm{gG}}^{ij}(l\mid z_{1}, z_{2}) = \int_{0}^{\chi _{\rm hor}} {\rm d}\chi'\\ & \frac{p_{n}^{i}(\chi'\mid \chi (z_{1})) q_{e}^{j}(\chi'\mid \chi (z_{2}))}{\chi'^{2}}  P_{\mathrm{g}\delta}\left (k = \frac{l+0.5}{\chi'}, z(\chi')\right ),
\end{split}
\end{equation}
while the contribution from the magnification-shear is:

\begin{equation}
\begin{split}
        & C_{\mathrm{mG}}^{ij}(l\mid z_{1}, z_{2}) = 2(\alpha^{i}-1)\int_{0}^{\chi _{\rm hor}} {\rm d}\chi'\\ & \frac{q_{n}^{i}(\chi'\mid \chi (z_{1})) q_{e}^{j}(\chi'\mid \chi (z_{2}))}{\chi'^{2}}  P_{\delta\delta}\left (k = \frac{l+0.5}{\chi'}, z(\chi')\right ).
\end{split}
\end{equation}

Finally, the contribution from the magnification-intrinsic shear is given by:

\begin{equation}
\begin{split}
        & C_{\mathrm{mI}}^{ij}(l\mid z_{1}, z_{2}) = 2(\alpha^{i}-1)\int_{0}^{\chi _{\rm hor}} {\rm d}\chi'\\ & \frac{p_{n}^{i}(\chi'\mid \chi (z_{1})) q_{e}^{j}(\chi'\mid \chi (z_{2}))}{\chi'^{2}}  P_{\delta \rm{I}}\left (k = \frac{l+0.5}{\chi'}, z(\chi')\right ).
\end{split}
\end{equation}

Thus, when constraining the \gls{gc} and \gls{ia} parameters, all the terms expressed in eq.~\ref{eq:all_terms_Cnn} and eq.~\ref{eq:contaminants_wgp} are taken into account. The effects of the contaminants in our analyses can be found in Appendix~\ref{sec:contaminants_contributions}.

\subsection{Likelihood analysis}\label{sec:likelihood}

With the theoretical predictions (Sections~\ref{sec:Correlation_photometric} and \ref{sec:Contaminants_correlation_photometric}) and the measurements (Section~\ref{sec:estimators}) of our observables, we carry out a likelihood analysis to constrain the galaxy bias and the \gls{ia} parameters. The likelihood ($L$) of a data vector ($D$) with respect to a given model ($M$) evaluated at the set of parameters ($\theta$), given a certain covariance matrix ($C$), can be expressed as:
\begin{equation}\label{eq:likelihood_fit}
    -2 \ln L(\theta) = \chi^{2}_{\mathrm{fit}} = \sum_{i=1}^{n}\sum_{j=1}^{n} \left [ D_{i} - M_{i} (\theta) \right ]^{T} C_{ij}^{-1} \left [ D_{j} - M_{j} (\theta) \right ],
\end{equation}
where $n$ is the number of data points and $\chi^{2}_{\mathrm{fit}}$ is the goodness of the fit.

However, given the limited area used for this analysis, we are restricted to a few \gls{jk} regions for each of the configurations we study. This leads to an also limited resolution in the covariance matrix, which is proportional to \citep{SVD_gaztañaga}:
\begin{equation}\label{eq:resolution_covariance}
    \Delta \bar{C} \sim \sqrt{\frac{2}{N_{\mathrm{JK}}}},
\end{equation}
where $\bar{C}_{ij}=C_{ij}/\sqrt{\sigma_{C,i}\sigma_{C,j}}$ is the correlation matrix, with $\sigma_{C, i}$ the square root of the diagonal of $C$ evaluated at position $i$. As a consequence, the smaller the number of \gls{jk} regions, the more noise in $\bar{C}$ might appear, which can lead to instabilities when computing the inverse of the covariance matrix in eq.~\ref{eq:likelihood_fit}. To solve this problem, we perform a \gls{svd} of the normalised covariance matrix:

\begin{equation}
    \bar{C}_{ij} = \left ( U_{ik} \right )^{T} D_{kl} V_{lj},
\end{equation}
where $D_{ij} = \lambda_{i}^{2} \delta_{ij}$ is a diagonal matrix with $\lambda_{i}^{2}$ values in the diagonal, which correspond to the singular values of the decomposition, and $U_{ij}$ and $V_{ij}$ are orthogonal matrices that decompose $\bar{C}$ into $D_{ij}$. The singular values represent the independent number of modes in the covariance matrix, and values too close to zero may cause noise in $\bar{C}$. When computing the inverse covariance in eq.~\ref{eq:likelihood_fit}, we keep the dominant singular values by setting the condition:
\begin{equation}
    \lambda_{i}^{2} > \sqrt{\frac{2}{N_{\mathrm{JK}}}},
\end{equation}
which corresponds to the resolution limit specified in eq.~\ref{eq:resolution_covariance}. Thus, the $\chi^{2}_{\mathrm{fit}}$ in eq.~\ref{eq:likelihood_fit} in the case of applying a \gls{svd} is redefined as:
\begin{equation}\label{eq:chi2_SVD}
    \chi^{2}_{\mathrm{fit, SVD}} = \sum_{i=1}^{n}\sum_{j=1}^{n} \bar{\Delta}_{i}^{T} \bar{C}_{\mathrm{SVD}, ij}^{-1} \bar{\Delta}_{j},
\end{equation}
where $\bar{C}_{\mathrm{SVD}}^{-1}$ is the inverse of the normalised covariance matrix after performing \gls{svd}, and we have defined the difference $\Delta_{i} \equiv D_{i} - M_{i} (\theta)$ and its normalisation as $\bar{\Delta}_{i} = \Delta_{i}/\sigma_{C, i}$. We further define the reduced $\chi^{2}_{\mathrm{fit, SVD}}$ as:
\begin{equation}\label{eq:reduced_chi2_SVD}
    \chi^{2}_{\nu, \mathrm{fit, SVD}} = \frac{\chi^{2}_{\mathrm{fit, SVD}}}{N_{\mathrm{d.o.f.}}},
\end{equation}
where $N_{\mathrm{d.o.f.}}$ corresponds to the number of degrees of freedom, which we define as the number of points after performing the \gls{svd} minus 3, the number of fitted parameters ($b_{1}$, $b_{2}$ and $A_{1}$).

In order to sample the posterior distribution, we employ \texttt{emcee} \citep{emcee}, a python implementation of Goodman and Weare's Affine Invariant Markov chain Monte Carlo (MCMC) ensemble sampler \citep{emcee_goodman}. We use the integrated autocorrelation time ($\tau_{f}$) to quantify the Monte Carlo error. The idea behind this quantity is that the samples of the chain are not independent, and one has to estimate the effective number of independent samples. This number can be quantified as $N/\tau_{f}$, with $N$ the total number of sampled points. From this, a convergence criterion can be established, where we impose that $N/\tau_{f}>100$ and that $\tau_{f}$ changes by less than 1\% every 100 iterations. Besides this convergence criterion, we set the maximum number of iterations to 10000 per “walker'', with a total of 32 “walkers''. However, in all the cases analysed in this study, the convergence is reached before the maximum number of iterations. Finally, when plotting the results from the chain, we set the \texttt{emcee} “thin'' parameter to $\tau_{f}/2$ to select independent points. We note that all these specifications may vary depending on the case under study, and that we follow the recommendations set by the available documentation from \texttt{emcee}\footnote{https://emcee.readthedocs.io/en/stable/}. As a consistency test, we ran some of our chains using Nautilus \citep{nautilus}, which is a nested sampler, and found the same constraints.

Table~\ref{tab:priors_NLA} shows the priors used for $b_{1}$, $b_{2}$ and $A_{1}$ when fitting the joint $w_{\rm{gg}} \cup w_{\rm{gp}}$ data vector. In the case of $b_{1}$ and $A_{1}$, we set similar flat priors as those described in \citet{Samuroff_photometric_correlations}. The differences are that the $b_{1}$ prior is reduced from [0, 3] to [0,2], since our samples usually present a galaxy bias closer to $b_{1}\sim1$, while the $A_{1}$ prior is increased from [-8, 8] to [-12, 12], to account for more extreme $A_{1}$ values. For the 2nd-order galaxy bias, $b_{2}$, we set a Gaussian prior with a mean of 0 and standard deviation of 0.5.

\begin{table}
\centering
\caption{Priors for the galaxy bias and the \gls{nla} model. $\mathcal U$ is a uniform prior, with $\mathcal{U}(\text{min},\text{max})$, while $\mathcal N$ represents a Gaussian with $\mathcal{N}(\mu,\sigma)$.}
\label{tab:priors_NLA}
\begin{tabular}{|l|l|}
\hline
\hline
Parameter & Prior   \\ \hline
$b_1$     &  $\mathcal U(0, 2)$       \\ 
$b_2$     & $\mathcal{N}(0, 0.5)$ \\
$A_1$      & $\mathcal U(-12, 12)$    \\ \hline
\end{tabular}
\end{table}

The scale cuts in our combined data vectors are fixed to $r_{p, \mathrm{min}}=2.0\mpc$ for $w_{\rm{gg}}$, which is typically the minimum separation we can reach with the non-linear galaxy bias model presented in Section~\ref{sec:galaxy_power_spectra}. For the case of $w_{\rm{gp}}$, we tried different options, which are reviewed in more detail in Appendix~\ref{sec:scale_cut_analysis}, but decided to use the same $r_{p, \mathrm{min}}=2.0\mpc$ as in $w_{\rm{gg}}$ for the main constraints in this paper. Note that this is a different approach from what is commonly used in the literature for the \gls{nla} model, where they usually set $r_{p, \mathrm{min}}=6.0\mpc$, such as in \citet{Observation_red_galaxies_Johnston} and \citet{KiDS_IA}, amongst others. Nevertheless, this scale cut is more justified by the limitation in modelling the $w_{\rm{gg}}$ estimator, which is estimated from linear galaxy bias in these previous works, rather than by a limitation of the \gls{nla} model itself. This is proven in Paviot et al. (in prep.), where they include an analysis of the dependence on the scale cuts when using linear and non-linear galaxy bias in the Flagship simulation. Additionally, given the reduced number of data points for the \gls{paus} measurements, a scale cut of $r_{p, \mathrm{min}}=2.0\mpc$ is preferable over a scale cut of $r_{p, \mathrm{min}}=6.0\mpc$, since it increases the available points to fit in our data vector, from 3 to 5, respectively. Finally, we do not set a $r_{p, \mathrm{max}}$ value in the modelling, so that we include the maximum projected separation that we can measure in the \gls{paus} wide fields, since it remains within a range that can be reliably modelled.

\section{Results}\label{sec:results}

The results we obtain from the \gls{gc} and \gls{ia} measurements are shown in this section. Each of the measurements is complemented by the modelled signal fitted to it, following the methodology presented in Sections~\ref{sec:Correlation_photometric}-\ref{sec:likelihood}. The results we present here correspond to the brighter sample (W1+G09+W3 with $i_{\mathrm{AB}}<22$), while the comparison of $A_{1}$ as a function of colour, luminosity, stellar mass and redshift, for both the brighter and fainter samples (W1+W3 with $i_{\mathrm{AB}}<22.5$), is deferred to Appendix~\ref{sec:brighter_vs_fainter}.

To study the dependence of \gls{ia} on physical properties, we split our measurements by colour, defining a red and a blue shape sample. Furthermore, together with the colour split, we study different scenarios where we separate the sample in 3 equipopulated luminosity, stellar mass or redshift bins. As a consequence of these divisions, we obtain a total of 20 different scenarios, for which we measure and model the photometric \gls{gc} and the \gls{ia}. Note that we first split the objects of the shape sample by colour and, later, by equipopulated bins, so as to obtain the same number of objects in each shape subsample. This may result in slightly different boundaries in the luminosity, stellar mass and redshift bins for red and blue objects. As for the dense samples, we also split them in luminosity, stellar mass and redshift bins, to account for their different redshift ranges. However, we do not separate them by colour, as this results in higher \gls{snr}. This does not represent a problem, since the purpose of measuring the \gls{gc} here is to obtain the galaxy bias of the dense sample, which is required in the $w_{\rm{gp}}$ estimator (eq.~\ref{eq:wgp_photometric}). Even though the dense samples are not separated by colour, the slightly different luminosity, stellar mass and redshift ranges for red and blue shape samples may cause the dense subsamples of both cases to differ, although, in general, the vast majority of objects coincide.

Table~\ref{tab:IA_cases} summarises some of the properties of the 20 different configurations by showing the number of objects in the dense (N$_{\mathrm{D}}$) and the shape (N$_{\mathrm{S}}$) samples, the number of \gls{jk} regions, the mean luminosity, stellar mass and redshift, the precision of the \gls{photo-z} as a function of $\sigma_{68}(\Delta z)$ (eq.~\ref{eq:delta_z}), the $\alpha$ parameter (eq.~\ref{eq:alpha_magnification}) and the reduced $\chi^{2}$ of \gls{snr} ($\chi^{2}_{\nu, \mathrm{SNR}}$) for the $w_{\rm{gp}}$ and $w_{\rm{gx}}$ estimates, defined below.

The number of objects in the dense samples is affected by the different luminosity, stellar mass and redshift ranges defined for red and blue shape samples. Thus, N$_{\mathrm{D}}$ varies for red and blue shape galaxies across luminosity, stellar mass and redshift bins. As for N$_{\mathrm{S}}$, it is lower for red than for blue objects, as seen in Fig.~\ref{fig:PAUS_color_selection}. Hence, we expect larger error bars in the measurements of red objects.

The number of angular \gls{jk} regions is the sum of the angular regions of the three fields, which corresponds to 18 angular \gls{jk} regions, as discussed in Section~\ref{sec:estimators}. In the case where we only perform the analysis as a function of colour, this is complemented by 3 cuts in equipopulated redshift bins, defining \gls{jk} regions in the radial direction as well, which results in a total of 54 \gls{jk} regions. Given that the \gls{ia} signal does depend on the redshift, the definition of \gls{jk} regions in the radial direction might affect the covariance matrix, since the different regions might not be independent but conserve some evolution with redshift. This is reviewed in Appendix~\ref{sec:error_estimation}, where we find that the \gls{jk} errors are consistent with the ensemble covariance matrix. The same redshift bins used to define \gls{jk} regions in the colour-based division are also employed to define our samples divided by redshift. Thus, each of these cases have 18 \gls{jk} regions. Finally, in the case where we split our samples either by colour and luminosity or colour and stellar mass bins, we need to define new radial binnings for the JK regions. The reason for this is that each luminosity and stellar mass bin exhibits a different redshift distribution, making it impossible to define the same equipopulated redshift bins. In this case, we decide to divide the redshift space for each scenario in 2 equipopulated redshift bins, accounting for 36 \gls{jk} regions in each case.

The $\sigma_{68}(\Delta z)$ is also shown for each configuration. As expected, in the case of the samples divided by redshift bins, the sample with the higher redshift cut presents a lower \gls{photo-z} precision. In the case of the luminosity and stellar mass bins, we note that they are closely related with redshift, with the lower bins presenting lower redshift means. Nevertheless, this does not necessarily translate into worse $\sigma_{68}(\Delta z)$ values, as these properties are also linked to apparent magnitude, so that fainter objects correspond to lower luminosity and stellar mass bins. As \gls{photo-z} precision also depends on the apparent magnitude, with fainter objects having less precise \gls{photo-z}, this counteracts the relation between luminosity or stellar mass and mean redshift. In that sense, for the case of red objects, the \gls{photo-z} precision increases as we move towards higher luminosity and stellar mass bins, while in the case of blue objects, the \gls{photo-z} precision is quite stable.

We define the reduced $\chi^{2}$ of \gls{snr} ($\chi^{2}_{\nu, \mathrm{SNR}}$) for the $w_{\rm{gp}}$ and $w_{\rm{gx}}$ estimates as:
\begin{equation}\label{eq:chi2_detection}
    \chi^{2}_{\nu, \mathrm{SNR}} = \frac{\sum_{i=1}^{n}\sum_{j=1}^{n}  \bar{D}_{i}^{T} \; \bar{C}_{\mathrm{SVD}, ij}^{-1} \; \bar{D}_{j}}{N_{\mathrm{d.o.f.}}-1},
\end{equation}
where $\bar{D}_{i} = D_{i}/\sigma_{C, i}$ is the normalised data vector ($D_{i}$) over the square root of the diagonal covariance matrix ($\sigma_{C, i}$), and $N_{\mathrm{d.o.f.}}$ is the number of degrees of freedom after performing the \gls{svd} (see Section~\ref{sec:likelihood}). The $\chi^{2}$ employed in this equation is similar to that presented in eq.~\ref{eq:chi2_SVD}, but this time the model is replaced by the null value, in order to check for detection of \gls{ia} signal or systematics. Furthermore, no scale cuts are applied in this case, since we are also interested in the \gls{ia} signal at small scales. It is important to note that this test helps to identify the detection of signal ($w_{\rm{gp}}$) or systematics ($w_{\rm{gx}}$), but it is not a conclusive test. The reason for this is that, in some cases, especially for the systematics, the value of $\chi^{2}_{\nu, \mathrm{SNR}}$ increases because of uncorrelated points in $w_{\rm{gx}}$\footnote{Strictly speaking, if the covariance matrix were exact, uncorrelated fluctuations would not bias $\chi^{2}_{\nu, \mathrm{SNR}}$. However, because our estimated covariance might only capture leading-order terms, residual higher-order contributions may cause such fluctuations to add to $\chi^{2}_{\nu, \mathrm{SNR}}$, especially in the case of systematics.}. Thus, to confirm detection of \gls{ia} or systematics, the signal has to be correlated for different projected separations. Otherwise, we risk misinterpreting noise in the measurements as signal or systematics. However, this test is useful for identifying non-detections, in the sense that a low value of $\chi^{2}_{\nu, \mathrm{SNR}}$ indicates correlations consistent with 0. Regarding the high values of $\chi^{2}_{\nu, \mathrm{SNR}}$ in $w_{\rm{gx}}$ in Table~\ref{tab:IA_cases}, we also measured $\chi^{2}_{\nu, \mathrm{SNR}}$ for $w_{\rm{gx}}$ in the MICE simulation and found that, for the area analysed in this work, some of the values were large for some \gls{paus}-like triplets, but they decreased and became consistent with 0 for larger areas. This supports the idea that finding large values of $\chi^{2}_{\nu, \mathrm{SNR}}$ in $w_{\rm{gx}}$ for \gls{paus} is not unexpected in some cases.

Based on the values of $\chi^{2}_{\nu, \mathrm{SNR}}$ for $w_{\rm{gp}}$, we find a mild detection of \gls{ia} for red objects and no detection for blue galaxies when we only split the sample by colour. The separations by luminosity and by stellar mass bins show an increase of $\chi^{2}_{\nu, \mathrm{SNR}}$ as we move towards brighter and more massive red galaxies, whereas no \gls{ia} signal is observed for blue objects. Finally, the split in redshift bins for red galaxies shows no \gls{ia} detection for the lowest redshift bin, while a clear \gls{ia} signal is seen for the intermediate and highest bins. For blue galaxies, there is again no \gls{ia} detection, except for the intermediate redshift bin, where a mild detection of \gls{ia} seems to be present, although further analyses may be needed for confirmation. As for the values of $\chi^{2}_{\nu, \mathrm{SNR}}$ for $w_{\rm{gx}}$, in general they indicate that we are not affected by systematics. We have analysed the cases where $\chi^{2}_{\nu, \mathrm{SNR}}>2$ for $w_{\rm{gx}}$ (which we do not show here for conciseness), and found that they are driven by uncorrelated points.

\begin{table*}
\centering
\caption{Number of objects in the dense (N$_{\mathrm{D}}$) and shape (N$_{\mathrm{S}}$) samples, number of \gls{jk} regions (N$_{\mathrm{JK}}$), mean luminosity, mean stellar mass, mean redshift, $\sigma_{68}$, magnification bias and $\chi^{2}_{\nu, \mathrm{SNR}}$ for red and blue galaxies for the \gls{ia} cases studied in this paper.}
\label{tab:IA_cases}
\begin{tabular}{|c|c|c|c|c|c|c|c|c|c|c|}
\hline
\hline
Sample & All & Lum. bin 1 & Lum. bin 2 & Lum. bin 3 & $M_{\star}$ bin 1 & $M_{\star}$ bin 2 & $M_{\star}$ bin 3 & $z$ bin 1 & $z$ bin 2 & $z$ bin 3 \\ \hline
N$_{\mathrm{D}}$ red & 460914 & 162321 & 98180 & 186320 & 307905 & 84652 & 64358 & 213232 & 117675 & 130007 \\ 
N$_{\mathrm{D}}$ blue & 460914 & 154322 & 167494 & 127300 & 113610 & 125747 & 218398 & 133762 & 159292 & 167860 \\ 
N$_{\mathrm{S}}$ red & 82269 & 27355 & 25719 & 25611 & 27301 & 26222 & 26942 & 27284 & 27440 & 27545 \\ 
N$_{\mathrm{S}}$ blue & 279336 & 91929 & 89094 & 90457 & 92075 & 92214 & 93257 & 91975 & 93510 & 93851 \\ 
N$_{\mathrm{JK}}$ & 54 & 36 & 36 & 36 & 36 & 36 & 36 & 18 & 18 & 18 \\ 
$\log_{10}$ \textless{}$L/L_{0}$\textgreater $\,$ red & -0.59 & -0.91 & -0.58 & -0.27 & -0.86 & -0.57 & -0.34 & -0.71 & -0.59 & -0.48 \\ 
$\log_{10}$ \textless{}$L/L_{0}$\textgreater $\,$ blue & -0.58 & -1.15 & -0.51 & -0.1 & -1.09 & -0.49 & -0.18 & -1.06 & -0.5 & -0.2 \\ 
$\log_{10}$ \textless{}$M_{\star}/\mathrm{M}_{\odot}$\textgreater $\,$ red & 10.62 & 10.25 & 10.68 & 10.98 & 10.17 & 10.66 & 11.04 & 10.35 & 10.63 & 10.88 \\ 
$\log_{10}$ \textless{}$M_{\star}/\mathrm{M}_{\odot}$\textgreater $\,$ blue & 9.88 & 9.19 & 9.97 & 10.47 & 9.1 & 9.93 & 10.59 & 9.29 & 9.97 & 10.37 \\ 
\textless{}$z$\textgreater $\,$ red & 0.55 & 0.48 & 0.59 & 0.63 & 0.43 & 0.57 & 0.67 & 0.36 & 0.55 & 0.75 \\ 
\textless{}$z$\textgreater $\,$ blue & 0.48 & 0.28 & 0.51 & 0.68 & 0.31 & 0.5 & 0.64 & 0.25 & 0.46 & 0.73 \\ 
$\sigma_{68}$ red & 0.008 & 0.026 & 0.014 & 0.005 & 0.018 & 0.013 & 0.006 & 0.005 & 0.006 & 0.014 \\ 
$\sigma_{68}$ blue & 0.014 & 0.016 & 0.02 & 0.013 & 0.017 & 0.014 & 0.013 & 0.005 & 0.015 & 0.017 \\ 
$\alpha(i_{\mathrm{AB}})$ red & 0.93 & 1.28 & 0.9 & 0.68 & 1.15 & 0.63 & 0.42 & 0.63 & 0.82 & 1.65 \\ 
$\alpha(i_{\mathrm{AB}})$ blue & 0.93 & 1.3 & 0.9 & 0.6 & 1.48 & 1.06 & 0.62 & 0.62 & 0.72 & 1.46 \\
$w_{\rm{gp}}$ $\chi^{2}_{\nu, \mathrm{SNR}}$ red & 2.25 & 2.08 & 3.00 & 3.62 & 1.50 & 2.00 & 3.13 & 0.65 & 6.85 & 3.31 \\
$w_{\rm{gp}}$ $\chi^{2}_{\nu, \mathrm{SNR}}$ blue & 0.56 & 0.81 & 1.55 & 0.71 & 0.91 & 1.09 & 0.55 & 0.68 & 2.78 & 1.68 \\
$w_{\rm{gx}}$ $\chi^{2}_{\nu, \mathrm{SNR}}$ red & 1.06 & 1.49 & 1.84 & 0.97 & 2.21 & 1.28 & 0.61 & 0.31 & 0.83 & 0.78 \\
$w_{\rm{gx}}$ $\chi^{2}_{\nu, \mathrm{SNR}}$ blue & 2.80 & 2.09 & 2.02 & 0.65 & 0.87 & 0.59 & 0.79 & 0.92 & 0.54 & 1.63 \\ \hline
\end{tabular}
\end{table*}

\subsection{Division by colour}

\begin{figure*}
    \centering
    \includegraphics[width=0.98\textwidth]{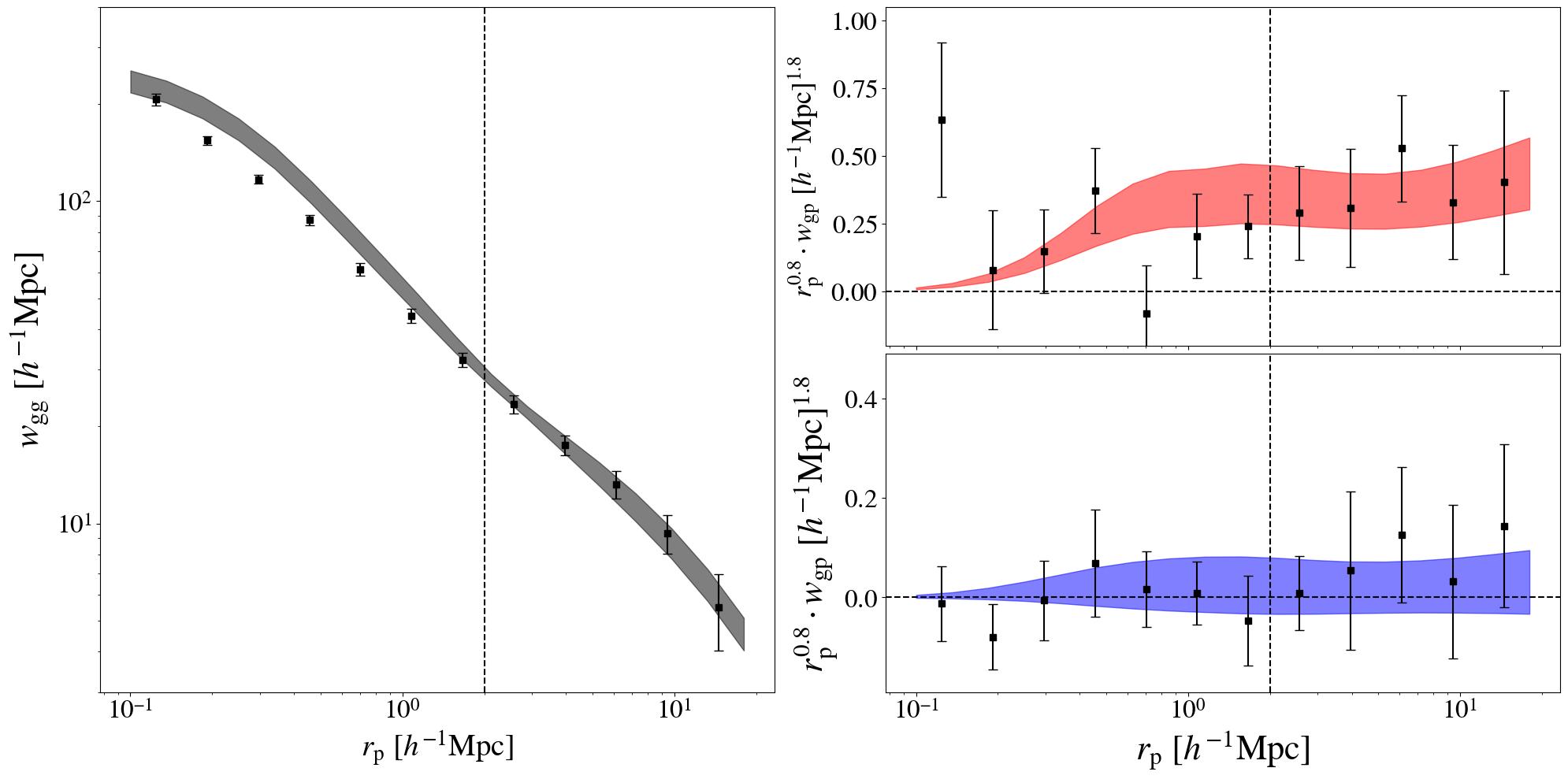}
    \caption{\gls{gc} (left, $w_{\rm{gg}}$) and \gls{ia} (right, $w_{\rm{gp}}$) measurements. The shaded areas show the 1$\sigma$ uncertainty in the best fit using the non-linear galaxy bias and the \gls{nla} models. The dashed vertical lines indicate the scale cuts. The \gls{gc} measurements correspond to the dense sample, while the \gls{ia} ones to the red (top) and blue (bottom) shape samples.}
    \label{fig:PAUS_wgg_wgp_color}
\end{figure*}

Fig.~\ref{fig:PAUS_wgg_wgp_color} shows the \gls{gc} (left) and the \gls{ia} (right) measurements as black dots, in terms of $w_{\rm{gg}}$ and $w_{\rm{gp}}$, respectively, with error bars corresponding to the square root of the diagonal of the \gls{jk} covariance. The \gls{gc} measurements correspond to the dense sample, while the \gls{ia} ones to the red (top) and blue (bottom) shape samples. These measurements are accompanied by shaded areas, which represent the 68\% best fit model to the measurements. The dashed vertical lines show the $r_{\rm {p}}=2\mpc$ scale cut below which we do not fit the model. We find a clear signal of \gls{ia} for red galaxies, while there is alignment consistent with 0 for blue ones, as seen from the values of $w_{\rm{gp}}$ $\chi^{2}_{\nu, \mathrm{SNR}}$ in Table~\ref{tab:IA_cases}. These findings are in line with those from previous analyses by \citet{IA_harry}, who only studied the W3 field.

\begin{figure}
    \centering
    \includegraphics[width=0.48\textwidth]{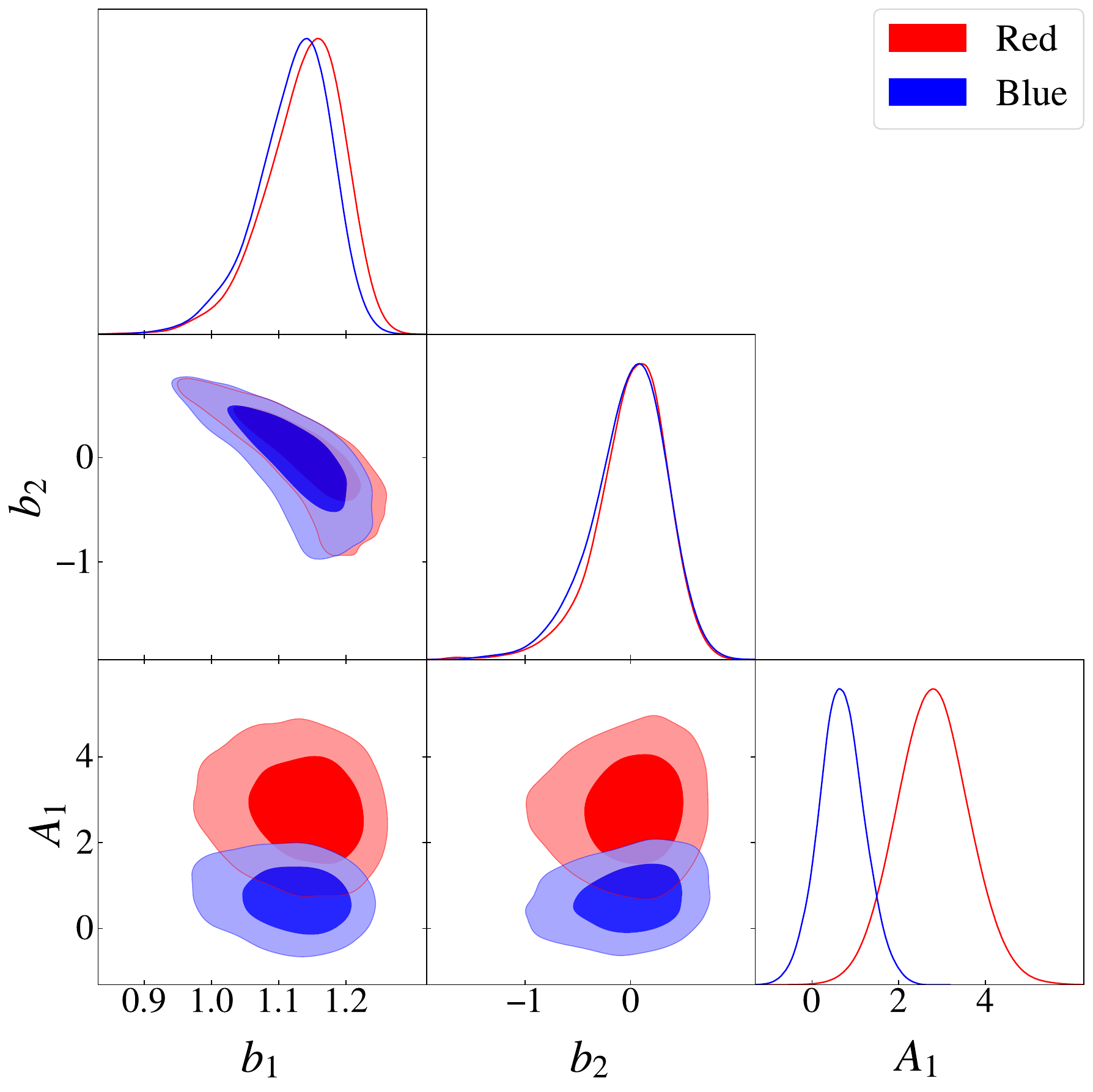}
    \caption{$1\sigma$ and $2\sigma$ contour plots for the galaxy bias, $b_1$ and $b_2$, and the \gls{ia}, $A_1$, parameters for red and blue galaxies.}
    \label{fig:PAUS_contour_plots_color}
\end{figure}

Fig.~\ref{fig:PAUS_contour_plots_color} shows the $1\sigma$ and $2\sigma$ contour plots for the galaxy bias parameters, $b_{1}$ and $b_{2}$, and for the \gls{ia} amplitude, $A_{1}$, for red and blue galaxies. In terms of the $A_{1}$ parameter, its amplitude peaks above 2 for the red sample, with $A_{1}=2.78^{+0.83}_{-0.82}$, while for the blue sample it is consistent with 0, as was also seen in Fig.~\ref{fig:PAUS_wgg_wgp_color}, with $A_{1}=0.68^{+0.53}_{-0.51}$. As discussed in Appendix~\ref{sec:scale_cut_analysis}, the $w_{\rm{gg}} \cup w_{\rm{gp}}$ fits for blue galaxies can be pushed towards smaller scales without an increase in $\chi^{2}_{\nu, \mathrm{fit, SVD}}$, yielding narrower constraints in $A_{1}$. In particular, we obtain a value of $A_{1}=0.53^{+0.32}_{-0.31}$ when setting $r_{p, \mathrm{min}}=1.0\mpc$ in $w_{\rm{gp}}$. In the case of the $b_{1}$ and $b_{2}$ terms, since we are not splitting the dense sample by colour, we obtain very similar values for both red and blue galaxies. The differences we observe might come from the fact that the \gls{snr} of the $w_{\rm{gp}}$ estimator is different for red and blue galaxies, given that there are 3–4 times more blue than red objects. So, even though the constraining power of the galaxy bias parameters mostly comes from $w_{\rm{gg}}$ (given its higher \gls{snr} with respect to $w_{\rm{gp}}$), the $w_{\rm{gp}}$ estimator also affects the constraints through eq.~\ref{eq:wgp_photometric}. Note that the galaxy bias parameter that eq.~\ref{eq:wgp_photometric} constrains is $b_{1}$, since $P_{\rm{gI}}(k,z) = b_{1}P_{\delta \rm{I}}(k,z)$ (eq.~\ref{eq:relation_pgI_PGI}). However, as an indirect consequence, the $b_{2}$ term might be affected through $w_{\rm{gg}}$ by the change in $b_{1}$. This effect will be slightly different depending on the number of objects in the shape sample. Another consideration that might affect the difference between the galaxy bias parameters for red and blue galaxies is related with the term $p_{e}(\chi'\mid \chi (z_{i}))$ in eq.~\ref{eq:cl_wgp_photometric}, which accounts for the error distribution associated to modelling correlation functions with \gls{photo-z}. Ideally, the effect of this term should not depend on colour, luminosity, stellar mass or redshift. However, since it involves knowledge of \glspl{spec-z} and the availability of these does depend on these properties, the constraining power on $b_{1}$ and $b_{2}$ may also be affected by this factor. 

\subsection{Division by colour, luminosity and stellar mass}\label{sec:divistion_color_luminosity}

In this section, we analyse the \gls{gc} and \gls{ia} signals by first splitting the samples by colour and luminosity bins. The luminosity is computed from the apparent $r_{\mathrm{AB}}$-band magnitude from \gls{cfhtlens}, in the case of the W1 and W3 fields, and from \gls{kids}, in the case of the G09 field. First, we compute the absolute magnitude from:

\begin{equation}\label{eq:abs_mag_definition}
    M_{r} = m_{r} - 5 \log_{10}(D_{\mathrm{L}}/10\mathrm{pc})+2.5 \log_{10} \frac{\phi_{r, z}}{\phi_{r, z=0}},
\end{equation}
where $M_{r}$ and $m_{r}$ correspond to the absolute and apparent magnitudes of the $r_{\mathrm{AB}}$-band, respectively. $D_{\mathrm{L}}$ is the luminosity distance and the last term corresponds to the $k$-correction, with $\phi_{r, z}$ and $\phi_{r, z=0}$ being the flux of the galaxy in the $r$-band at redshift $z$ (the redshift at which the object is located) and $z=0$, respectively. The $k$-correction is computed using the \texttt{BCNz} code \citep{BCNz_eriksen}, which outputs the flux of each galaxy for the redshift range $z=[0,2]$. 

A caveat in the computation of the absolute magnitude is the fact that we are using the $r_{\mathrm{AB}}$-band magnitude from different surveys, which may lead to differences in the number counts for the fields under study. This was the case for Fig.~3 from \citet{photo_z_wide_fields}, where there was the need to add an offset to the $i_{\mathrm{AB}}$ magnitude from \gls{kids} in order to match that of \gls{cfhtlens}. A similar scenario happens in this situation, where we need to add an offset of +0.4 in the $r_{\mathrm{AB}}$ absolute magnitude of the \gls{kids} field to match the distribution of the \gls{cfhtlens} fields. Moreover, it is worth noting that we do not account for the evolution of the stellar population with redshift in the computation of eq.~\ref{eq:abs_mag_definition}, usually referred to as ``e-correction'' \citep{Poggianti_e_correction}. As a result, we do not correct for the passive evolution that red galaxies undergo over time, which causes galaxies of a given mass to appear brighter at higher redshifts. This justifies the need to study the \gls{ia} dependence as a function of stellar mass, as will be done in Sec.~\ref{sec:divistion_color_luminosity}.

Once the absolute magnitude is computed through eq.~\ref{eq:abs_mag_definition}, we compute the normalised luminosity with respect to the absolute magnitude $M_{r, 0}=-22$, in order to compare with the literature, so that:

\begin{equation}\label{eq:luminosity_definition}
    M_r - M_{r, 0} = -2.5 \log_{10} \frac{L_{r}}{L_{0}}.
\end{equation}

We generate three equipopulated luminosity bins, after splitting by colour, which generate samples with means of $\log_{10} (L_{r}/L_{0})\sim-0.91 (-1.15)$, $\log_{10}(L_{r}/L_{0})\sim-0.58(-0.51)$ and $\log_{10}(L_{r}/L_{0})\sim-0.27(-0.1)$ for the red (blue) samples. We refer to these luminosity bins as luminosity bin 1, 2 and 3, respectively.

\begin{figure*}
    \centering
    \includegraphics[width=0.98\textwidth]{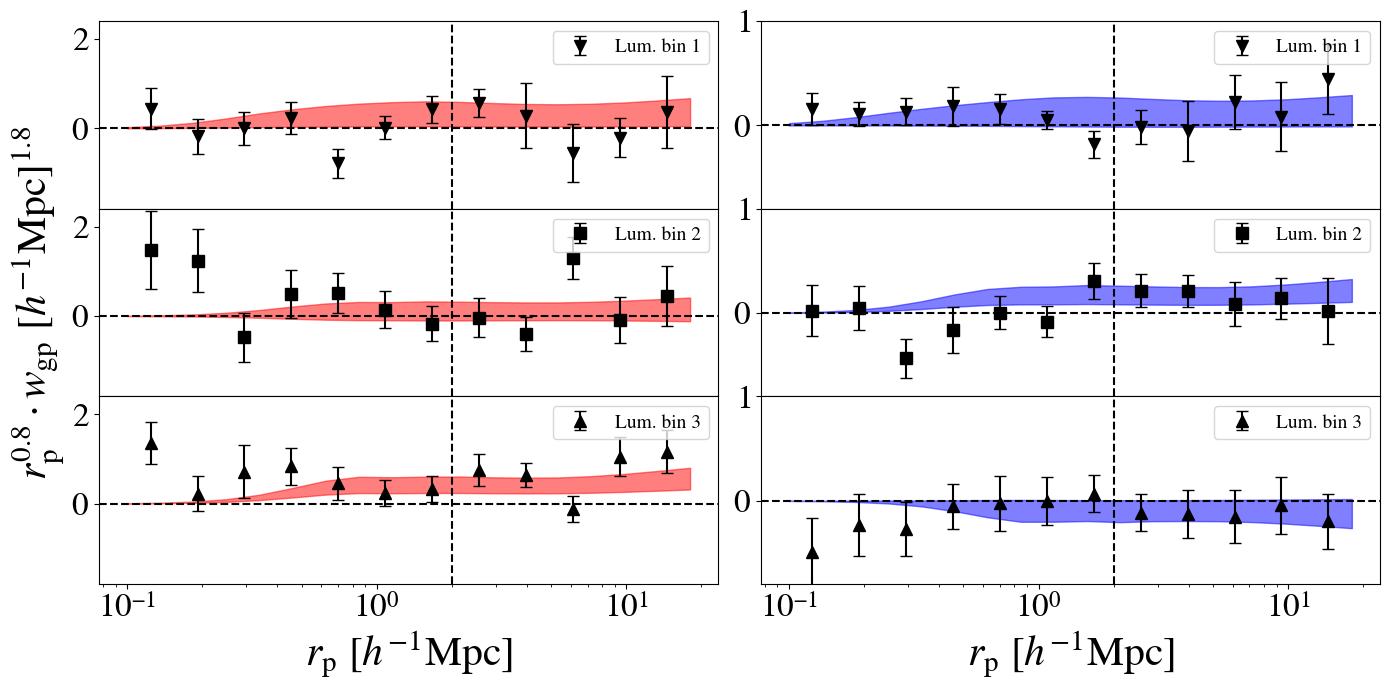}
    \caption{\gls{ia} measurements in terms of $w_{\rm{gp}}$ for the three luminosity bins, for red (left) and blue (right) objects. The shaded areas show the 1$\sigma$ uncertainty in the best fit using the \gls{nla} model. The dashed vertical lines indicate the scale cuts.}
    \label{fig:PAUS_wgg_wgp_color_luminosity}
\end{figure*}

Fig.~\ref{fig:PAUS_wgg_wgp_color_luminosity} shows the $w_{\rm{gp}}$ measurements and modelling, including the division into three luminosity bins. We focus here on the \gls{ia} measurements, instead of also showing the \gls{gc} measurements, as in Fig.~\ref{fig:PAUS_wgg_wgp_color}. However, we note that the \gls{gc} shows lower values for luminosity bin 1, corresponding to the faintest objects, than for bins 2 and 3, which present similar \gls{gc} values, with slightly lower values for luminosity bin 3. On the left-hand side plot, the evolution of the \gls{ia} measurements as a function of luminosity for red objects is indicated by upside down triangles, squares and face up triangles for the luminosity bins 1, 2 and 3, respectively. The same configuration, but for blue galaxies, is shown on the right-hand side plot. As in the case of Fig.~\ref{fig:PAUS_wgg_wgp_color}, the dashed vertical lines at $r_{\rm {p}}=2\mpc$ indicate the minimum separation we include in the modelling of our measurements. The shaded areas show the best fit model to the data, together with its $1\sigma$ uncertainty. In the case of the \gls{ia} measurements for red objects, the signal is consistent with 0 for the two faintest luminosity bins, whereas a positive alignment is observed for luminosity bin 3. This is in agreement with the $\chi^{2}_{\nu, \mathrm{SNR}}$ in Table~\ref{tab:IA_cases}, with a higher $\chi^{2}_{\nu, \mathrm{SNR}}$ in luminosity bin 2 than in luminosity bin 1 explained by the higher variance of the former. In the case of blue galaxies, the alignment is consistent with 0 for luminosity bin 1, although for the majority of the $r_{\rm {p}}$ separations there is a small trend of positive alignment. Luminosity bin 2 is also consistent with 0 for most scales, although there is a hint of positive alignment at $r_{\rm {p}}>2\mpc$, as indicated by the blue line that corresponds to the modelling. Finally, for the brightest luminosity bin, luminosity bin 3, there is a trend of negative alignment for most scales, although the upper limits of the measurements are also consistent with 0.

\begin{figure}
    \centering
    \includegraphics[width=0.48\textwidth]{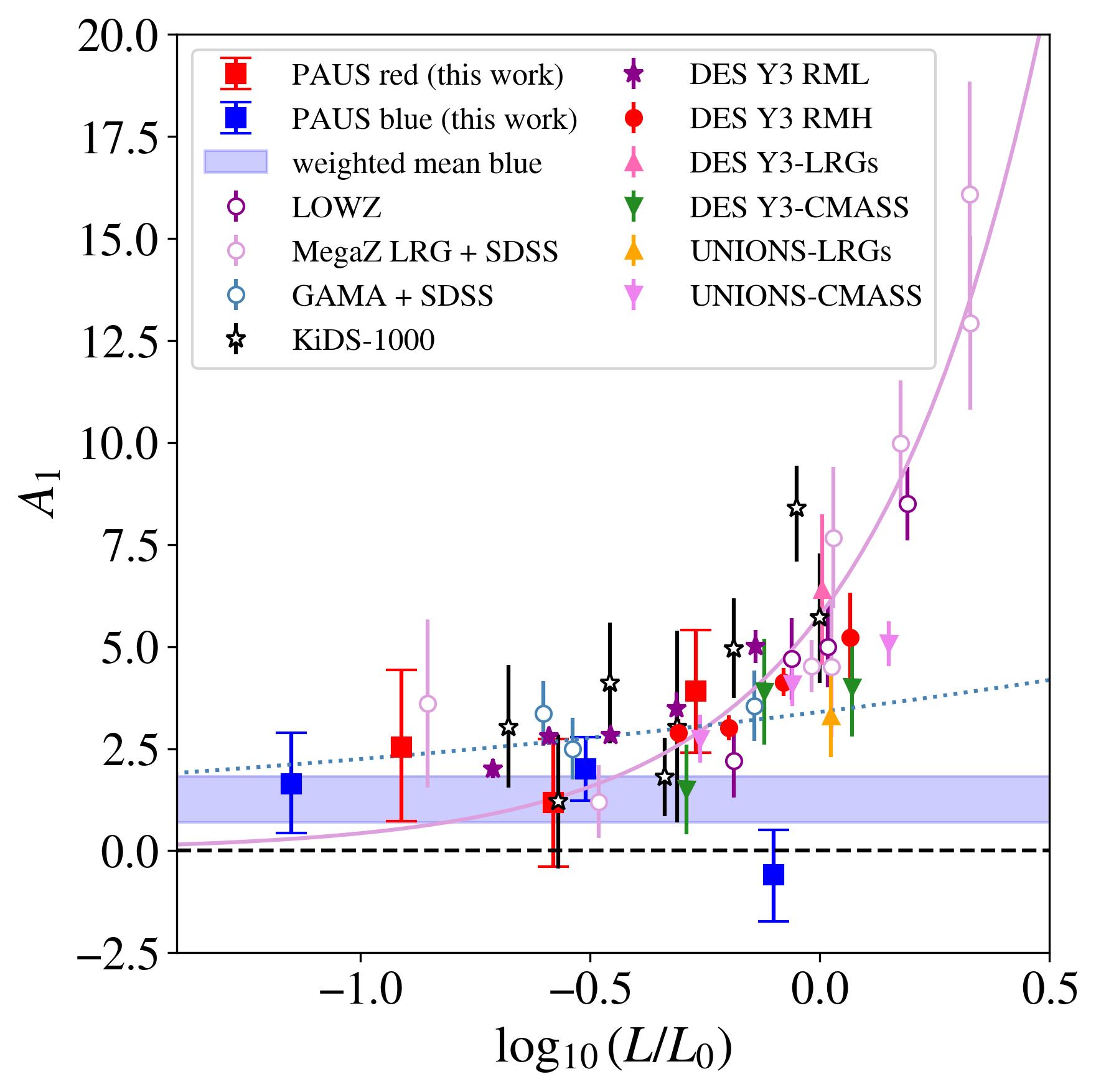}
    \caption[a]{The dependence of $A_{1}$ on luminosity for \gls{paus} red and blue galaxies (red and blue squares, respectively). The shaded blue band shows the weighted mean of the blue fits, accompanied by its $1\sigma$ uncertainty. Previous results from the literature are also shown: LOWZ (dark purple circles, \citealt{BOSS_LOWZ_2}), MegaZ LRG+SDSS (light purple circles, \citealt{Observation_red_galaxies_Joachimi}), GAMA+SDSS (blue circles, \citealt{Observation_red_galaxies_Johnston}), \gls{kids}-1000 (black stars, \citealt{KiDS_IA}), DES Y3 RML (which stands for redMaGiC low-z, purple stars, \citealt{Samuroff_photometric_correlations}), DES Y3 RMH (redMaGic high-z, red circles, \citealt{Samuroff_photometric_correlations}), and DES Y3-eBOSS LRGs (pink triangle up), DES Y3-CMASS (green triangle down), UNIONS-eBOSS LRGs (orange  triangle up) and UNIONS-CMASS (violet triangle down) from \cite{Unions_IA}. The dotted blue line shows a power law from fitting the GAMA+SDSS results (low luminosity), while the solid purple line indicates a fit to the MegaZ LRG+SDSS results (high luminosity). Figure adapted and extended to include the \gls{paus} measurements from \cite{Samuroff_photometric_correlations}.}
    \label{fig:PAUS_IA_bias_color_luminosity}
\end{figure}

The dependence of $A_{1}$ as a function of luminosity is shown in Fig.~\ref{fig:PAUS_IA_bias_color_luminosity}. For the \gls{paus} results (squares), the \gls{ia} amplitude of luminosity bin 1 for red objects is centred at $A_{1}\sim2$, while at $A_{1}\sim1$ for blue galaxies. For the case of the intermediate luminosity bin, the $A_{1}$ amplitude is centred on $A_{1}\sim1$ for the red sample, although consistent with 0, and at $A_{1}\sim1-2$ for the blue one. Finally, for the brightest luminosity bin, there is a clear \gls{ia} amplitude for red objects, centred on $A_{1}\sim4$, with the alignment being in agreement with 0 for blue galaxies. Together with the \gls{paus} results, we include constraints from previous analyses. These constraints correspond to studies analysing the \gls{ia}-luminosity evolution and come from LOWZ \citep{BOSS_LOWZ_2}, MegaZ LRG+SDSS \citep{Observation_red_galaxies_Joachimi}, red and blue GAMA+SDSS \citep{Observation_red_galaxies_Johnston}, \gls{kids}-1000 \citep{KiDS_IA}, DES Y3 RML (redMaGiC low-z, \citealt{Samuroff_photometric_correlations}), DES Y3 RMH (redMaGiC high-z, \citealt{Samuroff_photometric_correlations}), DES Y3-eBOSS LRGs, DES Y3-CMASS, UNIONS-eBOSS LRGs and UNIONS-CMASS (the last four coming from \citealt{Unions_IA}). We note that the DES Y3-eBOSS LRGs and DES Y3-CMASS results used here are taken from \citealt{Unions_IA}, rather than \citealt{Samuroff_photometric_correlations}, because the latter omits the mixed term of the clustering power spectrum and galaxy shape noise in its covariance matrix. As indicated in \citealt{Unions_IA}, this term becomes dominant at $r_{\rm {p}} \gtrsim 29\mpc$, and its omission leads to potential biases and an underestimation of the error bars in the fits from \citealt{Samuroff_photometric_correlations}. We also note that the fits of all these previous analyses take as minimum $r_{\rm {p}}$ scales larger than $6\mpc$, in contrast to our choice of $2\mpc$. These previous fits indicate an $A_{1}-L$ relation based on a broken power law, where the slope for $\log_{10}(L/L_{0})<-0.2$ is shallower than above it. This is indicated by the dotted blue line for low luminosities (which is fitted with the GAMA+SDSS results) and the solid purple line for high luminosities (fitted with the MegaZ LRG+SDSS measurements). This double power law is thought to be driven by the relation between luminosity-to-halo-mass relation, which is also described by a double power law. This was discussed in \cite{maria_cristina_fortuna_halo_mass}, where a single power law relation was found between \gls{ia} amplitude and halo mass. We do not try to fit a power law to our data for two reasons. First, the number of data points is reduced, with 3 luminosity bins, so that the power law one can obtain would not be well constrained. Second, for some points, the error bars are quite large, making a power law fit difficult. Thus, we opt for comparing our measurements with the literature. However, we note that the samples used in the literature correspond to red galaxies and vary in some physical properties with respect to the \gls{paus} samples, so that it is expected that there are differences in the $A_{1}-L$ relation they follow. 

One of the most important goals from the study of \gls{ia} with \gls{paus} is to extend the $A_{1}-L$ relation towards fainter objects. In this regard, the two fainter luminosity bins we use lie in the luminosity range dominated by the GAMA+SDSS power law (dotted blue line), even extending the luminosity range with respect to previous measurements down to $\log_{10}(L/L_{0})\sim -0.9$ for red objects and down to $\log_{10}(L/L_{0})\sim -1.15$ for blue ones. Conversely, the brighter luminosity bin lies at the intersection between the broken power law from previous studies. For the case of red galaxies, the lowest luminosity bin agrees with the power law from the GAMA+SDSS measurements, while the intermediate luminosity bin agrees more with the power law from the MegaZ LRG+SDSS measurements. Nevertheless, both power laws are less than $1\sigma$ away from the intermediate luminosity bin constraint. As for the highest luminosity bin, the \gls{paus} measurements agree with those from previous measurements. Even though we also include the measurements of blue galaxies in this plot, we do not expect them to follow the power laws from previous literature, since those were computed for red populations. In addition to the $A_{1}$ fits for blue galaxies, we show in Fig.~\ref{fig:PAUS_IA_bias_color_luminosity} their weighted mean, with its $1\sigma$ uncertainty, as a shaded blue band, where we find $A_1=1.26^{+0.57}_{-0.57}$, which is consistent with 0 at $\sim2\sigma$.

In order to complement the \gls{ia} dependence as a function of luminosity, we also study the \gls{ia} dependence as a function of stellar mass, where the latter is derived from \texttt{CIGALE}. As in the case of the luminosity, we divide the stellar mass range in 3 equipopulated bins, with means of $\log_{10}(M_{\star}/\mathrm{M}_{\odot})\sim10.17 (9.1)$, $\log_{10}(M_{\star}/\mathrm{M}_{\odot})\sim10.66 (9.93)$ and $\log_{10}(M_{\star}/\mathrm{M}_{\odot})\sim11.04 (10.59)$ for red (blue) galaxies. While the literature has mainly focused on the dependence of \gls{ia} on halo mass \citep{mass_dependence_IA_1, mass_dependence_IA_2, mass_dependence_IA_3, mass_dependence_IA_4, maria_cristina_fortuna_halo_mass}, finding that \gls{ia} increases with it, this property is not available in our dataset. Instead, since stellar mass is available and is correlated with halo mass \citep{relation_stellar_halo_mass_1, relation_stellar_halo_mass_2}, and also depends on galaxy properties linked to \gls{ia}, such as morphology and formation history, we choose to analyse the dependence of \gls{ia} on stellar mass. In that sense, there exist studies of the dependence of \gls{ia} on stellar mass in hydrodynamical simulations \citep{Chisari_hydro, Hilbert_2017, Bate_stellar_mass_2020, Shi_stellar_mass_2021} that, in general, also show increasing alignments with mass, while observational studies still remain scarce. 

As for the relation between luminosity and stellar mass, it is usually described by the stellar mass-to-light ratio ($M_{\star}/L$), so that brighter galaxies usually present higher stellar mass-to-light ratios than fainter ones. Nevertheless, this ratio is dependent on diverse properties, such as colour, luminosity and observed wavelength \citep{Bell_2001, Kauffmann_2003, Bell_2003}. Moreover, as previously discussed, the passive evolution that red galaxies undergo is not accounted for in the determination of the absolute magnitude (eq.~\ref{eq:abs_mag_definition}) and, thus, the \gls{ia} luminosity dependence does not account for the difference in luminosity of similar stellar mass galaxies at different redshifts. As a consequence, even though stellar mass and luminosity are strongly correlated, there is still some scatter between both quantities, which justifies the analysis of the \gls{ia} dependence with stellar mass, providing additional insight into the luminosity dependence.

\begin{figure*}
    \centering
    \includegraphics[width=0.51\textwidth]{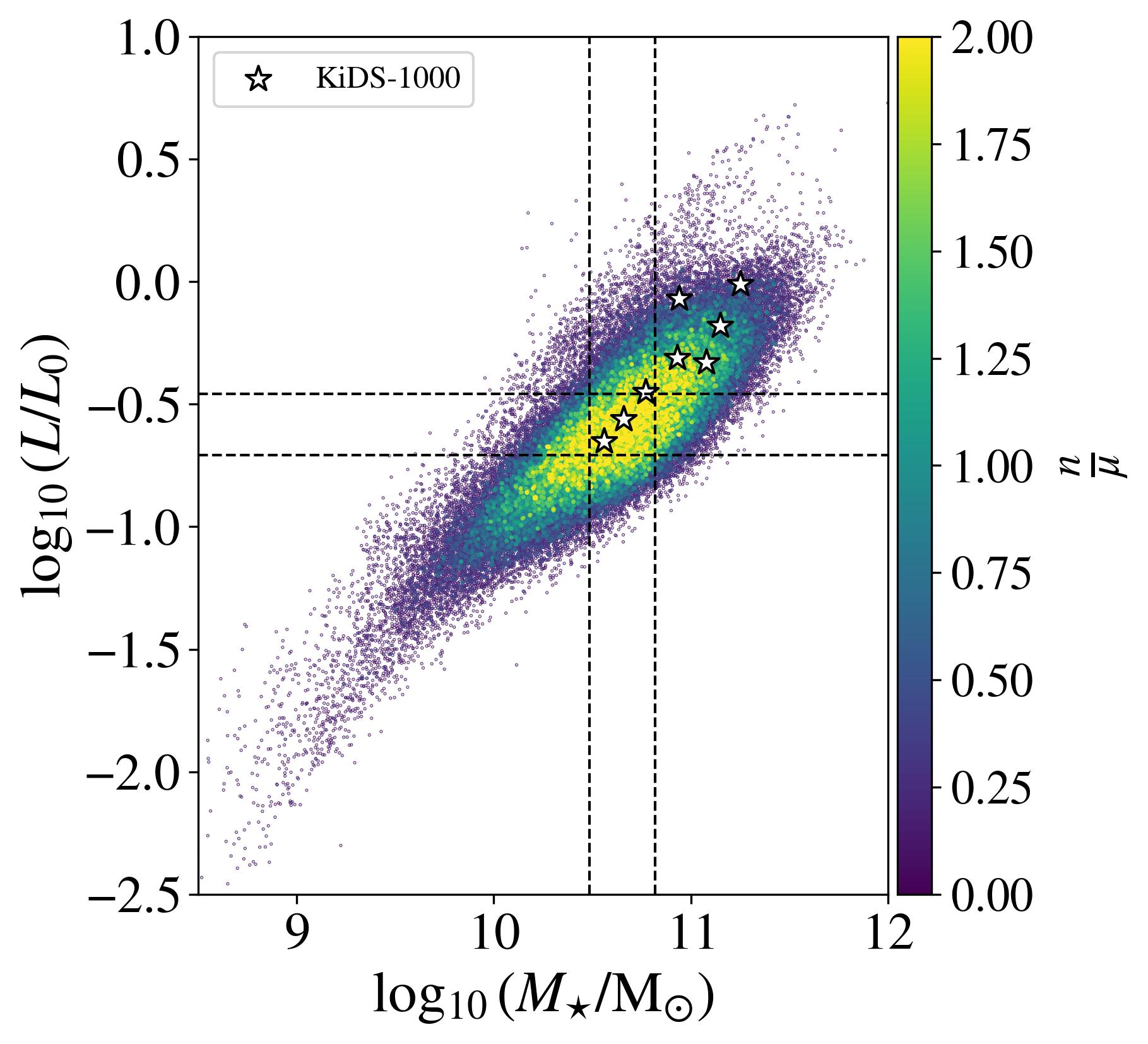}
    \includegraphics[width=0.47\textwidth]{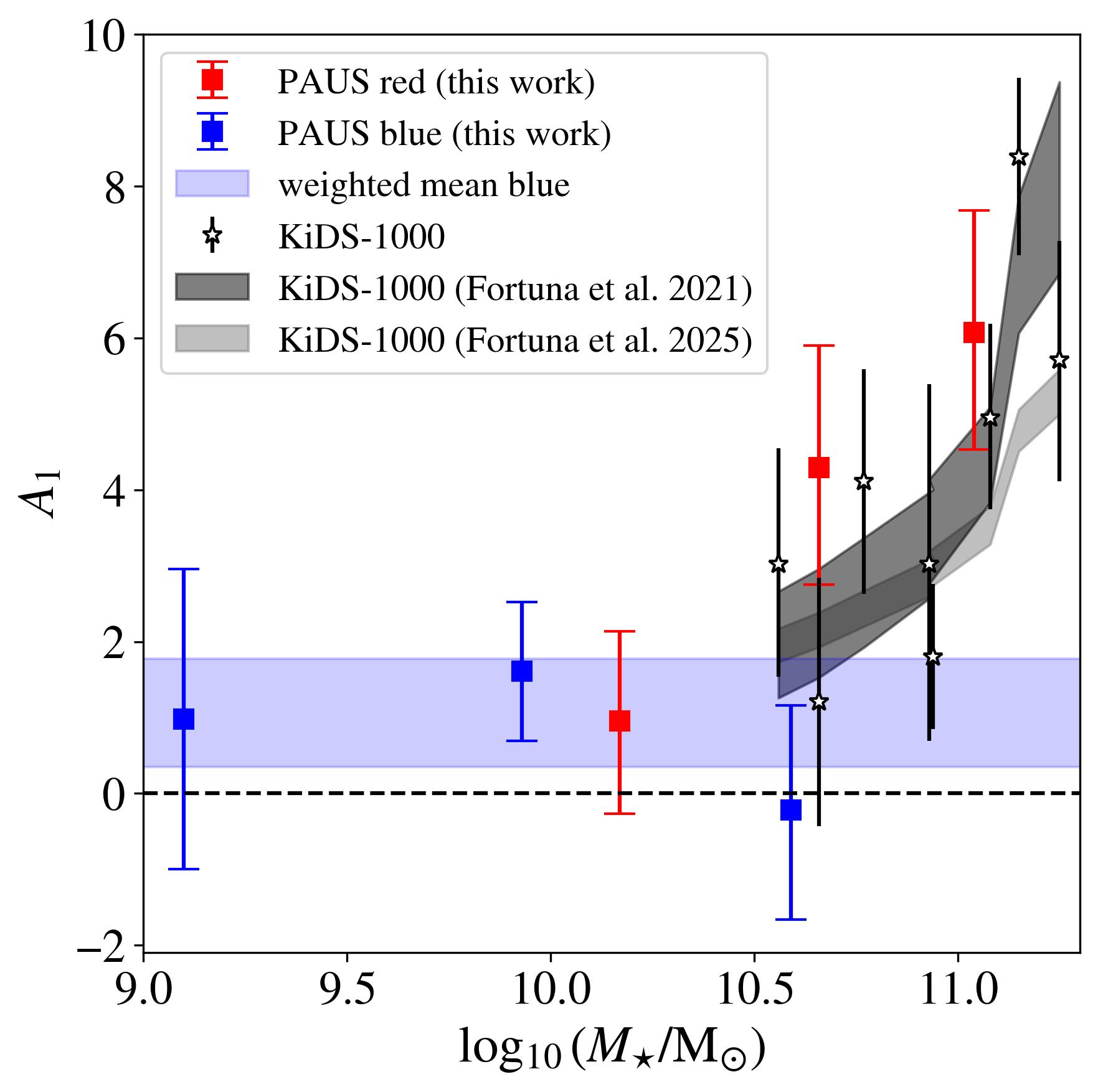}
    \caption{(Left): Luminosity versus stellar mass coloured by the number of objects in each pixel normalised by the median value. (Right): The dependence of $A_{1}$ as a function of stellar mass for \gls{paus} red and blue galaxies (red and blue squares, respectively). The shaded blue band shows the weighted mean of the blue fits, accompanied by its $1\sigma$ uncertainty. The black shading indicates the power law fit, with a $1\sigma$ uncertainty, obtained from the values in \citet{KiDS_IA}, using the halo masses from \citet{maria_cristina_fortuna_halo_mass}, while the grey shading indicates the power law derived by \citet{maria_cristina_fortuna_halo_mass}. In both plots, the stars indicate the points included in \citet{KiDS_IA}.}
    \label{fig:PAUS_IA_bias_color_stellar_mass}
\end{figure*}

The left-hand side of Fig.~\ref{fig:PAUS_IA_bias_color_stellar_mass} shows the luminosity versus the stellar mass of the PAUS wide fields for red objects\footnote{The scatter for the case of blue objects is larger, as expected from spiral galaxies \citep{Bell_2001}, but not shown here for conciseness, since the conclusions do not change.}, showing a clear relation between both quantities, albeit with some scatter. The dashed vertical lines indicate the stellar mass values used to define the equipopulated $M_{\star}$ bins, while the horizontal dashed lines indicate the luminosity values that define the equipopulated luminosity bins. A general trend indicates that the luminosity increases with stellar mass, as expected. Nevertheless, the division we perform assigns different luminosity bins to a single $M_{\star}$ bin and vice versa. Thus, if the \gls{ia} alignment is driven by mass, a cleaner relation is expected when splitting by stellar mass, rather than by luminosity. We also include on the left-hand side of Fig.~\ref{fig:PAUS_IA_bias_color_stellar_mass} the luminous and dense samples used in \citet{KiDS_IA} for comparison with our samples, where the stellar mass estimates are extracted from \citet{maria_cristina_fortuna_halo_mass}. These correspond to one of the faintest samples used to study \gls{ia} in the literature so far, and we can see that \gls{paus} enables us to even reach fainter and less massive objects. The right-hand side of Fig.~\ref{fig:PAUS_IA_bias_color_stellar_mass} shows the evolution of \gls{ia} as a function of stellar mass, where we observe that the \gls{ia} amplitude for the red samples increases with mass. Conversely, in comparison with Fig.~\ref{fig:PAUS_IA_bias_color_luminosity}, the evolution as a function of stellar mass is more clear, as opposed to the luminosity division, where we saw a decrease in the \gls{ia} amplitude for the intermediate luminosity bin. As for the blue objects, we are able to reach very low-mass galaxies, down to $\log_{10}(M_{\star}/\mathrm{M}_{\odot})\sim9$. As in Fig.~\ref{fig:PAUS_IA_bias_color_luminosity}, we include the weighted mean for blue galaxies as a shaded blue band, finding $A_1=1.06^{+0.71}_{-0.71}$, being consistent with 0 by less than $2\sigma$.

Most galaxies in our study fall within the luminosity range $L_r<3.2\times 10^{10}L_\odot h^{-2}$ \, ($\log_{10}(L/L_{0})\lesssim-0.15$), which corresponds to the break in the double power law relation between \gls{ia} and luminosity found by \citet{KiDS_IA}. Consequently, within this luminosity range, the relationship between luminosity and halo mass could be described by a single power law, as shown in Fig.~3 of \citet{maria_cristina_fortuna_halo_mass}. However, due to the complex relationship between stellar mass, halo mass and luminosity, extrapolating our results to stellar mass at higher luminosities is not straightforward. To verify whether the observed trends hold for more luminous galaxies beyond $L_r\sim3.2\times 10^{10}L_\odot h^{-2}$, we incorporate the dense and luminous samples from \citet{KiDS_IA}, also shown on the right-hand side plot of Fig.~\ref{fig:PAUS_IA_bias_color_stellar_mass}. We then fit a power law between $A_1$ and halo mass, using the $A_1$ values from the \citet{KiDS_IA} samples and the halo masses from \citet{maria_cristina_fortuna_halo_mass}, such that:

\begin{equation}
    A_1(M_{h}) = A_{M_{h}} \left ( \frac{M_h}{M_{h,0}} \right )^{\beta_{M_{h}}},
\end{equation}
where $M_{h}$ is the halo mass, $M_{h,0}=10^{13.2}h^{-1}$M$_{\odot}$ is the pivot mass and we obtain $A_{M_{h}}=5.79^{+0.69}_{{-0.69}}$ and $\beta_{M_{h}}=0.65^{+0.20}_{{-0.20}}$. Since \citet{maria_cristina_fortuna_halo_mass} provide a mapping between stellar mass and halo mass, we can express the power law in terms of stellar mass. This is shown in Fig.~\ref{fig:PAUS_IA_bias_color_stellar_mass} as a black shaded area, encompassing the $1\sigma$ uncertainty of the power law fit. Notably, this fit successfully describes the two most massive \gls{paus} bins. Finally, we also compare our $A_1$ fits as a function of stellar mass with the power law derived in \citet{maria_cristina_fortuna_halo_mass} (grey shaded band), which includes the samples from \citet{KiDS_IA} but also from more massive objects \citep{BOSS_LOWZ_2, Observation_red_galaxies_Joachimi, mass_dependence_IA_2}. In this case, the match with our fits is not that good, with the power law predicting less \gls{ia} than our $A_1$ fits. However, the range of masses used for that power law is quite broad and the error bars become very small, making the comparison more difficult when including less massive objects. We leave the derivation of halo masses for \gls{paus} for future work, so that a more realistic comparison with the power law as a function of halo mass can be performed and we can reach the lowest stellar mass bin we defined.

\subsection{Division by colour and redshift}\label{sec:divistion_color_redshift}

We now analyse the results obtained when splitting the samples by colour and redshift. We generate three equipopulated redshift bins, approximately covering the redshift range $z_{\mathrm{b}}\sim[0.1, 0.4]$, $z_{\mathrm{b}}\sim[0.4, 0.6]$ and $z_{\mathrm{b}}\sim[0.6, 1.0]$, with means of $z_{\mathrm{b}}\sim0.36(0.25)$, $z_{\mathrm{b}}\sim0.55(0.46)$ and $z_{\mathrm{b}}\sim0.75(0.73)$ for red (blue) galaxies. We refer to these redshift bins as $z_{\mathrm{b}}$ bin 1, $z_{\mathrm{b}}$ bin 2 and $z_{\mathrm{b}}$ bin 3, respectively.

\begin{figure*}
    \centering
    \includegraphics[width=0.98\textwidth]{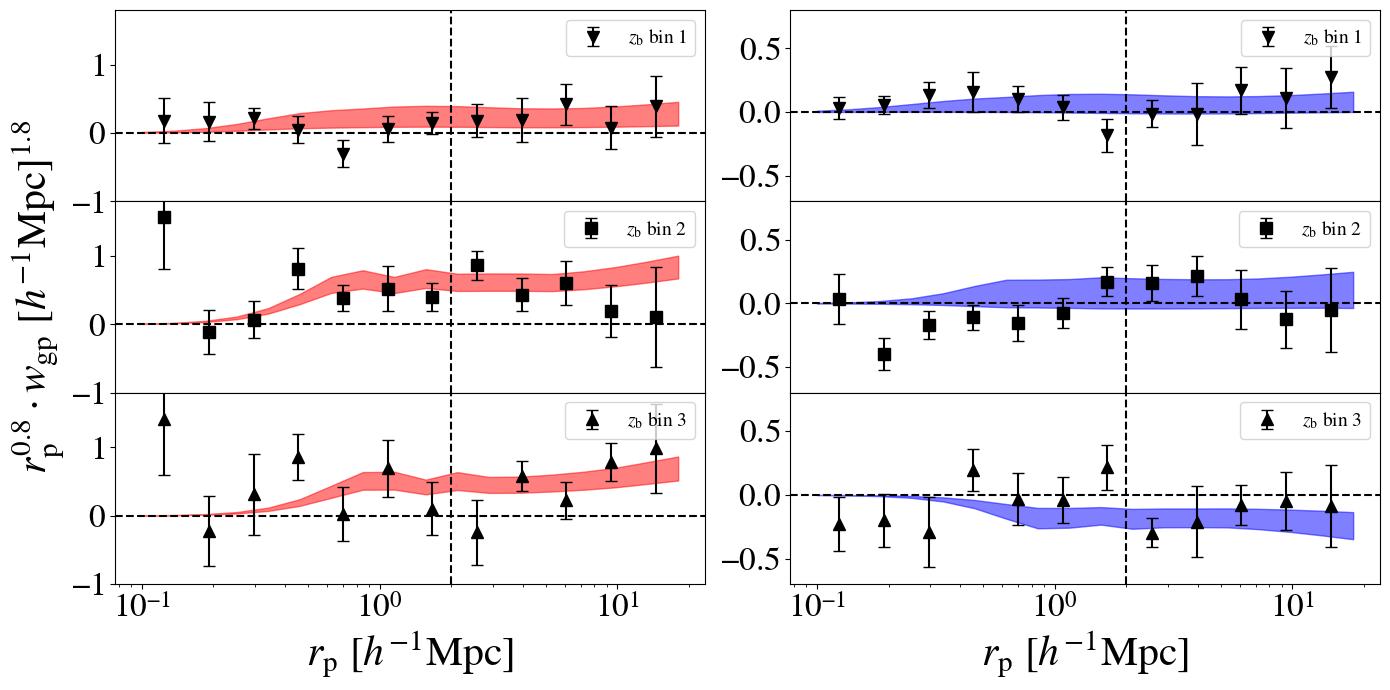}
    \caption{\gls{ia} measurements in terms of $w_{\rm{gp}}$ for the three redshift bins, for red (left) and blue (right) objects. The shaded areas show the 1$\sigma$ uncertainty in the best fit using the non-linear galaxy bias and the \gls{nla} models. The dashed vertical lines show the minimum $r_{\rm {p}}$ used in the modelling.}
    \label{fig:PAUS_wgg_wgp_color_redshift}
\end{figure*}

Fig.~\ref{fig:PAUS_wgg_wgp_color_redshift} shows the $w_{\rm{gp}}$ measurements and modelling for red and blue objects for the different redshift bins. The distribution is the same as in Fig.~\ref{fig:PAUS_wgg_wgp_color_luminosity}. Even though not shown in the figure, a general trend indicating an increase of \gls{gc} signal with redshift is observed. This behaviour is expected, since high-redshift samples tend to have higher \gls{gc} than lower redshift samples for a fixed apparent magnitude, as the galaxies are intrinsically brighter at higher redshift, although this is partially counteracted by the passive evolution of galaxies. In the case of the \gls{ia} measurements for red objects, the lower limits of the $w_{\rm{gp}}$ measurements for the lowest redshift bin are consistent with 0 in most of the $r_{\rm {p}}$ bins, in accordance to Table~\ref{tab:IA_cases}, with a subtle preference towards positive alignment for the central values of the measurements. For $z_{\mathrm{b}}$ bin 2 and $z_{\mathrm{b}}$ bin 3, there is a stronger positive alignment, which is similar between both redshift bins 2 and 3 for $r_{\rm {p}}>2\mpc$. Nevertheless, for $z_{\mathrm{b}}$ bin 2, the values at scales lower than $r_{\rm {p}}=2\mpc$ are higher than for $z_{\mathrm{b}}$ bin 3, which explains the higher $\chi^{2}_{\nu, \mathrm{SNR}}$ seen in Table~\ref{tab:IA_cases} for $z_{\mathrm{b}}$ bin 2. In the case of the blue samples, the $w_{\rm{gp}}$ measurement for the lowest redshift bin is consistent with 0 at all $r_{\rm {p}}$ separations. For $z_{\mathrm{b}}$ bin 2, the alignment is also consistent with 0 at $r_{\rm {p}}>2\mpc$, but with higher variance than in the lower redshift bin case. Nevertheless, for $r_{\rm {p}}<2\mpc$, there seems to be a consistent preference towards negative \gls{ia}, although very weak, which may explain the high $\chi^{2}_{\nu, \mathrm{SNR}}$ observed in Table~\ref{tab:IA_cases} for that case. Finally, for $z_{\mathrm{b}}$ bin 3, at $r_{\rm {p}}>2\mpc$ the signal slightly prefers negative \gls{ia}, with the top error bars in most cases still being consistent with 0. In the case of the signal at $r_{\rm {p}}<2\mpc$, the hint of negative alignment seen in the $z_{\mathrm{b}}$ bin 2 case is almost diluted.

\begin{figure}
    \centering
    \includegraphics[width=0.48\textwidth]{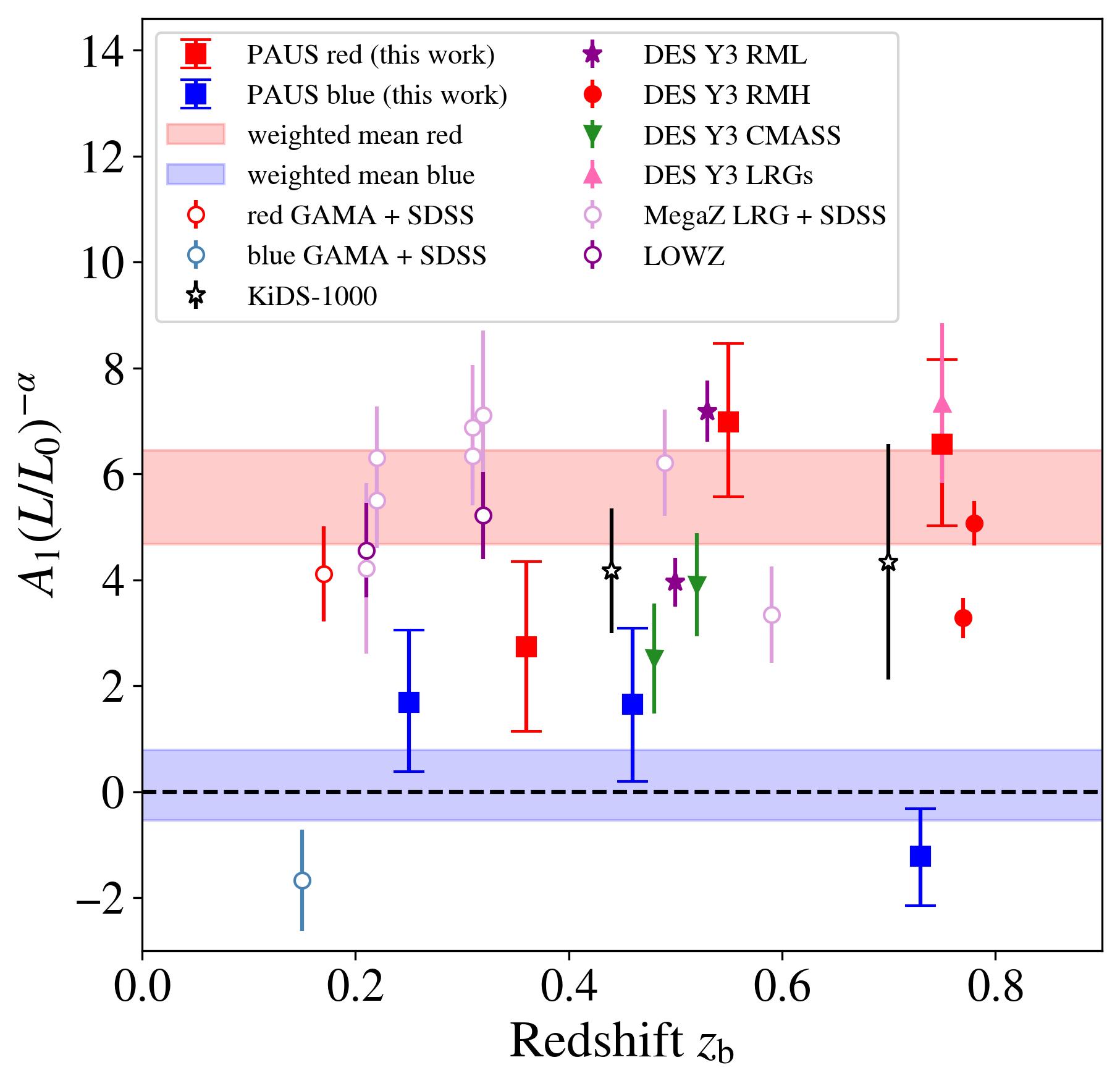}
    \caption{The dependence of $A_{1}$ on redshift for \gls{paus} red and blue galaxies (red and blue squares, respectively), after accounting for the luminosity evolution across redshift bins. The shaded red and blue bands show the weighted mean of the $A_{1}$ fits for red and blue objects, respectively, including their $1\sigma$ error. Previous results from the literature are also shown: GAMA+SDSS (red and blue circles, \citealt{Observation_red_galaxies_Johnston}), \gls{kids}-1000 (black stars, \citealt{KiDS_IA}), DES Y3 RML (purple stars), DES Y3 RMH (red circles), DES Y3-CMASS (green triangle down) and DES Y3-LRGs (pink triangle up) from \citealt{Samuroff_photometric_correlations}, MegaZ LRG+SDSS (pink circles, \citealt{Observation_red_galaxies_Joachimi}) and LOWZ (violet circles, \citealt{BOSS_LOWZ_2}). The power law index $\alpha$ depends on the luminosity of a given sample. For \gls{paus}, GAMA+SDSS and \gls{kids}-1000, $\alpha=0.18$; for DES Y3 RML, DES Y3 RMH and DES Y3-CMASS, $\alpha$ is either 0.18 or 1.13; and for DES Y3-LRGs, MegaZ LRG+SDSS and LOWZ, $\alpha=1.13$.}
    \label{fig:PAUS_IA_bias_color_redshift}
\end{figure}

The evolution of $A_{1}$ as a function of redshift is shown in Fig.~\ref{fig:PAUS_IA_bias_color_redshift} for the \gls{paus} red and blue objects (red and blue squares, respectively). In magnitude-limited samples, as it is the case for \gls{paus}, low-redshift objects exhibit lower mean luminosities than high-redshift ones. As a result, it is not straightforward to determine whether the observed evolution of \gls{ia} amplitudes across redshift-split subsets is driven by differences in mean redshift or mean luminosity. Thus, if the luminosity is not accounted for when studying the evolution of \gls{ia} with redshift, we expect lower \gls{ia} amplitudes for low-redshift samples. One approach to deal with this is to define a subsample within a narrow luminosity range, to later split it in redshift bins. Nevertheless, this method reduces the overall number of objects analysed. Here, we opt to implement a different technique, which consists of multiplying the $A_{1}$ fits, obtained from the redshift bins defined in this section, by the inverse of the power law that describes the \gls{ia} dependence with luminosity. With this approach, we take into account the different mean luminosities of each redshift bin and expect to remove, at first order, the luminosity dependence. Since the mean luminosities of the redshift bins we define are closer to the luminosity range covered by \citealt{Observation_red_galaxies_Johnston}, rather than that of \citealt{Observation_red_galaxies_Joachimi}, we use the power law index from the former, with $\alpha=0.18$. We consider that value when multiplying our $A_{1}$ fits by the inverse of the power law, for both red and blue galaxies\footnote{For blue galaxies, we do not expect the same evolution of $A_{1}$ with luminosity as for red galaxies. However, to homogenise the comparison, we also multiply by the inverse power law the $A_{1}$ values from blue galaxies.}, as it can be seen on the y-label of Fig.~\ref{fig:PAUS_IA_bias_color_redshift}. After accounting for the luminosity variation, we do not find a clear dependence of \gls{ia} with redshift, although it is true that we find more alignment for the two higher redshift bins in red galaxies. Nevertheless, the weighted mean of the $A_{1}$ fits from the \gls{paus} red objects, depicted as a red shaded band with its $1\sigma$ uncertainty, shows that our fits are well described by a constant value of $A_{1} (L/L_{0})^{-0.18}=5.56^{+0.89}_{-0.89}$ by less than $2\sigma$. A similar trend is observed for blue galaxies, where we find an \gls{ia} signal consistent with 0, as indicated by the blue shaded band, with $A_{1} (L/L_{0})^{-0.18}=0.12^{+0.67}_{-0.67}$. Together with the \gls{paus} results, we include fits from the literature coming from \citealt{Observation_red_galaxies_Johnston} (Samples G: $z<0.26$ red and blue from Table B.1) and \citealt{KiDS_IA} (Samples Z1 and Z2 from Table 1), which are in agreement with the shaded bands derived for the \gls{paus} red objects, and lie in a similar luminosity range, so that we employ $\alpha=0.18$ when multiplying by the inverse of the luminosity power law. We also include the fits from \citealt{Samuroff_photometric_correlations} (DES Y3 RML, DES Y3 RMH, DES Y3-CMASS and DES Y3-LRGs from Fig.~E1), which are in general consistent with the fits from \citealt{Observation_red_galaxies_Johnston} and \citealt{KiDS_IA}, but not necessarily with the red shaded band. These discrepancies likely arise from differences in the red galaxy classification, with respect to \gls{paus}, of both redMaGic \citep{redMaGiC_ref} and CMASS \citep{CMASS_sample}, especially in the latter, which includes bluer and fainter objects than redMaGiC. Also, as discussed in Section~\ref{sec:divistion_color_luminosity}, the error bars from the \citealt{Samuroff_photometric_correlations} fits are underestimated, so we would expect a better agreement with the red shaded band derived from \gls{paus} if the error bars were revised. The luminosity range of these \citealt{Samuroff_photometric_correlations} fits is diverse and we either employ $\alpha=0.18$ or $\alpha=1.13$ (the power law index derived from \citealt{Observation_red_galaxies_Joachimi}) depending on the lumninosity value of each sample. Finally, we include the results from MegaZ LRG+SDSS (\citealt{Observation_red_galaxies_Joachimi}, Table 3) and LOWZ (\citealt{BOSS_LOWZ_2}, Table 2), which are in agreement with the red shaded band from the \gls{paus} results and which correspond to high-luminosity samples, for which we consider $\alpha=1.13$ when accounting for the \gls{ia} luminosity evolution.

Table~\ref{tab:A1_IA_cases} summarises the $A_{1}$, $b_{1}$\footnote{The $b_{2}$ values are consistent with 0 for all cases and are not shown in the Table.} and $\chi_{\nu,\mathrm{fit, SVD}}^{2}$ (eq.~\ref{eq:reduced_chi2_SVD}) for the different scenarios studied, where $\chi_{\nu,\mathrm{fit, SVD}}^{2}$ is obtained from the joint fit of $w_{\rm{gg}}$ and $w_{\rm{gp}}$. Note that the $\chi_{\nu,\mathrm{fit, SVD}}^{2}$ varies from values below and above 1, with 1 being the expected value for a $\chi_{\nu}^{2}$ if the fit of a data vector to a model is correctly performed. Nevertheless, checking the $\chi_{\nu}^{2}$ values from previous literature analyses, such as the ones included in Fig.~\ref{fig:PAUS_IA_bias_color_luminosity} and Fig.~\ref{fig:PAUS_IA_bias_color_redshift}, we find $\chi_{\nu,\mathrm{fit, SVD}}^{2}$ values within the range of these previous studies. We also find similar values of $\chi_{\nu,\mathrm{fit, SVD}}^{2}$ for the MICE fits, indicating that it is unlikely that there is a problem with the \gls{paus} data when performing the fits. An explanation on why $\chi_{\nu}^{2}$ values below and above 1 may be acceptable can be found in \citet{reduced_chi2_do_dont}, where they analyse the effect of priors and non-linear models on the computation of $\chi_{\nu}^{2}$ and on the number of degrees of freedom. Also, in some cases, the $\chi_{\nu,\mathrm{fit, SVD}}^{2}>1$ since there is some noise and outlier points in the measurements that increase the values. Finally, in general, the $\chi_{\nu,\mathrm{fit, SVD}}^{2}$ values for blue galaxies are $\chi_{\nu,\mathrm{fit, SVD}}^{2}<1$ since the \gls{ia} signal is close to 0, so that the signal is easier to model.

\begin{table*}
\centering
\caption[$A_{1}$, $b_{1}$ and $\chi_{\nu,\mathrm{fit, SVD}}^{2}$ for red and blue galaxies.]{$A_{1}$, $b_{1}$ and $\chi_{\nu,\mathrm{fit, SVD}}^{2}$ for red and blue galaxies for the \gls{ia} cases studied in this paper.}
\label{tab:A1_IA_cases}
\begin{tabular}{|c|c|c|c|c|c|c|}
\hline
\hline
Sample & $A_{1}$ red & $A_{1}$ blue & $b_{1}$ red & $b_{1}$ blue & $\chi_{\nu,\mathrm{fit, SVD}}^{2}$ red & $\chi_{\nu,\mathrm{fit, SVD}}^{2}$ blue \\ \hline
All & 2.78$_{-0.82}^{+0.83}$ & 0.68$_{-0.51}^{+0.53}$ & 1.14$_{0.07}^{0.05}$ & 1.13$_{0.06}^{0.05}$ & 0.30 & 0.25 \\ 
Lum. bin 1 & 2.54$_{-1.82}^{+1.90}$ & 1.64$_{-1.21}^{+1.24}$ & 1.00$_{0.10}^{0.08}$ & 1.01$_{0.10}^{0.08}$ & 1.05 & 1.03 \\ 
Lum. bin 2 & 1.18$_{-1.57}^{+1.55}$ & 2.00$_{-0.78}^{+0.78}$ & 1.27$_{0.09}^{0.08}$ & 1.28$_{0.06}^{0.05}$ & 3.44 & 0.31 \\ 
Lum. bin 3 & 3.91$_{-1.50}^{+1.50}$ & -0.59$_{-1.14}^{+1.10}$ & 1.28$_{0.06}^{0.05}$ & 1.33$_{0.06}^{0.05}$ & 3.01 & 0.14 \\ 
$M_{\star}$ bin 1 & 0.95$_{-1.21}^{+1.19}$ & 0.98$_{-1.98}^{+1.98}$ & 1.12$_{0.07}^{0.05}$ & 0.85$_{0.17}^{0.12}$ & 0.29 & 1.00 \\ 
$M_{\star}$ bin 2 & 4.30$_{-1.55}^{+1.61}$ & 1.61$_{-0.91}^{+0.91}$ & 1.43$_{0.07}^{0.06}$ & 1.26$_{0.06}^{0.05}$ & 0.82 & 0.39 \\ 
$M_{\star}$ bin 3 & 6.07$_{-1.54}^{+1.61}$ & -0.22$_{-1.44}^{+1.38}$ & 1.78$_{0.08}^{0.07}$ & 1.46$_{0.06}^{0.06}$ & 0.28 & 0.26 \\ 
$z$ bin 1 & 2.04$_{-1.20}^{+1.20}$ & 1.09$_{-0.84}^{+0.88}$ & 1.01$_{0.10}^{0.08}$ & 0.97$_{0.13}^{0.10}$ & 0.30 & 0.54 \\ 
$z$ bin 2 & 5.47$_{-1.10}^{+1.16}$ & 1.35$_{-1.19}^{+1.16}$ & 1.17$_{0.12}^{0.10}$ & 1.14$_{0.10}^{0.08}$ & 1.61 & 0.74 \\ 
$z$ bin 3 & 5.38$_{-1.26}^{+1.31}$ & -1.11$_{-0.86}^{+0.82}$ & 1.38$_{0.08}^{0.08}$ & 1.39$_{0.09}^{0.08}$ & 1.96 & 0.77 \\ \hline
\end{tabular}
\end{table*}

\section{Conclusions}\label{sec:conclusions}

The \glsentrylong{ia} (\gls{ia}) of galaxies remain a major systematic in the era of precision cosmology, mimicking the weak gravitational lensing signal and biasing cosmological analyses. Therefore, it is of utmost importance to characterise \gls{ia} for different kinds of galaxy populations, in order to be able to correct for them.

In this paper, we measure the photometric \gls{gc} and \gls{ia} signals from the \gls{paus} wide fields in the redshift range $0.1<z_{\mathrm{b}}<1.0$, down to $i_{\mathrm{AB}}<22$, and approximately covering $\sim 400000$ objects with galaxy shapes. We analyse the \gls{gc} and \gls{ia} by performing sample divisions based on colour, luminosity, stellar mass and redshift.

The \gls{gc} and \gls{ia} signals are measured by projecting 3-dimensional correlation functions, employing the \gls{paus} \gls{photo-z} estimated in \citet{photo_z_wide_fields} and the galaxy shapes extracted from the \gls{kids} and \gls{cfhtlens} external catalogues, and calibrated to account for the \gls{psf} and the multiplicative bias. The colour separation between red (passive)  and blue (active) galaxies is defined by combining a NUV$rK$ diagram, which traces the \gls{ssfr} and dust attenuation, and a spectral type parameter from the \gls{kids} and \gls{cfhtlens} catalogues.

We model the \gls{gc} and \gls{ia} signals in order to obtain constraints on the galaxy bias and \gls{ia} parameters. For that, we use correlation functions that account for signal dilution due to the lower precision of \gls{photo-z}, while also incorporating contaminant terms from magnification and shear.

We include consistency tests to ensure that both our measurements and constraints are robust. For this purpose, we employ the MICE simulation, generating a galaxy mock that both resembles the galaxy populations obtained in \gls{paus} and reproduces the \gls{paus}-like \gls{photo-z}. The consistency tests involve validating the random catalogues used in the measurements, checking that the use of \gls{photo-z}, instead of \gls{spec-z}, does not bias our constraints, and assessing that the error estimation in our \gls{paus} measurements is robust.

The analysis is performed for three scenarios. First, for a colour-based division analysis, we measure \gls{ia} for red galaxies and find \gls{ia} amplitudes consistent with $A_{1}=2.78_{-0.82}^{+0.83}$, while, for blue galaxies, we observe a null \gls{ia} signal, with $A_{1}=0.68_{-0.51}^{+0.53}$.

Second, besides the colour split, we bin by luminosity and stellar mass, defining three equipopulated regions in both parameters. Our results extend the $A_{1}$ dependence towards lower luminosities and lower mass objects, with respect to the literature, while still being consistent with previous work at higher luminosities and masses. For the case of the luminosity dependence, for red galaxies, there is an overall trend of \gls{ia} increase with it, although there is a decrease of \gls{ia} amplitude in the intermediate luminosity bin. For the case of the stellar mass dependence, for red galaxies, we see a clearer increase of \gls{ia} with mass. For blue galaxies, \gls{ia} are consistent with 0 for both luminosity and stellar mass dependence.

Finally, in addition to the colour-based division, we split the sample into three equipopulated redshift bins. Red galaxies show slight alignment in the lowest redshift bin, while the intermediate and highest redshift bins yield a similar \gls{ia} amplitude of $A_{1}\sim5$. However, variations in \gls{ia} with redshift are affected by differences in the luminosity distributions across redshift bins, which we try to correct by applying a luminosity-dependent scaling to the IA fits, based on the inverse of the assumed power law relation. After applying this correction, the dependence with redshift for red objects is compatible with a constant value. For blue galaxies, measurements generally show no signal.

The results presented in this work help constrain an unexplored regime dominated by low-luminosity and low-mass galaxies. As upcoming stage-IV surveys will observe fainter and less massive galaxies than stage-III surveys, it is essential to accurately constrain \gls{ia} for these relevant galaxy populations. \gls{paus} enables this by providing informative \gls{ia} priors, ensuring unbiased cosmological analyses.

\section*{Acknowledgements}

The authors would like to thank Benjamin Camacho, Fabian Hervas Peters, Harry Johnston, Simon Samuroff, Kai Hoffmann, Jonathan Blazek, Elizabeth Gonzalez, Diego García Lambas and Martin Kilbinger for useful conversations about this project. 

DNG, MC and EG acknowledge support from the Spanish Ministerio de Ciencia e Innovacion (MICINN), project PID2021-128989NB. DNG and H. Hoekstra acknowledge support from the European Research Council (ERC) under the European Union’s Horizon 2020 research and innovation program with Grant agreement No. 101053992. MS acknowledges support by the State Research Agency of the Spanish Ministry of Science and Innovation under the grants 'Galaxy Evolution with Artificial Intelligence' (PGC2018-100852-A-I00) and 'BASALT' (PID2021-126838NB-I00) and the Polish National Agency for Academic Exchange (Bekker grant BPN/BEK/2021/1/00298/DEC/1). This work was partially supported by the European Union's Horizon 2020 Research and Innovation program under the Maria Sklodowska-Curie grant agreement (No. 754510). H. Hildebrandt is supported by a DFG Heisenberg grant (Hi 1495/5-1), the DFG Collaborative Research Center SFB1491, an ERC Consolidator Grant (No. 770935), and the DLR project 50QE2305. BJ acknowledges support by the ERC-selected UKRI Frontier Research Grant EP/Y03015X/1. ME acknowledges funding by MCIN with funding from European Union NextGenerationEU (PRTR-C17.I1) and by Generalitat de Catalunya. JC acknowledges support from the grant PID2021-123012NA-C44 funded by MCIN/AEI/ 10.13039/501100011033 and by “ERDF A way of making Europe”. FJC acknowledges support from the Spanish Plan Nacional project PID2022-141079NB-C31. CP acknowledges support from the Spanish Plan Nacional project PID2022-141079NB-C32. PR acknowledges the support by the Tsinghua Shui Mu Scholarship, the funding of the National Key R\&D Program of China (grant no. 2023YFA1605600), the science research grants from the China Manned Space Project with No. CMS-CSST2021-A05 and the Tsinghua University Initiative Scientific Research Program (No. 20223080023). The PAU Survey is partially supported by MINECO under grants CSD2007-00060, AYA2015-71825, ESP2017-89838, PGC2018-094773, PGC2018-102021, PID2019-111317GB, SEV-2016-0588, SEV-2016-0597, MDM-2015-0509 and Juan de la Cierva fellowship and LACEGAL and EWC Marie Sklodowska-Curie grant No 101086388 and no.776247 with ERDF funds from the EU Horizon 2020 Programme, some of which include ERDF funds from the European Union. IEEC and IFAE are partially funded by the CERCA and Beatriu de Pinos program of the Generalitat de Catalunya. Funding for PAUS has also been provided by Durham University (via the ERC StG DEGAS-259586), ETH Zurich, Leiden University (via ERC StG ADULT-279396 and Netherlands Organisation for Scientific Research (NWO) Vici grant 639.043.512), University College London and from the European Union's Horizon 2020 research and innovation programme under the grant agreement No 776247 EWC. The PAU data center is hosted by the Port d'Informaci\'o Cient\'ifica (PIC), maintained through a collaboration of CIEMAT and IFAE, with additional support from Universitat Aut\`onoma de Barcelona and ERDF. The authors acknowledge the PIC services department team for their support and fruitful discussions.

\section*{Data Availability}

The PAUS wide field catalogue can be accessed via \href{https://cosmohub.pic.es/home}{CosmoHub} (\citealt{CosmoHub1, CosmoHub2}), through this link: \href{https://cosmohub.pic.es/catalogs/319}{https://cosmohub.pic.es/catalogs/319}. Access to the PAUS public data, including the raw and the reduced images, can be found at \href{https://pausurvey.org/public-data-release/}{https://pausurvey.org/public-data-release/}. The authors will share other data products, such as correlation functions or \gls{gc} and \gls{ia} parameter constraints, upon reasonable requests.



\bibliographystyle{mnras}
\bibliography{999_references} 




\appendix

\section{Comparison of the \texorpdfstring{$A_{1}$}{A1} fits for the brighter and fainter samples}\label{sec:brighter_vs_fainter}

\begin{figure*}
    \centering
    \includegraphics[width=0.98\textwidth]{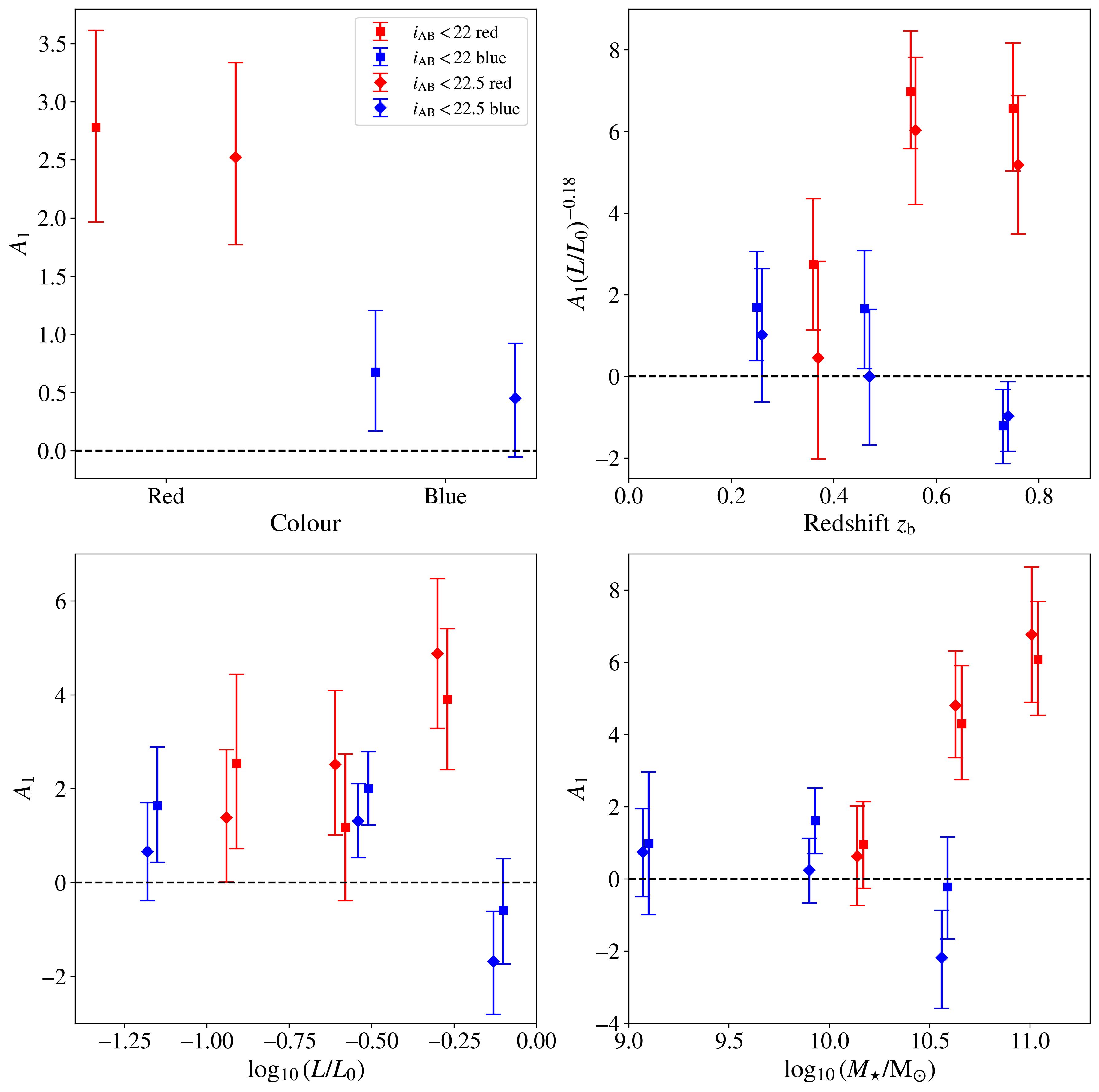}
    \caption{Comparison of the $A_{1}$ values for the brighter (W1+G09+W3 fields with $i_{\rm AB}<22$, squares) and fainter (W1+W3 fields with $i_{\rm AB}<22.5$, diamonds) samples. The four splits analysed in this work are depicted: galaxy colour (top left), galaxy colour and redshift (top right), galaxy colour and luminosity (bottom left), and galaxy colour and stellar mass (bottom right).}
    \label{fig:comparison_brighter_fainter}
\end{figure*}

Here, we include a comparison of the $A_{1}$ amplitudes obtained for both the brighter (W1+G09+W3 fields with $i_{\rm AB}<22$, squares) and the fainter (W1+W3 fields with $i_{\rm AB}<22.5$, diamonds) samples analysed in this paper. This can be seen in Fig.~\ref{fig:comparison_brighter_fainter} for the division in galaxy colour (top left), galaxy colour and redshift (top right), galaxy colour and luminosity (bottom left) and galaxy colour and stellar mass (bottom right). We note that the mean redshift, luminosity and stellar mass values for the brighter and fainter samples are virtually the same, since the brighter sample is the one that we divide in equipopulated bins to define the range limits. Thus, to avoid overlapping, we apply a small shift in the mean redshift, luminosity or stellar mass values when plotting the $A_{1}$ values.

For the case of the separation by galaxy colour, the \gls{ia} amplitudes for both brighter and fainter samples are consistent with one another, with a slight decrease of the amplitude for the fainter sample. This is an indication that the \gls{ia} is quite stable when going from $i_{\mathrm{AB}}<22$ to $i_{\mathrm{AB}}<22.5$ when only splitting by colour. It also implies that the combination of fields is done in a robust way since, if we exclude the G09 field from the analysis, as it is the case for the fainter sample, the results remain almost unaltered. Thus, the alignment in all the three fields is equivalent, as expected for similar galaxy populations.

For the case of the separation by galaxy colour and redshift, the $A_{1}$ amplitudes for the brighter and fainter samples are consistent with one another for red galaxies, where we observe that, although the mean of the fit is higher for the brighter samples, the constraints are between errors. For the case of blue galaxies, this is also the case, with the $A_{1}$ amplitude consistent between brighter and fainter samples and with 0 in all cases.

The evolution of $A_{1}$ as a function of galaxy colour and luminosity indicates that, in all cases, the amplitudes from both brighter and fainter samples are consistent with one another, although their central values do not agree. We note that the decrease in the $A_{1}$ amplitude for the intermediate luminosity bin in the red brighter sample, seen in Section~\ref{sec:divistion_color_luminosity}, is not observed for the fainter sample, which shows a more clear evolution with luminosity. For the case of the $A_{1}$ evolution with galaxy colour and stellar mass, we also find consistent values for the brighter and fainter samples.

\section{Alternative colour classifications}\label{sec:colour_splits_literature}

\begin{figure*}
    \centering
    \includegraphics[width=0.88\textwidth]{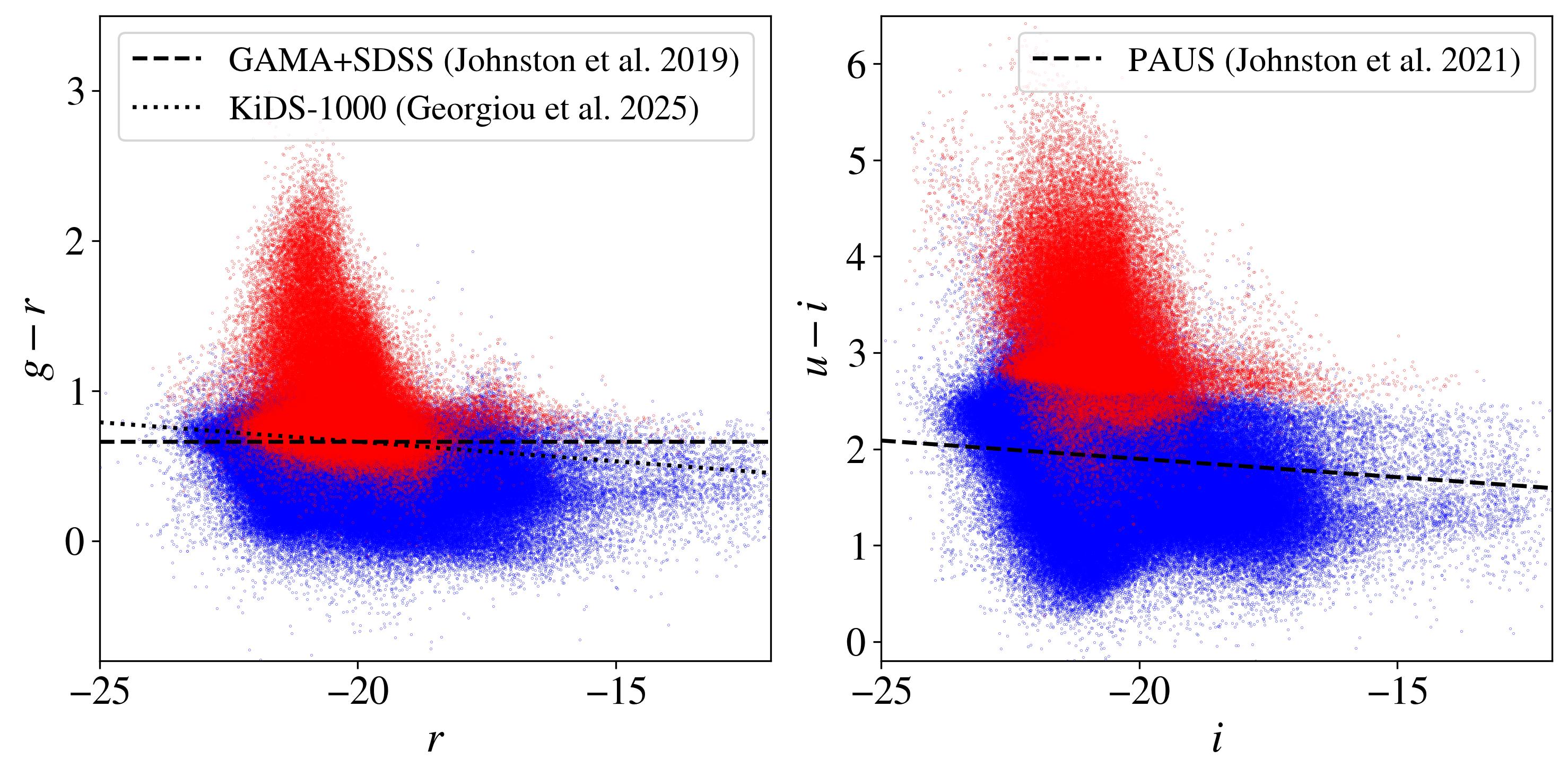}
    \caption{$g-r$ vs. $r$ (left) and $u-i$ vs. $i$ (right) colour-magnitude diagrams showing the red and blue objects obtained from the NUV$rK$ colour split and the $T_{\mathrm{\texttt{BPZ}}}$ parameter (Section~\ref{sec:colour_split_PAUS}). The splits performed by GAMA+SDSS (\citealt{Observation_red_galaxies_Johnston}, dashed line) and KiDS-1000 (\citealt{Georgiou_2025}, dotted line) are depicted on the left plot, while the split performed by the previous \gls{paus} \gls{ia} analysis (\citealt{IA_harry}, dashed line) is depicted on the right plot.}
    \label{fig:comparison_CMD}
\end{figure*}

Here, we compare the red and blue galaxy classification used in this work with alternative approaches from the literature. Fig.~\ref{fig:comparison_CMD} shows the red and blue populations, defined by the NUV$rK$ colour split and the $T_{\mathrm{\texttt{BPZ}}}$ parameter employed in this work (see Section~\ref{sec:colour_split_PAUS}), displayed in the colour-magnitude diagrams $g-r$ vs. $r$ (left) and $u-i$ vs. $i$ (right), where $g$, $r$ and $i$ correspond to rest-frame absolute magnitudes $k$-corrected to $z=0$. In the left plot, we show as a dashed line the separation employed by the GAMA+SDSS \citep{Observation_red_galaxies_Johnston} analysis, where red galaxies are defined as those with $g-r>0.66$. A dotted line shows the separation used in KiDS-1000 \citep{Georgiou_IA}, with $g-r>0.14-0.026r$. In the right plot, we show the cut adopted in the previous \gls{paus} \gls{ia} analysis, where they defined a cut of $u-i>1.138-0.038i$. 

Visual inspection reveals that most galaxies defined as red in this work are also classified as red by the literature. However, there is a larger mismatch for the classification of blue galaxies, with a substantial fraction of galaxies classified as blue by our colour split but considered red by the literature, especially for the previous \gls{paus} analysis. As a result, our colour classification returns a higher percentage of blue galaxies ($\sim$79\%), compared to the cuts employed in the literature, which yield $\sim$61\% (GAMA+SDSS), $\sim$64\% (KiDS-1000) and $\sim$34\% (previous \gls{paus} analysis) of blue galaxies. The fact that we still observe a null \gls{ia} signal in our blue sample suggests that some galaxies classified as red by the literature may belong to the blue population, potentially lowering the observed \gls{ia} signal in their red samples.

Table~\ref{tab:red_blue_objects_all} quantifies the agreement between the classification used in this work and the colour-magnitude classifications from GAMA+SDSS (left table, named J2019), KiDS-1000 (centre table, named G2025) and by the previous \gls{paus} \gls{ia} analysis (right, named J2021). We note that the sum of rows and columns, which corresponds to the number of red or blue galaxies obtained from the colour split from a given survey, does not coincide across tables. This discrepancy arises because some galaxies lack flux measurement in certain bands and are therefore discarded when performing the colour cut. 

\begin{table*}
\centering
\caption{Number of galaxies classified as red and blue in this work in combination with the classification obtained from different colour splits: GAMA+SDSS (J2019, left table), KiDS-1000 (G2025, centre table) and the previous \gls{paus} \gls{ia} analysis (J2021, right table).}
\label{tab:red_blue_objects_all}

\renewcommand{\arraystretch}{1.2} 
\begin{tabular}{@{}ccc@{}} 
\begin{tabular}{|c|c|c|}
\hline
 & Red J2019 & Blue J2019 \\
\hline
Red (this work) & 106765 & 15127 \\
Blue (this work) & 106251 & 366194 \\
\hline
\end{tabular}
&
\begin{tabular}{|c|c|c|}
\hline
 & Red G2025 & Blue G2025 \\
\hline
Red (this work) & 104913 & 16979 \\
Blue (this work) & 87261 & 385184 \\
\hline
\end{tabular}
&
\begin{tabular}{|c|c|c|}
\hline
 & Red J2021 & Blue J2021 \\
\hline
Red (this work) & 96689 & 358 \\
Blue (this work) & 244626 & 223395 \\
\hline
\end{tabular}
\end{tabular}
\end{table*}

\section{Randoms}\label{sec:randoms}

When measuring \gls{gc} and \gls{ia} from eq. \ref{eq:galaxy_clustering_measurement} and \ref{eq:intrinsic_alignment_measurement}, we need a set of random catalogues to help us define the mean density of galaxies. Building these random catalogues is not a trivial exercise, since we want to obtain catalogues where the \gls{gc} is null but that, at the same time, follow the number count distribution as a function of redshift. 

Since in this paper we study projected correlation functions, we want random catalogues both in the angular and the radial directions. To obtain randoms in the angular direction is not a difficult task if we assume that our data has a homogeneous distribution in RA and Dec. This can be done in this analysis, since we are cutting at an $i_{\mathrm{AB}}$ magnitude much lower than the limiting magnitudes of the reference catalogue surveys. If that was not the case, we should develop more advanced techniques, such as the ones described in \cite{SOM_angular_randoms}, where they develop randoms in the angular direction using \glspl{som}. However, in this paper, the challenging task comes from developing a set of random catalogues in the radial direction. Previous attempts to do this, such as \cite{Shaun_randoms, Farrow_randoms, IA_harry}, have applied an algorithm that estimates the radial distribution of the randoms by computing a maximum volume for each galaxy, centred on its current position, around which random points can be distributed. This method is designed to fit the luminosity and overdensity functions as a function of redshift for magnitude-limited samples.

Here, we choose to develop a new technique that relies on the MICE simulation (see Section~\ref{subsec:MICE}). It consists of comparing the \gls{gc} and \gls{ia} signals for two different sets of random catalogues. In the first case, the radial distribution of the random catalogue is generated by sampling over the radial distribution of the full octant of MICE. This corresponds to the average distribution of matter without the effect of local clustering, as a result of the large size of the MICE area. This way, we obtain a representation of a random catalogue that does not carry a \gls{gc} signal. In the second case, the radial distribution of the random catalogue is obtained by smoothing the radial distribution of the triplets of MICE \gls{paus}-like patches (see Section~\ref{subsec:MICE}) with a tophat filter of a given number of Mpc. This case is the one we can reproduce when measuring the observables with the \gls{paus} data, since the first case is only accessible with the MICE simulation. 

Once the two sets of random catalogues are defined, the \gls{gc} and the \gls{ia} signals are measured using both of them and the difference between the signals is analysed. An indication that both sets of random catalogues are similar is that the signals are consistent between one another. We test different levels of radial smoothing for the second case, going from a smoothing of 20 Mpc until 420 Mpc (which corresponds to the radial distance at redshift 0.1, that is, the minimum redshift we consider when measuring our correlation functions), finding the best match with a smoothing of 420 Mpc. 

\begin{figure}
    \centering
    \includegraphics[width=0.48\textwidth]{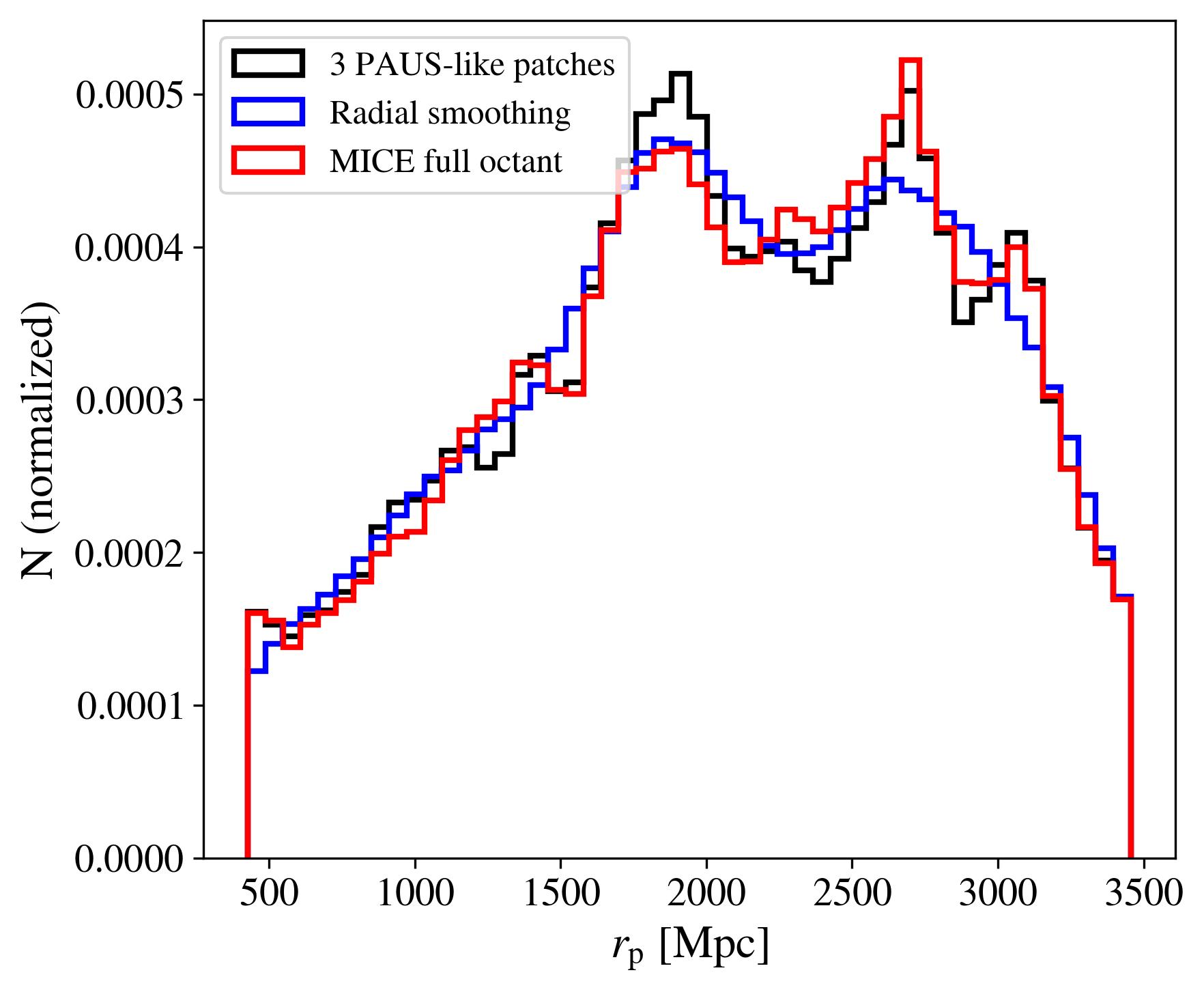}
    \caption{Normalised radial distribution of 3 \gls{paus}-like patches in MICE (black), the random catalogues following the radial distribution of the 3 patches with a smoothing of 420 Mpc (blue) and the radial distribution of the full octant of MICE (red).}
    \label{fig:MICE_radial_distribution_randoms}
\end{figure}

Fig.~\ref{fig:MICE_radial_distribution_randoms} shows the comparison of the normalised radial distribution of three cases: the combination of three randomly selected \gls{paus}-like patches in MICE, the randoms generated when smoothing these three \gls{paus}-like patches with a tophat filter of 420Mpc and the randoms generated from sampling over the MICE full octant. Note how similar the radial distributions are for the two random generation approaches, indicating that a smoothing of 420Mpc yields random catalogues analogous to those when sampling the MICE full octant radial distribution.

\begin{figure*}
    \centering
    \includegraphics[width=0.98\textwidth]{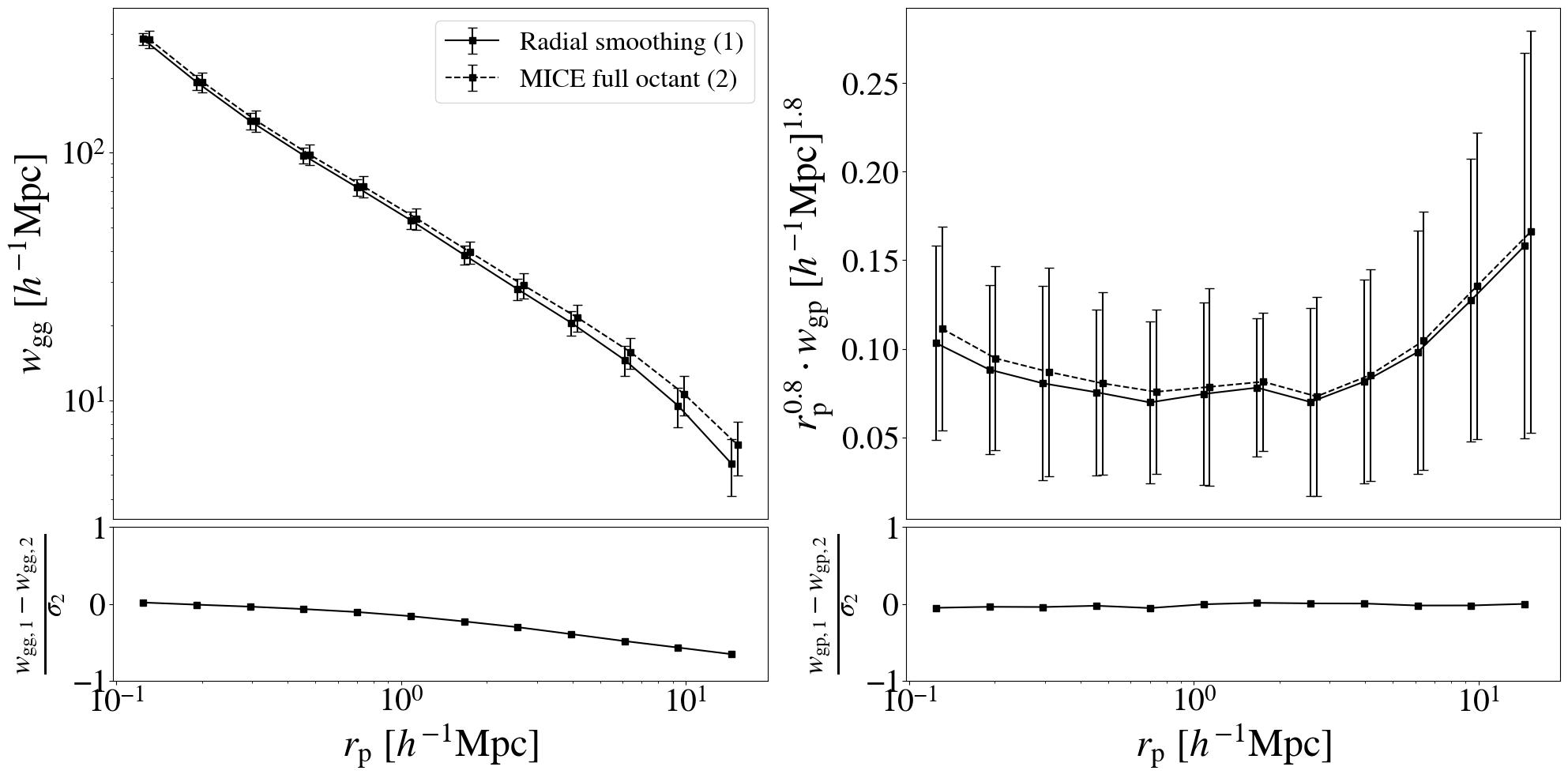}
    \caption{Comparison of the \gls{gc} (left) and \gls{ia} (right) measurements in MICE with the two versions of random catalogues. The measurements are consistent for both types of randoms, with a difference of less than 1 $\sigma$ in the \gls{gc} case and negligible for \gls{ia}, as seen in the bottom panel, which shows the difference between the measurements divided by the error in the full octant case.}
    \label{fig:MICE_comparison_wgg_randoms}
\end{figure*}

\begin{figure}
    \centering
    \includegraphics[width=0.48\textwidth]{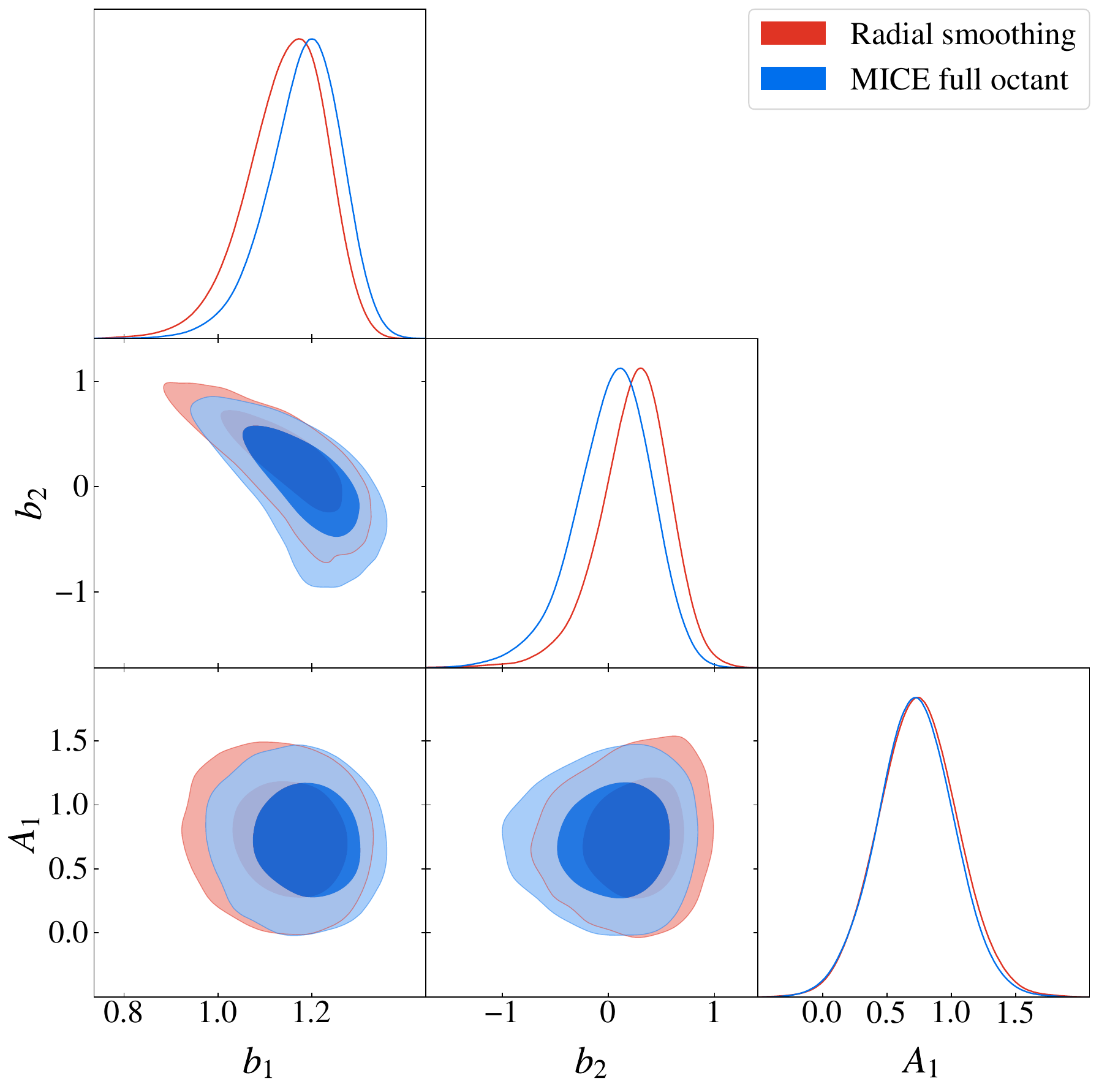}
    \caption[Galaxy bias and \gls{ia} parameters contour plots with different random catalogues.]{Galaxy bias and \gls{ia} parameters contour plots obtained from fitting the \gls{gc} and the \gls{ia} signals in MICE, using the smoothed version of the randoms (red) and the randoms from the MICE full octant (blue).}
    \label{fig:MICE_contour_plot_wgg_randoms}
\end{figure}

The comparison of the projected \gls{gc} and \gls{ia} signals for the different versions of randoms is depicted in Fig.~\ref{fig:MICE_comparison_wgg_randoms}. On the one hand, the top plots of the figure show the comparison of the mean $w_{\rm{gg}}$ (left) and $w_{\rm{gp}}$ (right) signals, obtained from the combination of the 60 triplets of MICE \gls{paus}-like patches. On the other hand, the bottom plots show the difference of the signals divided by the error, which has been computed as the square root of the diagonal of the ensemble covariance of the 60 triplets (see Appendix~\ref{sec:error_estimation} for more details about the ensemble covariance). It can be seen that the \gls{gc} and \gls{ia} signals obtained for both random catalogues are consistent with a difference below 1$\sigma$. Two important remarks can be extracted from this figure. First, the \gls{gc} signal is more affected than the \gls{ia} signal when using different random catalogues. This is expected, since the estimator of $\xi_{\rm{gp}}$ (eq.~\ref{eq:intrinsic_alignment_measurement}) correlates shapes with positions, while the estimator of $\xi_{\rm{gg}}$ (eq.~\ref{eq:galaxy_clustering_measurement}) correlates positions with positions, which is more sensitive to the random catalogue used. Second, although the signals are consistent, the case of the smoothed randoms with a tophat filter yields slightly less signal than the randoms from the MICE full octant. The reason for this is that the smoothed randoms may still carry some clustering, which reduces the total signal from \gls{gc} and \gls{ia}. However, we do not expect this to affect our constraints, as shown in Fig.~\ref{fig:MICE_contour_plot_wgg_randoms}.

The importance of constructing suitable random catalogues comes from the fact that we want to recover the true galaxy bias parameters, since they are fundamental in order to obtain the true \gls{ia} parameters (see eq.~\ref{eq:relation_pgI_PGI} and eq.~\ref{eq:cl_wgp_photometric}). Following with the discussion on the difference between the \gls{gc} and \gls{ia} signals, in Fig.~\ref{fig:MICE_contour_plot_wgg_randoms} we show the constraints on the galaxy bias and the \gls{ia} parameters for different sets of random catalogues. Blue contours show the case when using the randoms sampled from the MICE full octant distribution, while red contours show the case for the randoms obtained when smoothing the radial distribution with a tophat filter. The recovered parameters are consistent with each other, specially for the \gls{ia} amplitude, $A_1$. The \gls{svd} introduced in Section~\ref{sec:likelihood} has also been applied to this case, as well as in Appendix~\ref{sec:Comparison_zb_zs} and \ref{sec:error_estimation}.

\section{Comparison of the \texorpdfstring{photo-$z$}{a} and \texorpdfstring{spec-z}{b} signals in MICE}\label{sec:Comparison_zb_zs}

In this section, we compare the correlation functions in MICE when using \glspl{photo-z} or \glspl{spec-z} and test if the constrained \gls{gc} and \gls{ia} parameters are consistent for both cases.

It is expected that both the \gls{gc} and the \gls{ia} signals are smeared out along the radial direction when using \glspl{photo-z} instead of \glspl{spec-z}. This is explained by the fact that the positions of the objects are quantified with a lower precision when using \gls{photo-z}, and so the correlation between the objects decreases. One way to correct for that is to increase the \gls{los} integration in eq.~\ref{eq:wab_measurement}, in order to recover the pairs of correlated objects that have been spread over this radial direction. However, it is important not to define an excessively wide radial range, since this might turn into an increase of the noise due to the inclusion of uncorrelated pairs of objects. 

We define two radial binnings, one for the spectroscopic and one for the photometric case. On the one hand, we take the spectroscopic radial binning from previous works, such as \citet{Georgiou_IA} and \citet{Observation_red_galaxies_Johnston}, where they define it in the range -60$\mpc$ to 60$\mpc$ in steps of 4$\mpc$. On the other hand, we define the photometric radial binning as in eq.~\ref{eq:pi_binning}, following \citet{IA_harry}. In order to test if the spectroscopic and photometric radial binnings recover the same signal, we measure and compare the \gls{gc} and the \gls{ia} signals in the MICE simulation for both options, following a similar procedure as in previous Section~\ref{sec:randoms}. 

\begin{figure*}
    \centering
    \includegraphics[width=0.98\textwidth]{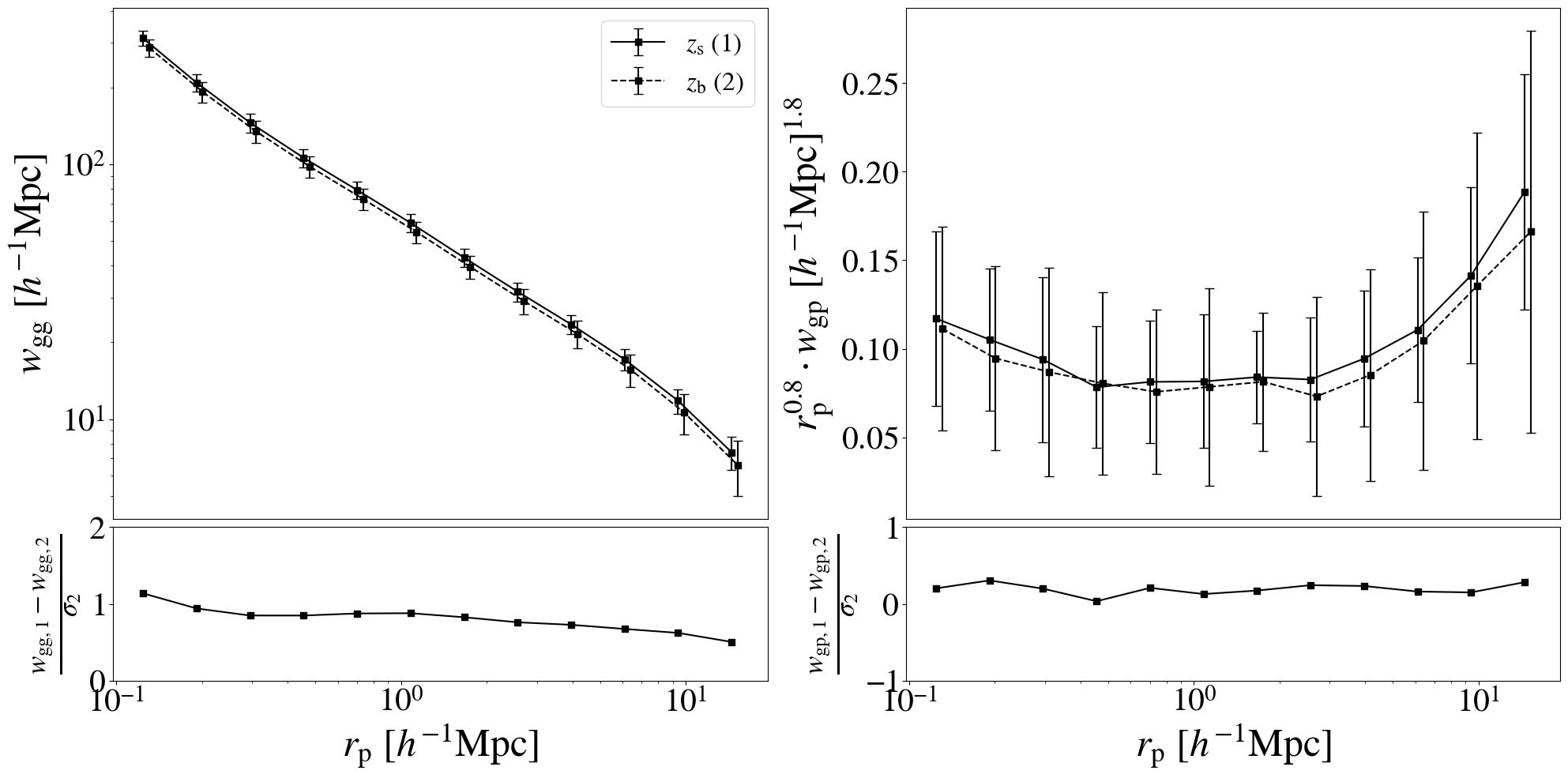}
    \caption{Comparison of the \gls{gc} (left) and \gls{ia} (right) measurements in MICE for different estimates of the radial positions, \glspl{spec-z} and \glspl{photo-z}. Both estimates have consistent measurements, with a maximum difference of less than 1 $\sigma$ at small projected separations in the \gls{gc} case and negligible for \gls{ia}, as seen in the bottom panel, which shows the difference between the measurements divided by the error in the \gls{photo-z} case.}
    \label{fig:MICE_comparison_wgg_wgp_photo_spec}
\end{figure*}

The top plots of Fig.~\ref{fig:MICE_comparison_wgg_wgp_photo_spec} show the comparison of the \gls{gc} (left) and \gls{ia} (right) signals obtained when using \glspl{spec-z} and \glspl{photo-z} in the MICE simulation. In turn, the bottom plots of the figure show the difference between the signals divided by the error from the ensemble covariance (bottom). On the one hand, for the \gls{gc} signal, the difference that results between both redshift estimates is of $\sim1\sigma$ at small scales, while it approaches $\sim 0.4\sigma$ at larger scales. Note how the signal measured using \glspl{spec-z} is slightly larger, as expected because of the smearing out caused by \gls{photo-z}. However, both results can be considered consistent. On the other hand, for the case of \gls{ia}, the difference in terms of $\sigma$ is almost negligible. This seems to indicate, at least in the MICE simulation, that \gls{ia} are less affected than \gls{gc} when using less precise redshift estimates. 

\begin{figure}
    \centering
    \includegraphics[width=0.48\textwidth]{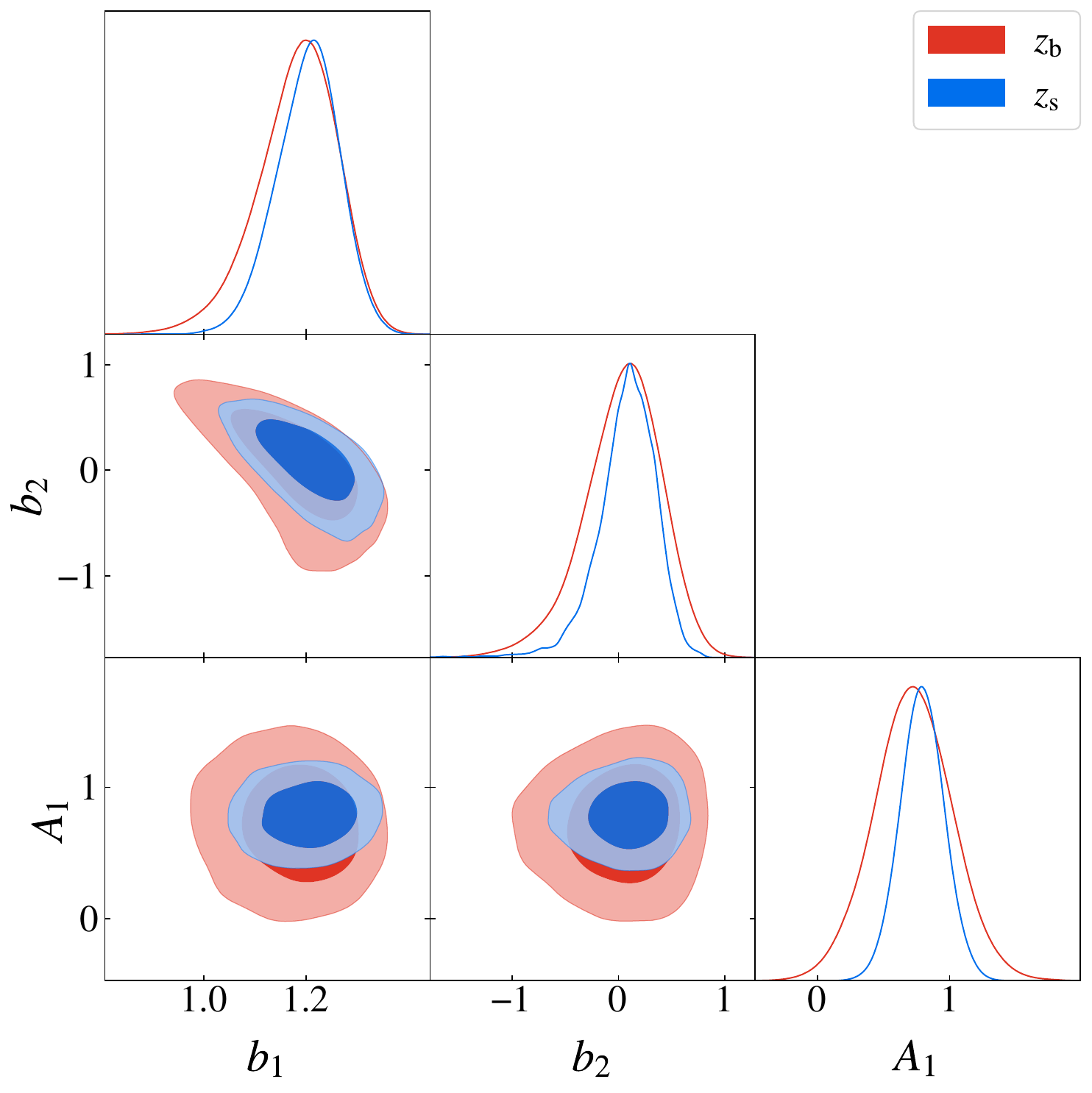}
    \caption{Galaxy bias and \gls{ia} parameters contour plots obtained from fitting the \gls{gc} and the \gls{ia} signals in MICE, using \glspl{photo-z} (red) and \glspl{spec-z} (blue).}
    \label{fig:MICE_contour_plot_wgg_photo_spec}
\end{figure}

Fig.~\ref{fig:MICE_contour_plot_wgg_photo_spec} shows the \gls{gc} bias and \gls{ia} parameters obtained when fitting both the spectroscopic and the photometric cases. Note that the constraining power in the case of employing \glspl{spec-z} is higher, given that the errors associated to spectroscopic correlation functions are lower. This leads to an enlargement of the posterior distributions in the case of photo-$z$, with respect to spec-$z$, of 80\% for $b_{1}$, 70\% for $b_{2}$ and 56\% for $A_{1}$. Still, the recovered values are consistent. This consistency test also serves as an indication that the pipeline we designed to model photometric correlation functions is correct.

Since the measurements in the data are also split in redshift bins, we repeated the tests in Fig.~\ref{fig:MICE_comparison_wgg_wgp_photo_spec} and Fig.~\ref{fig:MICE_contour_plot_wgg_photo_spec}, but dividing the redshift space in three bins. Even though we do not explicitly show the results for conciseness, the recovered galaxy bias and \gls{ia} parameters agree for the three redshift ranges defined. However, we note again that the constraints between the spectroscopic and the photometric scenarios are different in terms of area, with higher redshift bins leading to broader constraints in the case of \glspl{photo-z}, due to the decrease of \gls{photo-z} precision with redshift.

\section{Error estimation}\label{sec:error_estimation}

As discussed in Section~\ref{sec:estimators}, we estimate the errors for our correlation functions using the \gls{jk} method (eq.~\ref{eq:JK_error}). However, considering that the number of \gls{jk} regions is not very large, due to the limited area in \gls{paus}, we also need to validate our errors with the MICE simulation. In order to do that, we compare the errors derived from the \gls{jk} method with the errors from the ensemble covariance of the MICE simulation (eq.~\ref{eq:ensemble_covariance}), which are a representation of the true errors of the full octant:

\begin{equation}\label{eq:ensemble_covariance}
    \rm{Cov}^{\rm{ens}}_{\mathrm{ab}, i, j} = \sum_{k = 1}^{N} \frac{(w_{\mathrm{ab}, k, i} - \bar{w}_{\mathrm{ab},i})  (w_{\mathrm{ab}, k, j} - \bar{w}_{\mathrm{ab}, j})}{N},
\end{equation}
where $k$ corresponds to each combination of 3 \gls{paus}-like patches, $i$ and $j$ denote the $r_{\rm {p}}$ projected position of the data vector, N is the number of combinations and $\bar{w}_{\mathrm{ab},i}$ is the mean of the correlation function at the position $i$. We note that we compute the joint covariance of $w_{\rm{gg}}$ and $w_{\rm{gp}}$, so that $w_{\mathrm{ab}, k, i}$ and $\bar{w}_{\mathrm{ab},i}$ are the concatenation of $w_{\rm{gg}}$ and $w_{\rm{gp}}$. Given that each of the 3 \gls{paus}-like patches have an associated \gls{jk} covariance, we compare the ensemble covariance with the mean of the \gls{jk} covariances of the 60 3 \gls{paus}-like patches, which we name as “mean \gls{jk} covariance''.

\begin{figure*}
    \centering
    \includegraphics[width=0.48\textwidth]{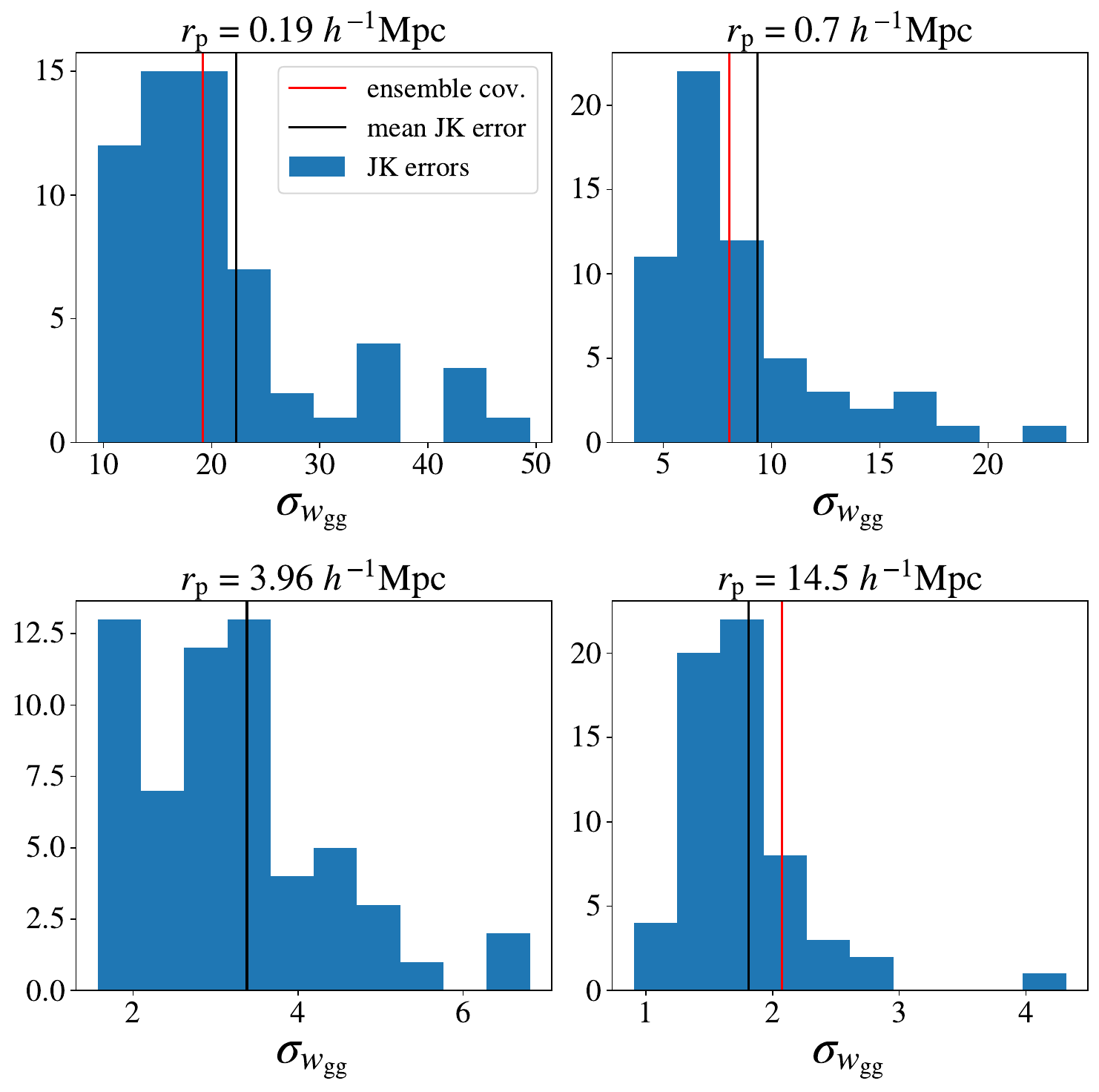}
    \includegraphics[width=0.48\textwidth]{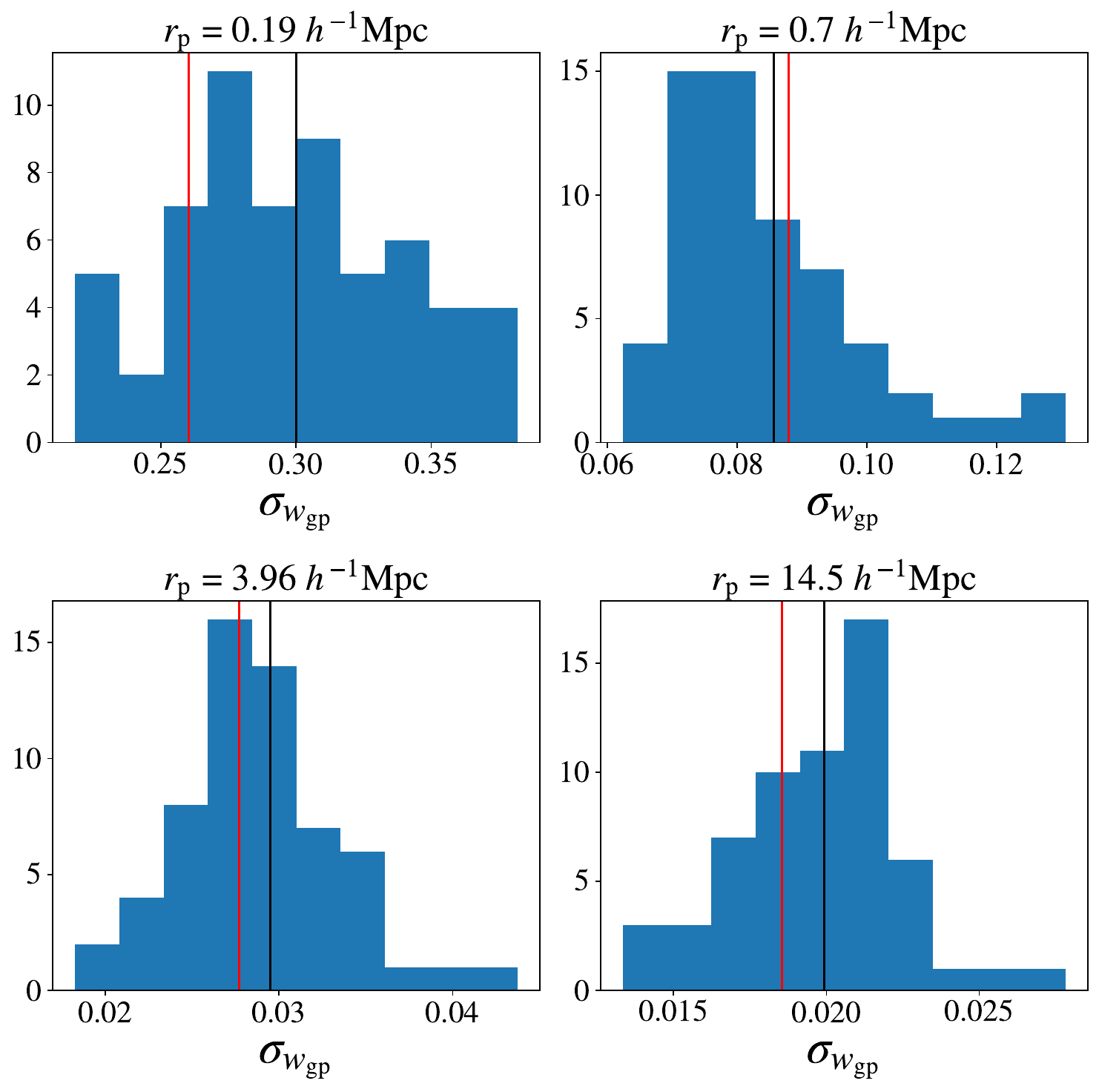}
    \caption{Distribution of the individual \gls{jk} errors of the 60 triplets of MICE \gls{paus}-like patches at different separations, for $w_{\rm{gg}}$ (4 left plots) and $w_{\rm{gp}}$ (4 right plots). Red and black vertical lines show the errors of the ensemble covariance and the mean \gls{jk} error at that separation, respectively.}
    \label{fig:MICE_error_distribution}
\end{figure*}

Fig.~\ref{fig:MICE_error_distribution} shows a comparison between the errors obtained from the ensemble and the mean \gls{jk} covariances, for the $w_{\rm{gg}}$ (left) and the $w_{\rm{gp}}$ (right) estimators. This is shown at 4 different separations in terms of $r_{\rm {p}}$, so as to capture the evolution as a function of the separation between galaxies. The histograms in the plots show the distribution of the individual \gls{jk} covariances (in terms of the square root of the diagonal of the covariance) for the 60 triplets of the MICE \gls{paus}-like patches. The vertical red and black lines show the errors of the ensemble and the mean \gls{jk} covariances, respectively. For the case of $w_{\rm{gg}}$, at small scales ($r_{\rm {p}}<1\mpc$), the mean \gls{jk} errors are overestimated with respect to the ensemble covariance. However, both errors agree at intermediate scales and the mean \gls{jk} errors are underestimated at large scales. In general, the distribution of the individual \gls{jk} errors for $w_{\rm{gg}}$ is located at values lower than the ensemble covariance, indicating that the mean \gls{jk} error may be driven by large individual \gls{jk} errors at small scales. For the case of $w_{\rm{gp}}$, the mean \gls{jk} errors are usually larger than the ensemble covariance, although with different variations. However, in all the $w_{\rm{gg}}$ and $w_{\rm{gp}}$ cases, the error estimates agree, since the ensemble covariance is well inside the range of the individual \gls{jk} errors.

\begin{figure*}
    \centering
    \includegraphics[width=0.98\textwidth]{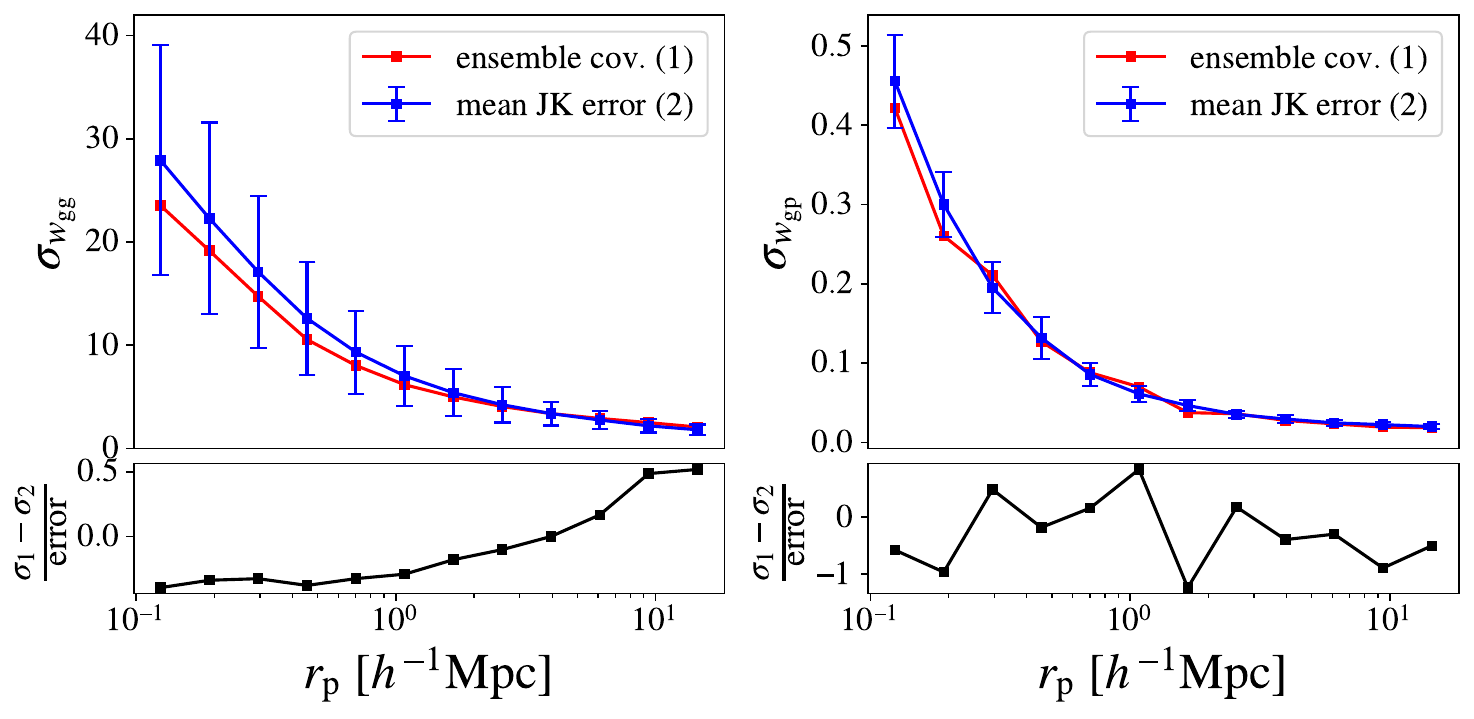}
    \caption{(Top): Errors of the ensemble covariance (red) and the mean \gls{jk} error (blue) as a function of the $r_{\rm {p}}$ separation. The error bars of the mean \gls{jk} error correspond to the standard deviation of the individual \gls{jk} errors with respect to the mean. (Bottom): Difference between the errors of the ensemble covariance and the mean \gls{jk} error, normalised by the standard deviation of the mean \gls{jk} error.}
    \label{fig:MICE_error_ratio}
\end{figure*}

The fact that both error estimates agree well can also be seen in Fig.~\ref{fig:MICE_error_ratio}. The top panel shows the errors of the ensemble covariance and the mean \gls{jk} error as a function of the $r_{\rm {p}}$ separation, for $w_{\rm{gg}}$ (left) and $w_{\rm{gp}}$ (right). The error bars depicted for the mean \gls{jk} error correspond to the standard deviation of the individual \gls{jk} errors with respect to the mean. The bottom panel shows the difference between the errors of the ensemble covariance and the mean \gls{jk} error, normalised by the standard deviation of the mean \gls{jk} error. From this figure, the agreement between both error estimates is verified for all $r_{\rm {p}}$ separations, with a maximum of a $\sim 1\sigma$ difference for the case of the $w_{\rm{gp}}$ estimator and below $\sim 0.5\sigma$ for $w_{\rm{gg}}$. As in Appendix~\ref{sec:Comparison_zb_zs}, we studied the comparison between the ensemble and the mean \gls{jk} covariances as a function of redshift. In all the cases, the errors of the mean \gls{jk} and the ensemble covariances present a similar behaviour as Fig.~\ref{fig:MICE_error_ratio}, with differences up to $\sim 1 \sigma$. 

\begin{figure}
    \centering
    \includegraphics[width=0.48\textwidth]{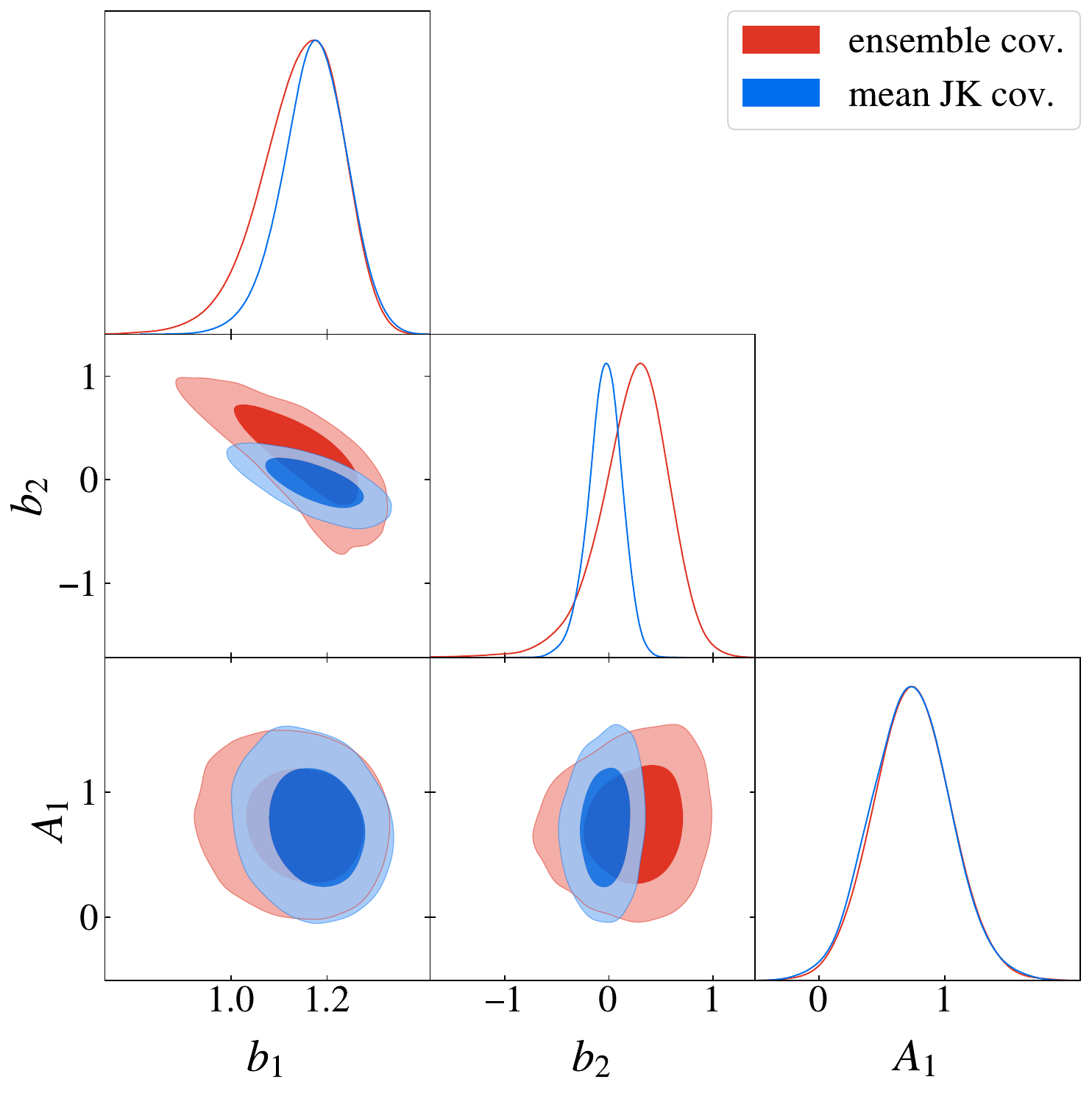}
    \caption{Galaxy bias and \gls{ia} parameters contours obtained when fitting the mean $w_{\rm{gg}} \cup w_{\rm{gp}}$ data vector with the ensemble covariance (red) and the mean \gls{jk} covariance (blue). The galaxy bias and \gls{ia} parameters obtained from both methods are very similar, specially for $A_{1}$.}
    \label{fig:MICE_contour_plot_wgg_different_errors}
\end{figure}

So far, we have shown how the diagonal terms of the mean \gls{jk} and the ensemble covariances relate to each other. Nevertheless, since the modelling of the \gls{paus} data uses the whole covariance matrix, we are also interested in how the off-diagonal terms affect the modelling of our observables. Besides, even though there is a maximum of $\sim 1\sigma$ differences between both error estimates, we are interested in how these may affect our constraints. In that sense, Fig.~\ref{fig:MICE_contour_plot_wgg_different_errors} shows the galaxy bias and the \gls{ia} constraints obtained when fitting the mean $w_{\rm{gg}} \cup w_{\rm{gp}}$ data vector obtained from the tiplets of MICE \gls{paus}-like patches, using the mean \gls{jk} and the ensemble covariances, where a \gls{svd} is performed to both covariance estimates. In the case of the mean \gls{jk} covariance, it is computed after individually performing the \gls{svd} to each 3 \gls{paus}-like patch \gls{jk} covariance. The galaxy bias constraints agree quite well for the $b_{1}$ parameter, while there are some discrepancies in $b_{2}$, although the constraints still overlap and this bias only represents a second-order effect on the \gls{gc}. As for the \gls{ia} parameter, the agreement between both covariance estimates is total. It is important to note that, in the \gls{paus} data, we only have one realisation of the results, while in the MICE simulation we have 60 triplets of \gls{paus}-like patches, from which we compute the mean \gls{jk} covariance. Thus, even though the mean \gls{jk} and the ensemble covariances yield very similar constraints, it might happen that the \gls{paus} data covariance differs more from the ensemble covariance, as is the case for some of the individual combinations seen in the histograms of Fig.~\ref{fig:MICE_error_distribution}. Nevertheless, in that same figure, the majority of the individual \gls{jk} errors agree quite well with the mean of the covariances, so the probability of significantly different errors in the data is expected to be low.

\section{Contaminants contribution to wgp}\label{sec:contaminants_contributions}

\begin{figure}
    \centering
    \includegraphics[width=0.4\textwidth]{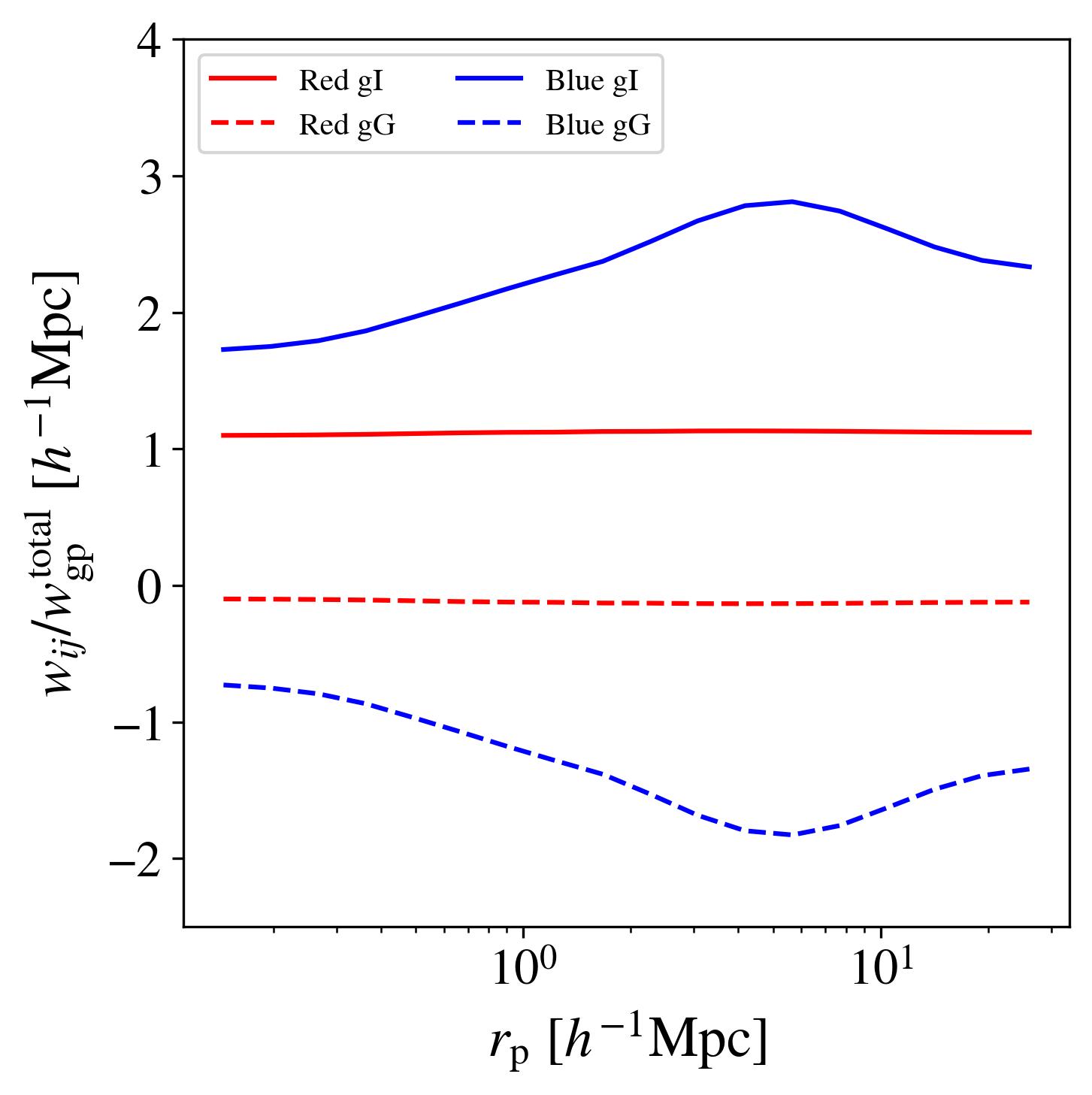}
    \caption{Fraction of contamination to $w_{\rm{gp}}$ for red and blue galaxies from the gI (solid line) and gG (dashed line) terms. The magnification terms, mI and mG, are not included here since they are consistent with 0.}
    \label{fig:PAUS_wgp_color_contaminants_fitted_with_NLA}
\end{figure}

Fig.~\ref{fig:PAUS_wgp_color_contaminants_fitted_with_NLA} shows how the contaminants affect the $w_{\rm{gp}}$ measurements, indicated as the ratio of each contaminant over the total $w_{\rm{gp}}$ signal. We consider the different terms that affect the position-shape correlation function, which are, as indicated in eq.~\ref{eq:contaminants_wgp}, gI, gG, mI and mG. Note that these terms do not need to have the same sign, which makes it possible for the ratio of one of the terms to be negative or positive and to be greater or less than 1. In fact, in general, the galaxy-shear term has opposite sign from the galaxy-intrinsic term, which leads to a decrease in the total value of $w_{\rm{gp}}$. This is a strong reason on why it is important to include contaminants in our modelling. The contaminants on $w_{\rm{gg}}$ are also included in the fitting of the data but are not shown here, since they contribute to the position-position correlation function at a sub-percent level. For both red and blue galaxies, the contaminant terms depending on magnification, mI and mG, are negligible, given that $\alpha(i_{\mathrm{AB}})$ for this case is close to 1, as seen in Table~\ref{tab:IA_cases}. For this reason, we do not include them in Fig.~\ref{fig:PAUS_wgp_color_contaminants_fitted_with_NLA}. The galaxy-intrinsic term, gI (solid line), is the most important one for both galaxy colours, as it is expected for measurements that are designed to focus on the \gls{ia} effect, since these consider objects close in redshift. The galaxy-shear term, gG (dashed line), is the main contaminant to the \gls{ia} measurements. In the case of red objects, the ratio of gG over $w_{\rm{gp}}$ is -0.2 (while the ratio of gI is 1.2), which accounts for a $\sim10\%$ of the signal. In the case of blue galaxies, the contribution from the galaxy-shear term is even higher, arriving to a $\sim35\%$, since the \gls{ia} signal is very low. Note that the percentages we described correspond to the ratio taken over all the $r_{\rm {p}}$ range. Nevertheless, the contribution from the gI and gG terms vary over that range, as seen in Fig.~\ref{fig:PAUS_wgp_color_contaminants_fitted_with_NLA}. The contribution of the contaminants to the other configurations studied in this paper was also assessed, although not shown here for conciseness. As in the separation by colour, the mI contribution is almost negligible, while the mG term has contributions larger than 1\% in the following cases, where $\alpha(i_{\mathrm{AB}})$ from Table~\ref{tab:IA_cases} is not close to 1: the blue brightest luminosity bin, with a 5.5\% contribution; the red and blue most massive bins, with 1\% and 5.7\% contributions, respectively; and the red and blue highest redshift bins, with 2.5\% and 5.4\% contributions, respectively. As for the main contaminant term, gG, it ranges between $\sim5\%-20\%$ for red objects and $\sim15\%-65\%$ for blue objects, where the larger contribution comes again from the fact that the \gls{ia} signal for blue galaxies is low.

\section{Scale cut analysis}\label{sec:scale_cut_analysis}

Here, we study the dependence on the scale cuts when fitting the $w_{\rm{gg}} \cup w_{\rm{gp}}$ data vector with the non-linear galaxy bias, for \gls{gc}, and the \gls{nla} model, for \gls{ia}. As indicated in Section~\ref{sec:likelihood}, we set $r_{p, \mathrm{min}}=2.0\mpc$ for both $w_{\rm{gg}}$ and $w_{\rm{gp}}$. However, in this appendix, we explore the scale cuts further by setting different values of $r_{p, \mathrm{min}}$ in the case of $w_{\rm{gp}}$, while fixing $r_{p, \mathrm{min}}=2.0\mpc$ for $w_{\rm{gg}}$. The justification of this approach is that the non-linear galaxy bias breaks down below $r_{p, \mathrm{min}}=2.0\mpc$. Nevertheless, in the case of the \gls{nla} model, previous studies set $r_{p, \mathrm{min}}=6.0\mpc$ based on observations, while Paviot et al. (in prep.) shows that this mainly depends on the fact that linear galaxy bias is used. Given the high number density of \gls{paus}, which allows correlations to be measured down to smaller scales than with other spectroscopic surveys, we want to explore the range of validity of the \gls{nla} model by exploring different scale cuts. The scale cuts we test for $w_{\rm{gp}}$ range from the common $r_{p, \mathrm{min}}=6.0\mpc$ down to the uncommon $r_{p, \mathrm{min}}=0.1\mpc$, which is the minimum projected separation defined in this work. We analyse the variations in $\chi_{\nu,\mathrm{fit, SVD}}^{2}$ (eq. \ref{eq:reduced_chi2_SVD}), together with the change in the $A_{1}$ constraints, as a function of $r_{p, \mathrm{min}}$.

\begin{figure}
    \centering
    \includegraphics[width=0.4\textwidth]{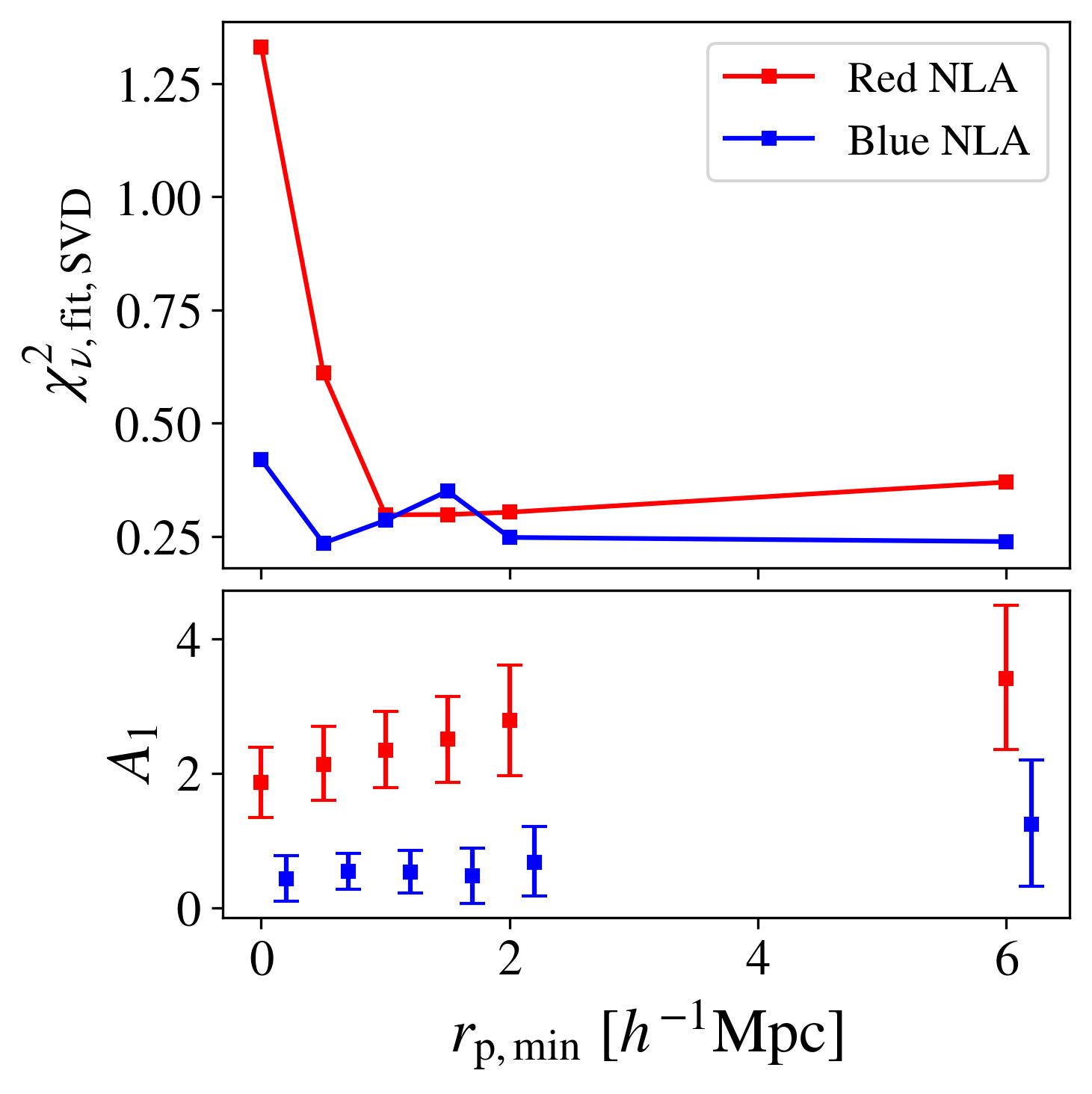}
    \caption{$\chi_{\nu,\mathrm{fit, SVD}}^{2}$ (top) and $A_{1}$ (bottom) as a function of the minimum $r_{\rm {p}}$ separation considered in the modelling of $w_{\rm{gp}}$ for red and blue galaxies.}
    \label{fig:chi2_fit_NLA_color}
\end{figure}

Fig.~\ref{fig:chi2_fit_NLA_color} shows the dependence on the scale cuts when splitting the \gls{paus} samples in red and blue galaxies. Both types of galaxies are evaluated at the same $r_{p, \mathrm{min}}$ values, but the $A_{1}$ amplitudes for blue galaxies are plotted at slightly larger values, to avoid overlapping. We highlight that we focus on the evolution of the $\chi_{\nu,\mathrm{fit, SVD}}^{2}$ value, rather than on its absolute value, given the remark made at the end of Section~\ref{sec:divistion_color_redshift} about the values below and above 1 in the $\chi_{\nu,\mathrm{fit, SVD}}^{2}$. For the case of red objects, the $\chi_{\nu,\mathrm{fit, SVD}}^{2}$ starts increasing at $r_{p, \mathrm{min}}<1.0\mpc$, from a stable value of $\chi_{\nu,\mathrm{fit, SVD}}^{2}~\sim 0.35$ up to $\sim1.4$. The values of $A_{1}$ continuously decrease as we include smaller $r_{p, \mathrm{min}}$ values, while its error bars are also reduced, given the larger number of available data points. In the case of blue objects, the $\chi_{\nu,\mathrm{fit, SVD}}^{2}$ is quite stable, but this is most likely due to the lack of signal for these types of galaxies.

We have also analysed the evolution of $\chi_{\nu,\mathrm{fit, SVD}}^{2}$ for the other cases included in this paper, although not shown here for conciseness. In general, we find that it is not possible to reach values of $r_{p, \mathrm{min}}<1.0\mpc$ without the $\chi_{\nu,\mathrm{fit, SVD}}^{2}$ increasing. Nevertheless, in some cases, the fact of reducing $r_{p, \mathrm{min}}$ allows some points with high noise to enter in the $\chi_{\nu,\mathrm{fit, SVD}}^{2}$, increasing its value for that particular $r_{p, \mathrm{min}}$ and then reducing it when further decreasing $r_{p, \mathrm{min}}$, since a point with less noise enters the computation. Thus, it is not straightforward to validate the use of low $r_{p, \mathrm{min}}$ values with this approach.


\bsp	
\label{lastpage}
\end{document}